\documentclass[12pt]{iopart}
\usepackage{amssymb}
\usepackage{latexsym}
\usepackage{bm}
  \usepackage{graphicx}
  \usepackage{cancel}

\newcommand{\la}{\left\langle}
\newcommand{\ra}{\right\rangle}
\newcommand{\be}{\begin{equation}}
\newcommand{\ee}{\end{equation}}
\newcommand{\bse}{\begin{subequations}}
	\newcommand{\ese}{\end{subequations}}
\newcommand{\bea}{\begin{eqnarray}}
\newcommand{\eea}{\end{eqnarray}}
\newcommand{\ba}{\begin{array}}
\newcommand{\ea}{\end{array}}
\newcommand{\ben}{\begin{enumerate}}
	\newcommand{\een}{\end{enumerate}}

\begin{document}

\title{Variable energy flux in turbulence}

\author{Mahendra K. Verma}

\address{Department of Physics, Indian Institute of Technology Kanpur, Kanpur 208016, India}
\ead{mkv@iitk.ac.in}
\vspace{10pt}
\begin{indented}
\item[]May 2020
\end{indented}

\begin{abstract}
	In three-dimensional hydrodynamic turbulence forced at large length scales, a constant energy flux $ \Pi_u $ flows from large scales to intermediate scales, and then to small scales.  It is well known that for multiscale energy injection and dissipation, the  energy flux $\Pi_u$ varies with scales.  In this review we describe this principle and show how this general  framework is useful for describing a variety of turbulent phenomena.  Compared to Kolmogorov's spectrum, the energy spectrum steepens in turbulence involving quasi-static  magnetofluid,  Ekman friction,  stable stratification, magnetohydrodynamics, and solution with dilute polymer.  However, in turbulent thermal convection,  in unstably stratified turbulence such as Rayleigh-Taylor turbulence, and in shear turbulence, the energy spectrum has an opposite behaviour due to  an increase of energy flux with wavenumber.   In addition, we briefly describe the role of variable energy flux in quantum turbulence, in binary-fluid turbulence including time-dependent Landau-Ginzburg and Cahn-Hillianrd equations, and in  Euler turbulence. 
\end{abstract}


 \section{Introduction}
\label{sec:Introduction}

Turbulence is observed in most natural flows, for example, in atmospheres and interiors of planets and stars, in oceanic flows, and in stellar and galactic winds.   Many engineering flows, as in air conditioners and combustion engines, as well as most kitchen flows are turbulent.  These complex flows have  multiple components that could be a combination of velocity, temperature, density, and magnetic fields.  The complexities of above turbulent flows appear daunting, yet, the mathematical models and tools developed over the last two centuries provide reasonable understanding  of such flows.  In this review article we will describe one such tool called {\em variable energy flux}.

The nonlinear interactions among the Fourier modes of a turbulent flow cause energy exchange among the modes~\cite{Kolmogorov:DANS1941Structure,Kolmogorov:DANS1941Dissipation,Monin:book:v1,Monin:book:v2,Leslie:book,Lesieur:book:Turbulence,McComb:book:Turbulence,Frisch:book,Davidson:book:Turbulence,Sagaut:book,Pope:book, Orszag:CP1973,Mccomb:RPP1995,Verma:book:ET}.  In one of the pioneering works, Kolmogorov~\cite{Kolmogorov:DANS1941Structure,Kolmogorov:DANS1941Dissipation}   showed that when hydrodynamic turbulence is forced at large scales, the large-scale kinetic energy is transferred to intermediate scales, called \textit{inertial range}, and then to small scales.  The flow is \textit{homogeneous} and \textit{isotropic}  in the inertial range.  For hydrodynamic turbulence, we define {\em kinetic energy flux} $\Pi_u(k_0)$, which is the net energy transfer from the Fourier modes of a wavenumber sphere of radius $k_0$ to the Fourier modes outside the sphere.   Kolmogorov~\cite{Kolmogorov:DANS1941Structure,Kolmogorov:DANS1941Dissipation}   argued that in the inertial range,   an absence of forcing and weak dissipation  leads to a constant $\Pi_u(k_0)$.  In addition, the inertial-range  kinetic energy spectrum, $E_u(k)$, varies as $k^{-5/3}$, also called {\em Kolmogorov's energy spectrum}~\cite{Leslie:book,Lesieur:book:Turbulence,McComb:book:Turbulence}.

Kolmogorov's theory of turbulence describes the energy spectrum and flux of hydrodynamic turbulence without any external force in the intermediate scales.   However  this assumption is not valid for many complex flows where the forcing and/or dissipation are active in the  intermediate scales.  For example, gravity acts at all scales in buoyant flows (stably stratified turbulence and thermal convection);  the Lorentz force  acts at all scales in MHD turbulence;    Ekman friction dissipates kinetic energy at all scales.   Due to the additional forcing and/or dissipation, the inertial-range energy flux of the  such flows varies with wavenumber.  In addition, the inertial-range kinetic energy spectrum  differs from Kolmogorov's $k^{-5/3}$ spectrum.   Interestingly, complex variations in energy spectrum and flux can be quantified using an equation for the variable energy flux:  $d\Pi_u(k) /dk = \mathcal{F}_u(k)-D_u(k)$, where $\mathcal{F}_u(k)$ is the kinetic energy injection rate by the external force at wavenumber $k$, and $D_u(k)$ is the dissipation rate at $k$~\cite{Frisch:book,Davidson:book:Turbulence,Pope:book,Verma:book:ET}.  The framework of {\em variable energy flux}, which is the theme of this review, helps us understand a wide range of turbulent phenomena. We introduce these topics in this section and detail them in subsequent chapters.

\begin{figure}[htbp]
	\begin{center}
		\includegraphics[scale = 0.7]{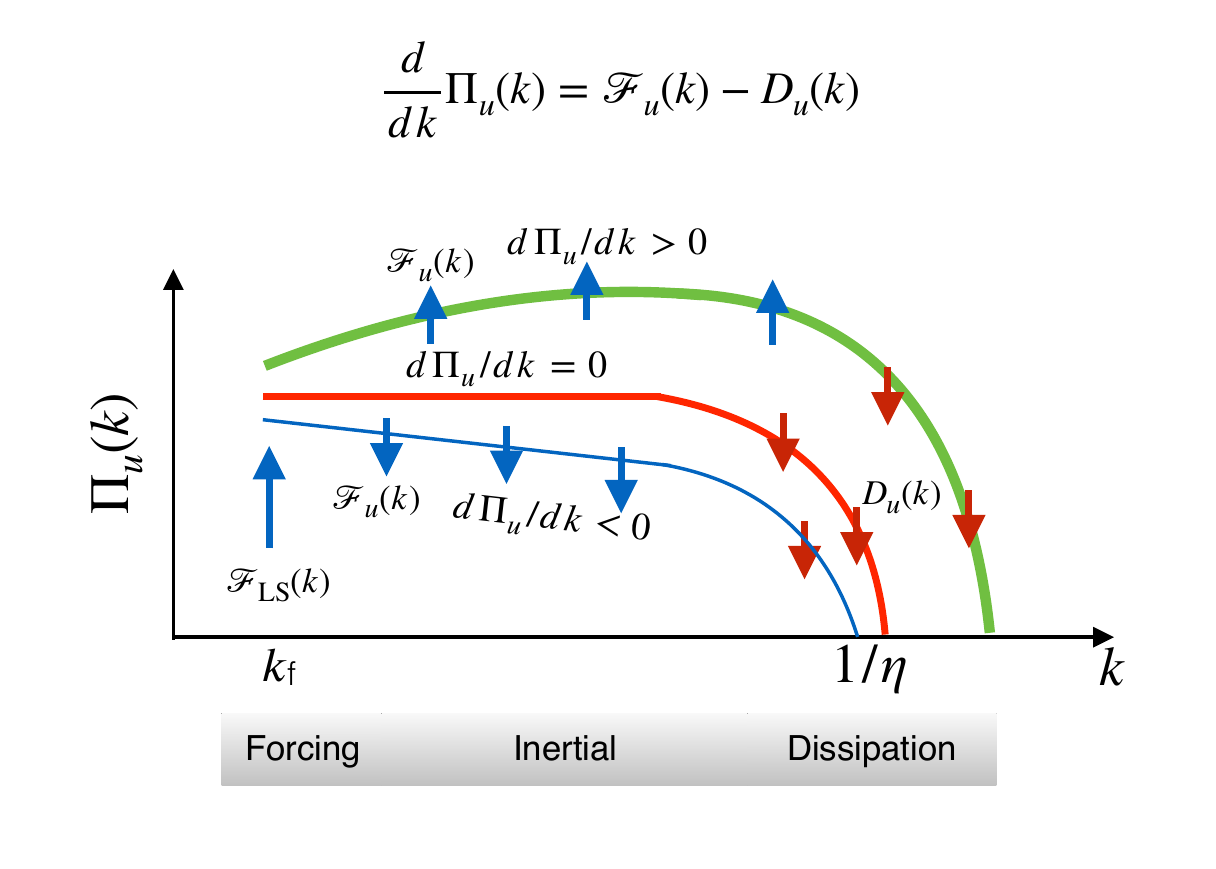}
	\end{center}
	\caption{ A figure illustrating variable energy flux. The blue and red arrows depict $\mathcal{F}_u(k)$ and $D_u(k)$ respectively. The red curve represents Kolmogorov's model for which $\Pi_u(k) = $ constant in the inertial range because $\mathcal{F}_u(k) =0$ and $D_u(k) = 0$.  The green curve represents the case when $\mathcal{F}_u(k) > 0$ and $d\Pi_u(k)/dk > 0$, while the blue curve represents flows with $\mathcal{F}_u(k) < 0$ and $d\Pi_u(k)/dk < 0$.}
	\label{fig:intro}
\end{figure}

Gravity acts at all scales in stably stratified turbulence and  in thermal convection and  generates variable kinetic energy flux in these systems.   For the stably stratified turbulence with moderate stratification,  Bolgiano~\cite{Bolgiano:JGR1959} and Obukhov~\cite{Obukhov:DANS1959} showed that $\mathcal{F}_u(k) < 0$ and $d\Pi_u(k)/dk < 0$ due to a conversion of kinetic energy to potential energy by buoyancy.  In particular, $\Pi_u(k) \sim k^{-4/5}$ and $E_u(k) \sim k^{-11/5}$ (different from Kolmogorov's $k^{-5/3}$ spectrum).   See  Fig.~\ref{fig:intro} for an illustration.

The physics of thermal convection, however, is quite different from the stably stratified turbulence even though the equations for the two systems are the same.   This is because thermal convection is unstable, while  stably stratified turbulence is stable.  In the inertial range of turbulent convection, thermal plumes drive the  velocity field. Hence,  $\mathcal{F}_u(k) > 0$ leading to  $d\Pi_u(k) /dk > 0$.   See  Fig.~\ref{fig:intro} for an illustration.  Detailed studies, however, show that for small and moderate Prandtl numbers, $\mathcal{F}_u(k)$ is primarily concentrated at small wavenumbers, as in Kolmogorov's theory of hydrodynamic turbulence.  Consequently,  the kinetic energy spectrum and flux of turbulent convection are similar to those predicted by Kolmogorov's turbulence theory~\cite{Kumar:PRE2014,Verma:NJP2017,Verma:book:BDF}.  These observations indicate the usefulness of variable energy flux in modelling buoyant flows.

Variable energy flux is also useful for describing magnetohydrodynamic (MHD) turbulence and dynamo.  Here, the Lorentz force, which is active at all scales, transfers  energy from the velocity field to the magnetic field.  These energy transfers are responsible for the enhancement of  magnetic field in astrophysical objects via dynamo mechanism~\cite{Moffatt:book,Brandenburg:PR2005,Verma:book:ET}.  These  transfers also lead to a reduction in kinetic energy flux and an enhancement of magnetic energy flux with wavenumber~\cite{Verma:book:ET}.  
In quasi-static MHD turbulence, Joule dissipation is significant at all scales. Consequently,  the inertial-range  kinetic energy flux decreases with $k$, and $E_u(k)$ is steeper than Kolmogorov's $ k^{-5/3} $ spectrum~\cite{Moreau:book:MHD,Knaepen:ARFM2008,Verma:ROPP2017}.   Anas and Verma~\cite{Anas:PRF2019} showed that the variable energy formalism successfully explains the spectral steepening observed in numerical simulations and experiments~\cite{Moreau:book:MHD,Knaepen:ARFM2008,Verma:ROPP2017}.    For very strong magnetic field, the Joule dissipation  steepens $E_u(k)$ even further and yields an exponential spectrum, which is $ \exp(-c k) $, where $ c $ is a constant.

 In shear turbulence,  shear acts at small wavenumbers and injects energy to the flow.  This injected energy leads to an increase in  the kinetic energy flux with $k$.  This variation in the kinetic energy flux may be responsible for the $1/f$ noise reported for many systems~\cite{Tchen:JRNBS1952,Matthaeus:PRL1986,Pereira:PRE2019}.  On the other hand, Ekman friction acts at all scales and depletes the kinetic energy flux in the inertial range  leading to a steeper kinetic energy spectrum than  $k^{-5/3}$~\cite{Boffetta:EPL2005,Verma:EPL2012}.  

In the viscous range of hydrodynamic turbulence, $d\Pi_u(k)/dk = -2\nu k^2 E_u(k)$, where $\nu$ is the kinematic viscosity.  The viscous dissipation leads to steep decline  in   $\Pi_u(k)$ and  $E_u(k)$. Reseachers  \cite{Pao:PF1965,Chen:PRL1993,Martinez:JPP1997,Pope:book} have attempted to model the energy spectrum in this range.  In particular,  Pao~\cite{Pao:PF1965} derived that in the inertial-dissipation range of 3D hydrodynamic turbulence,  $ \Pi_u(k) $  and $ E_u(k) k^{5/3} $ vary as $ \exp\{ -c (k/k_d )^{4/3} \}$, where $ c $ is a constant, and $ k_d $ is the Kolmogorov wavenumber.     Pao's model for the 3D inertial-dissipation range has been extended to 2D hydrodynamic turbulence~\cite{Verma:book:ET,Gupta:PRE2019}.   Falkovich~\cite{Falkovich:PF1994} and Verma and Donzis~\cite{Verma:JPA2007} showed that  the energy flux plays an important role in the bottleneck effect.   In this review we argue that the bottleneck effect may  possibly be due to a sudden suppression of the energy flux in the dissipation range.    

In most complex flows, one or several secondary fields are coupled to the velocity field.  Some examples of secondary fields: the density field in buoyant flows, the temperature field in thermal convection,  the  magnetic field in MHD turbulence, and the  conformation tensor of polymers in polymeric flows.  The nonlinear term  associated with the secondary field also induces scalar energy transfer or scalar energy flux.   For example, the potential energy fluxes  of stably stratified turbulence and turbulent thermal convection are constant.  In addition, a coupling between the velocity field with the secondary field often yields  energy exchange between the velocity field and the secondary field, as well as variability in the secondary energy flux~\cite{Dar:PD2001,Verma:PR2004,Alexakis:PRE2005}.  This phenomena is related to the turbulent drag reduction in  polymeric turbulence~\cite{deGennes:book:Intro,Valente:JFM2014,Valente:PF2016}  and in MHD turbulence~\cite{Verma:PP2020}.

The enstrophy ($ \int d{\bf r} \frac{1}{2}|\bm{\omega}|^2 $, where $ \bm{\omega} $ is the vorticity field) and kinetic helicity ($ \int d{\bf r}  \frac{1}{2}| [{\bf u} \cdot \bm{\omega} ] $) are important quantities of hydrodynamic turbulence. The fluxes of these quantities exhibit interesting properties.  For example, the enstrophy flux has a similar structure as those of  kinematic dynamo~\cite{Moffatt:book,Verma:book:ET}.  In MHD turbulence, the flux of magnetic helicity  too exhibit interesting properties~\cite{Brandenburg:PR2005}.

Many turbulent systems, including buoyancy-driven turbulence, MHD turbulence, and rotating turbulence are anisotropic~\cite{Davidson:book:TurbulenceRotating,Verma:book:ET}. Under strong external field, stably stratified turbulence, MHD turbulence, and rotating turbulence become quasi-two-dimensional with $|{\bf u}_\perp| \gg u_\parallel$, where ${\bf u}_\perp $ and $ u_\parallel$ are the perpendicular and parallel components of the velocity field in relation to the anisotropy direction.   On the other hand, in thermal convection,  $|{\bf u}_\perp| \ll u_\parallel$.    In these flows, the pressure acts as a mediator for the energy exchange between the parallel and perpendicular components of the velocity field~\cite{Verma:ROPP2017,Verma:book:ET,Sharma:PF2018}.   The energy fluxes of ${\bf u}_\perp $ and $ u_\parallel$   provide useful insights into the anisotropic nature of such flows.    The secondary fields too  have similar anisotropic fluxes, but these quantities have not been computed so far.

Quantum systems, such as superfluids and Bose-Einstein condensate, too exhibit turbulent behaviour for a parameter range.  Researchers have studied energy spectra and fluxes of these systems.  For example, in Helium-4, normal and superfluid components interact with each other that leads to  variability in their energy fluxes (e.g., \cite{Lvov:JLTP2006,Roche:EPL2009,Wacks:PRB2011,Madeira:ARCMP2020} and references therein).   The scenario is more complicated in Helium-3 that lacks normal component; here phonon coupling at small scales is expected to provide the dissipation~\cite{Krstulovic:PRE2011_GP,Tsatsos:PR2016}.   

The energy flux is a useful diagnostic tool for other nonequilibrium systems as well. For example, in  binary-mixture turbulence,  the energy flux provides valuable insights into the field dynamics, especially  phase separation and coarsening~\cite{Berti:RRL2005,Perlekar:SR2017}.   Researchers have also employed time-dependent Ginzburg-Landau and Cahn-Hilliard  equations to model the coarsening process~\cite{Cahn:JCP2004,Puri:book_edited}, where the energy flux is proving to be a very useful tool~\cite{Berti:RRL2005,Perlekar:SR2017}.

 Lee ~\cite{Lee:QAM1952} and  Kraichnan~\cite{Kraichnan:JFM1973} showed that inviscid hydrodynamic turbulence $ (\nu=0) $ exhibits equilibrium behaviour (also called \textit{absolute equilibrium}).  The energy flux  for this case  vanishes due to the detailed energy balance among the Fourier modes~\cite{Frisch:book}.  The evolution of such systems depends quite critically on the initial condition.  For example, Cichowlas et al.~\cite{Cichowlas:PRL2005} showed that Taylor-Green vortex as an initial condition yields a mixture of $ k^{-5/3} $ and $ k^2 $ spectra.  On the contrary, for white noise as initial condition, the system exhibits $ k^2 $ spectrum for the whole range of wavenumbers~\cite{Verma:PTRSA2020,Verma:arxiv2020_equilibrium}.  The former system exhibits variable energy flux, but the latter system (equilibrium configuration) has no energy flux.  The absolute equilibrium theory of hydrodynamics has been extended to MHD turbulence~\cite{Frisch:JFM1975,Stribling:PP1991}, quantum  turbulence~\cite{Davis:PRL2001,Krstulovic:PRL2011}, Burger turbulence~\cite{Frisch:PRL2008}, and other forms of turbulence~\cite{Frisch:book,Lesieur:book:Turbulence}.

So far, the  energy flux has been defined for  the wavenumber space.  Note, however, that Kolmogorov~\cite{Kolmogorov:DANS1941Structure,Kolmogorov:DANS1941Dissipation}  formulated a relationship between the energy flux and the real-space third-order structure function, which is related to the velocity difference between two points.   This theory of Kolmogorov  has been generalized to passive scalar turbulence~\cite{Yaglom:DANS1949}, MHD turbulence~\cite{Politano:EPL1998}, rotating turbulence, thermal convection~\cite{Ching:book}, etc.   Biferale and Procaccia~\cite{Biferale:PR2005}, Arad et al.~\cite{Arad:PRL1998}, and  Danaila et al.~\cite{Danaila:PD2012b}  have attempted to generalize the above  Kolmogorov's theory to anisotropic systems. Since structure functions and associated energy flux are covered in great detail in many books~\cite{Frisch:book,Lesieur:book:Turbulence} and reviews~\cite{Sreenivasan:RMP1999,Sreenivasan:ARFM1991}, they are not covered in detail in this review.

The aforementioned turbulent systems have been studied in great detail in the past, including in books~\cite{Monin:book:v1,Monin:book:v2,Leslie:book,Lesieur:book:Turbulence,McComb:book:Turbulence,Frisch:book,Davidson:book:Turbulence,Sagaut:book,Pope:book,Verma:book:ET} and review articles~\cite{Alexakis:PR2018}.  The equation of variable energy flux, $d\Pi_u(k) /dk = \mathcal{F}_u(k)-D_u(k)$, too appears in several textbooks, for example~\cite{Lesieur:book:Turbulence,McComb:book:Turbulence,Frisch:book,Davidson:book:Turbulence}.     In this review article, we  illustrate how various turbulent phenomena can be connected  via variable energy flux.  We also present the scaling laws and energy fluxes of many turbulent systems thematically in the framework of variable energy flux.  This perspective  provides valuable and unique insights.  For example, using variable energy flux, it has been  shown that the  dynamics of turbulent thermal convection is very different from that of stably stratified turbulence, contrary to a popular view that the Bolgiano-Obukhov  scaling for  stably stratified turbulence \cite{Bolgiano:JGR1959,Obukhov:DANS1959} also works for turbulent thermal convection~\cite{Lvov:PRL1991,Lvov:PD1992,Rubinstein:NASA1994}.  The contrast in the  energy fluxes of the two systems played a key role in the resolution of this critical puzzle.

The structure of the review is as follows.   Section~\ref{sec:flux} introduces  the energy transfers  and energy flux in hydrodynamic turbulence.  The formalism of variable energy flux  is presented in Sec.~\ref{sec:VF}.  In this section, we present various examples,  including variable energy fluxes in the dissipation range of hydrodynamic turbulence, in quasi-static MHD turbulence, and in shear turbulence.  In Sec.~\ref{sec:secondary} we derive the energy flux for the secondary field that is advected by velocity field and show how it could become variable when a multiscale force is applied to  the secondary field. Here, we also describe the  energy exchange between the secondary field and the velocity field, as well as those between the field components in anisotropic turbulence.   Section \ref{sec:buoyant} contains discussions on  turbulence in  stably stratified flows and thermal convection. Sections~\ref{sec:mhd} describes variable  energy fluxes in MHD turbulence and in a turbulent flow with dilute polymer.  Here, we discuss several exact relations among the fluxes of MHD turbulence.   Sections~\ref{sec:enstrophyHk} describes the  fluxes associated with enstrophy and kinetic helicity,     while Sec.~\ref{sec:2D} contains discussions on 2D turbulence.   Variable energy fluxes in dissipation-less systems are discussed  in Sec.~\ref{sec:Euler}, while those in quantum turbulence and binary-mixture turbulence are discussed in Sec.~\ref{sec:QT}.  Section~\ref{sec:struct_fn} summarises Kolmogorov's four-fifth law for hydrodynamic turbulence, as well as the laws for passive-scalar turbulence and MHD turbulence.   We conclude in Sec.~\ref{sec:conclusions}.

 \section{Energy flux in hydrodynamics}
\label{sec:flux}

In this section, we introduce the kinetic energy flux of hydrodynamics.  We start with the basic equations of hydrodynamics in real and Fourier spaces.

\subsection{Basic equations of hydrodynamics}

The Navier-Stokes (NS) equations given below describe an incompressible flow~\cite{Landau:book:Fluid}:
\bea
\frac{\partial {\bf u}}{\partial t} + ({\bf u} \cdot \nabla) {\bf u}  = 
- \frac{1}{\rho} \nabla p +  {\bf F}_u  + {\bf F}_\mathrm{LS}+  \nu \nabla^2 {\bf u}, \label{eq:gov:NS_ur} \\
 \nabla \cdot \mathbf{u} =  0, \label{eq:gov:div_u_eq0}
\eea
where  ${\bf u}({\bf r},t)$ is the velocity field, $p({\bf r},t)$ is the pressure field, ${\bf F}_\mathrm{LS}$ is the large-scale external force,  ${\bf F}_u$ is the force field such as buoyancy, and $\nu$ is the kinematic viscosity.  Under incompressible limit, the fluid density $\rho$ can be treated as a constant.   In this review, without loss of generality, $\rho$ is taken to be unity.   We distinguish ${\bf F}_\mathrm{LS}$ and  $ {\bf F}_u$ to clearly demarcate  the energy transfers from these forces.  Note that ${\bf F}_u$ could be a function of the velocity field; for example, Ekman friction is $-\alpha {\bf u}$, where $\alpha$ is a constant.  For an incompressible flow, the pressure field is determined using
  \bea 
p = -\nabla^{-2} \nabla\cdot \left[ (\mathbf{u}\cdot\nabla)\mathbf{u} - {\bf F}_u - {\bf F}_\mathrm{LS}  \right].
\label{eq:gov:pressure_Poisson}
\eea
The ratio of the nonlinear term $ ({\bf u} \cdot \nabla) {\bf u}$ and the viscous term $ \nu \nabla^2 {\bf u}$ is called {\em Reynolds number} $\mathrm{Re}$, which is  $ {UL}/{\nu} $, where $U,L$ are the large-scale velocity and length  respectively.

In three-dimensional (3D) inviscid ($ \nu=0 $)  hydrodynamics, for a periodic or vanishing  boundary condition, the total kinetic energy, $E_u = \frac{1}{2}  \int d{\bf r} u^{2}$, and the total kinetic helicity, $H_K = \frac{1}{2}  \int d{\bf r} ( {\bf u} \cdot \bm{\omega})$, are conserved~\cite{Lesieur:book:Turbulence, Frisch:book, Davidson:book:Turbulence}.   Here, $ \bm{\omega} = \nabla \times \bf{u} $ is the vorticity field.    In two-dimensional (2D) hydrodynamics, the total kinetic energy and the total enstrophy $\frac{1}{2} \int d{\bf r}  |\bm{\omega}|^2$ are conserved. Note that the physics of turbulence in 2D and 3D are quite different.    In this review, we will  focus on the fluxes of kinetic energy and associated secondary energy in 3D flows.  The fluxes of other quantities, such as enstrophy and kinetic helicity, will be discussed  briefly. 

The multiscale energy transfers and fluxes are conveniently described  using the velocity Fourier modes.   For compactness, we denote the Fourier transform of ${\bf u(r)}$ using  ${\bf u(k)}$; here ${\bf r}$ and ${\bf k}$ denote  the real  and Fourier space coordinates respectively.  In a Fourier space convolution, the other wavenumbers are denoted by ${\bf p}$ and ${\bf q}$.    The wavenumbers  are discrete for  a flow confined in a finite box, but they form a continuum for a flow in an infinite box. 

The Navier-Stokes equations are transformed in Fourier space as~\cite{Lesieur:book:Turbulence, Frisch:book, Davidson:book:Turbulence,Verma:book:BDF}
\bea
\frac{d}{d t}  {\bf u} (\mathbf{k}) + {\bf N}_u ({\bf k})
= -i {\bf k} p (\mathbf{k}) + {\bf F}_u({\bf k})  + {\bf F}_\mathrm{LS}({\bf k})-  \nu k^{2}  {\bf u}(\mathbf{k}) , \label{eq:gov:uk}\\
{\bf k\cdot u} (\mathbf{k}) = 0 \label{eq:gov:k_uk_zero},
\eea    
where
\bea 
{\bf N}_u ({\bf k}) & =& i  \sum_{\bf p}  {\bf \{ k \cdot u(q) \} u(p) } \label{eq:fourier:Nuk} 
\eea
is the Fourier transform of the nonlinear term ${\bf (u \cdot \nabla) u}$. Here ${\bf q=k-p}$.   The equation for the pressure mode $p(\mathbf{k})  $ is
\bea 
p(\mathbf{k}) =  \frac{i}{k^2}  {\bf k} \cdot \{ {\bf N}_u({\bf k}) - {\bf F}_u({\bf k}) -  {\bf F}_\mathrm{LS}({\bf k}) \}.
\label{eq:gov:Pk}
\eea

We define {\em modal kinetic energy} for wavenumber ${\bf k}$ as $ E_u({\bf k}) =  \frac{1}{2} |{\bf u}({\bf k})|^2 $.   Note that Parseval's theorem yields the following relation for the total  kinetic energy:
\bea
E_u = \frac{1}{2} \langle |{\bf u}({\bf r})|^2 \rangle =  \frac{1}{\mathrm{Vol}} \int d{\bf r} \frac{1}{2} |{\bf u}({\bf r})|^2 = 
\sum_{\bf k} \frac{1}{2} |{\bf u}({\bf k})|^2,
\eea
where ${\mathrm{Vol}}$ is the volume of the box.  We derive the following dynamical equation for $E_u({\bf k}) $ by performing a scalar product of  Eq.~(\ref{eq:gov:uk}) with ${\bf u^*(k)}$ and adding the resulting equation with its complex conjugate~\cite{Lesieur:book:Turbulence, Frisch:book, Davidson:book:Turbulence,Verma:book:BDF}:
\bea
\frac{d}{dt} E_u(\mathbf{k}) & = &  T_u({\bf k}) + \mathcal{F}_u{\bf k})  + \mathcal{F}_\mathrm{LS}{\bf k}) + D_u({\bf k}) \nonumber \\
& = &   \sum_{\bf p} \Im \left[ {\bf \{  k \cdot u(q) \}  \{ u (p) \cdot  u^*(k) \} }  \right] + \Re[ {\bf F}_u({\bf k}) \cdot  {\bf u^*(k)}]  \nonumber \\
&& + \Re[ {\bf F}_\mathrm{LS}({\bf k}) \cdot  {\bf u^*(k)}] - 2 \nu k^{2} E_u({\mathbf k}) ,  
\label{eq:gov:Euk}
\eea
where ${\bf q=k-p}$, and  $\Re[ . ]$, $\Im[ . ]$ stand respectively for the real and imaginary parts of the argument.   In the above equation, $T_u({\bf k})$ is the  nonlinear energy transfer  from all the  Fourier modes to ${\bf u(k)}$; $\mathcal{F}_u({\bf k}), \mathcal{F}_\mathrm{LS}{\bf k})$ are the respective energy supply rates from $ {\bf F}_u $ and $ {\bf F}_\mathrm{LS} $ to ${\bf u(k)}$; and $D_u({\bf k})$ is the viscous dissipation rate of ${\bf u(k)}$.

  The nonlinear interactions of Eq.~(\ref{eq:gov:Euk})  induce complex energy transfers among the Fourier modes.  However, a peep into a single wavenumber triad provides interesting insights into the nature of nonlinear interactions, which will be described below.
  
  \subsection{Triadic energy transfers and energy flux in hydrodynamics}
  \label{subsec:flux}
  
  Kraichnan~\cite{Kraichnan:JFM1959}  focussed on a pair of interacting wavenumber  triads, ${\bf (k, p, q)}$ and  ${\bf (-k, -p, -q)}$ with a condition that ${\bf k= p+ q} $. The corresponding Fourier modes are ${\bf u(k), u(p), u(q)}$ and their complex conjugates:  ${\bf u^*(k), u^*(p), u^*(q)}$.      For convenience, we set $\nu=0$ to suppress the viscous dissipation rate, which is a trivial linear term of Eq.~(\ref{eq:gov:Euk}). In addition, we assume that ${\bf F}_u = 0$ and ${\bf F}_\mathrm{LS}=0$.
  
 The wavenumbers ${\bf k, p, q}$  of Eq.~(\ref{eq:gov:Euk}) are not symmetric  (note ${\bf k = p+ q}$).  The formulas for the energy transfers are best expressed using a symmetric set  $({\bf k',p,q})$ obeying a constraint, ${\bf k'+p+q}=0$.  Note that ${\bf k'=-k}$.   For this triad, the dynamical equation for the modal energy $E_u({\bf k'}) $  is
  \bea
  \frac{d}{d t} E_u({\bf k'}) & = & -\Im\left[ {\bf  \{ k' \cdot u(q) \} \{ u(p)\cdot u(k') \} +\{ k' \cdot u(p) \} \{ u(q) \cdot u(k') \} }\right]  \nonumber \\
  & =  & S^{uu}({\bf k'|p,q}).  \label{eq:ET:Suu(X|Y,Z)} 
  \eea
  The equations for $E_u({\bf p}) $ and $E_u({\bf q}) $ are written in a similar manner.  
  
   The function $S^{uu}({\bf k'|p,q})$, called the {\em combined energy transfer}, represents the net kinetic energy transfer from modes ${\bf u(p)}$ and ${\bf u(q)}$ to ${\bf u(k')}$.   Note that the energy of the  mode ${\bf u(k')}$ grows at a  rate of $dE_u({\bf k'})/dt$.    Using Eq.~(\ref{eq:ET:Suu(X|Y,Z)}) and the incompressibility condition, ${\bf k' \cdot u(k')} = 0$, we derive that~\cite{Kraichnan:JFM1959,Lesieur:book:Turbulence}
  \bea
  S^{uu}({\bf k'|p,q}) + S^{uu}({\bf p|q,k'}) + S^{uu}({\bf q|k',p}) =0
  \eea
 that leads to the conservation of the total energy within a triad.

  Even though   Kraichnan's combined energy transfer formula  has been  widely used,  it does not provide individual energy transfers among the Fourier modes.  This task was first achieved by Dar et al.~\cite{Dar:PD2001} who derived the {\em mode-to-mode energy transfer}    from mode ${\bf u(p)}$ to mode ${\bf u(k')}$  with the mediation of mode ${\bf u(q)}$  as~\cite{Verma:PR2004}. 
  \bea
  S^{uu}({\bf k'|p|q}) = -\Im\left[ {\bf  \{ k' \cdot u(q) \} \{ u(p)\cdot u(k') \}} \right].
  \label{eq:ET:Suu_XYZ}
  \eea
  The formula contains a scalar product between the receiver  mode ${\bf u(k')}$ and the giver mode ${\bf u(p)}$, and another scalar product between the receiver wavenumber ${\bf k'}$ and the mediator mode   ${\bf u(q)}$.  The successive arguments of   $ S^{uu}$ are receiver, giver, and mediator wavenumbers respectively.  In terms of wavenumbers ${\bf k,p,q}$  with ${\bf k=p+q}$, the above formula is written as
  \bea
  S^{uu}({\bf k|p|q}) = \Im \left[ {\bf  \{  k \cdot u(q) \} \{ u(p) \cdot u^*(k) \} } \right] .
  \label{eq:ET:Suu_kpq}
  \eea

  Note that the mode-to-mode energy transfer functions satisfy the following properties:
  \bea
 S^{uu}({\bf k'|p|q})+S^{uu}({\bf k'|q|p}) =  S^{uu}({\bf k'|p,q})  , \label{eq:ET:Suu(X|Y|Z)_sum} \\
  S^{uu}({\bf k'|p|q}) = - S^{uu}({\bf p|k'|q}) . \label{eq:ET:Suu_plus_minus}
  \eea   
  The latter property follows from the incompressibility condition: ${\bf k' \cdot u(k')} =0$.   Dar et al.~\cite{Dar:PD2001} and Verma~\cite{Verma:PR2004} showed that Eq.~(\ref{eq:ET:Suu_XYZ}) satisfies Eqs.~(\ref{eq:ET:Suu(X|Y|Z)_sum}, \ref{eq:ET:Suu_plus_minus}), but it is not a unique solution to Eqs.~(\ref{eq:ET:Suu(X|Y|Z)_sum}, \ref{eq:ET:Suu_plus_minus}). A  circulating energy transfer that traverses along wavenumbers ${\bf p \rightarrow k' \rightarrow q \rightarrow p }$ could be added to the respective mode-to-mode energy transfers of Eq.~(\ref{eq:ET:Suu_XYZ}) without violating  Eqs.~(\ref{eq:ET:Suu(X|Y|Z)_sum}, \ref{eq:ET:Suu_plus_minus}).  The circulating transfer enters and leaves a mode, hence they do not alter the energy flux, which is a measurable quantity.  Therefore, the circulating transfer could be safely ignored.  Later, using tensor analysis and symmetry arguments,  Verma~\cite{Verma:book:ET} showed that Eq.~(\ref{eq:ET:Suu_XYZ}) provides a unique formula for the mode-to-mode energy transfer.
  
  A fluid flow has many Fourier modes.  Hence, the net energy transfer to the mode ${\bf u(k)}$ is a sum of energy transfers from all other modes.   In terms of mode-to-mode energy transfer, the net energy transfer to ${\bf u(k)}$ is~\cite{Dar:PD2001,Verma:PR2004}  
  \bea
  \frac{d}{dt} E_u(\mathbf{k}) & = &     \sum_{\bf p}  S^{uu}({\bf k|p|q})+ \mathcal{F}_u({\bf k})  + \mathcal{F}_\mathrm{LS}({\bf k})  - D_u({\bf k}), 
  \label{eq:ET:Euk}
  \eea
  where  ${\bf q=k-p}$.     A comparison of  Eq.~(\ref{eq:ET:Euk}) and Eq.~(\ref{eq:gov:Euk}) shows that
  \bea
  T_u({\bf k})  = \sum_{\bf p}  S^{uu}({\bf k|p|q}),
  \eea
  which is the net energy transfer to mode ${\bf u(k)}$ by nonlinearity.   Also note  that for any wavenumber region $A$,
  \bea
  \sum_{{\bf p} \in A}  \sum_{{\bf k} \in A} S^{uu}({\bf k|p|q}) =0.
  \label{eq:ET:Skpq_regionA}
  \eea
  The above relation hinges on Eq.~(\ref{eq:ET:Suu_plus_minus}).

  A very useful quantity in turbulence is the energy flux $\Pi_u(k_0)$, which is the total energy transfer from all the  modes inside a wavenumber sphere of radius $k_0$ to all the modes outside the sphere.  Starting from Kraichnan~\cite{Kraichnan:JFM1959}, researchers  have derived a number of formulas for $\Pi_u(k_0)$~\cite{Kolmogorov:DANS1941Structure,Kolmogorov:DANS1941Dissipation,  Kraichnan:JFM1959, Lesieur:book:Turbulence, Davidson:book:Turbulence,Dar:PD2001,Verma:PR2004,Frisch:book,Alexakis:PRL2005}.   Here we present the flux formula in terms of  mode-to-mode energy transfer, which is
  \bea
  \Pi_u(k_{0})=\sum_{p \le k_{0}} \sum_{k>k_{0}} S^{uu}({\bf k|p|q}) 
  = \sum_{p \le k_{0}} \sum_{k>k_{0}} \Im \left[ {\bf  \{  k \cdot u(q) \} \{ u(p) \cdot u^*(k) \} } \right] 
  \label{eq:ET:flux_Dar}
  \eea
  with ${\bf q=k-p}$.    In  the sum, the giver wavenumber ${\bf p}$ is inside the wavenumber sphere, while the receiver wavenumber ${\bf k}$ is outside the sphere.  For the above formula, it does not matter whether the mediator wavenumber ${\bf q}$ is inside or outside the sphere.  This liberty does not exist for the combined energy transfer formula in which both ${\bf u(p)}$ and ${\bf u(q)}$ supply energy to ${\bf u(k')}$.   The ambiguity of ${\bf u(q)}$ in the combined energy transfer poses a serious challenge for computing shell-to-shell energy transfer~\cite{Dar:PD2001,Verma:PR2004}. 
  
\begin{figure}[htbp]
  	\begin{center}
  		\includegraphics[scale=1]{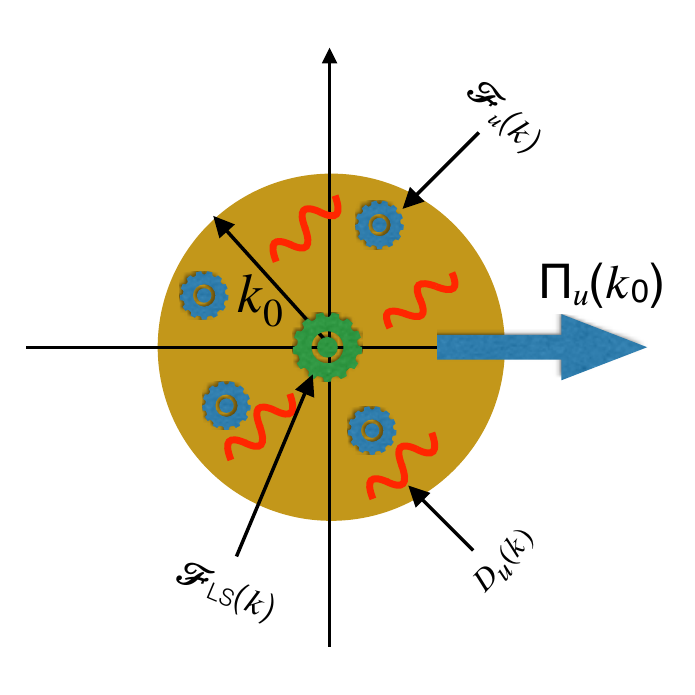}
  		\caption{Illustration of the energy flux $\Pi_u(k_0)$,  the kinetic energy injection rates  by  $\mathcal{F}_u({\bf k}) $ and  $\mathcal{F}_\mathrm{LS}({\bf k}) $ (wheels), and the  local viscous dissipation rate $ D_u({\bf k})$ (wavy lines). The viscous dissipation is  negligible at small wavenumbers. }
  		\label{fig:ET:flux_sphere}
  	\end{center}
  \end{figure}

  We  derive another set of formulas for the energy flux using $T_u({\bf k})$.   We sum the terms of Eq.~(\ref{eq:ET:Euk}) over all the modes inside the sphere that yields~\cite{Kraichnan:JFM1959,Lesieur:book:Turbulence,Frisch:book,Davidson:book:Turbulence}
  \bea
  \frac{d}{dt} \sum_{k \le k_0} E_u({\bf k})  =  \sum_{k \le k_0} T_u({\bf k}) +  \sum_{k \le k_0} \mathcal{F}_u({\bf k}) +  \sum_{k \le k_0} \mathcal{F}_\mathrm{LS}({\bf k}) - \sum_{k \le k_0} D_u({\bf k}). 
  \label{eq:ET:sum_Ek_less}
  \eea 
  To the  wavenumber sphere of radius $k_0$,   $\sum_{k \le k_0} \mathcal{F}_u({\bf k})$ and $  \sum_{k \le k_0} \mathcal{F}_\mathrm{LS}({\bf k}) $ are the net energy supply rates by $ {\bf F}_u$ and   $  {\bf F}_\mathrm{LS}$ respectively; and $\sum_{k \le k_0} D_u({\bf k})$ is the net  viscous dissipation rate in the sphere (see Fig.~\ref{fig:ET:flux_sphere}).   The sum  $\sum_{k \le k_0} T_u({\bf k})$  is the net energy transfer due to the nonlinear term  from the modes outside the sphere to the modes inside the sphere.  Hence, by definition, 
  \bea
  \Pi_u(k_0) = - \sum_{k \le k_0} T_u({\bf k}). \label{eq:ET:Pi_Tk_less}
  \eea
  Using Eqs.~(\ref{eq:ET:sum_Ek_less}, \ref{eq:ET:Pi_Tk_less}) we derive
  \bea
  \Pi_u(k_0) = - \frac{d}{dt} \sum_{k \le k_0} E_u({\bf k})  +   \sum_{k \le k_0} \mathcal{F}_u({\bf k}) +  \sum_{k \le k_0} \mathcal{F}_\mathrm{LS}({\bf k}) - \sum_{k \le k_0} D_u({\bf k}).
  \label{eq:Pi_dEbydt_min_visc}
  \eea
  Similarly, using the following equation, 
  \bea
  \frac{d}{dt} \sum_{k > k_0} E_u({\bf k})  =  \sum_{k > k_0} T_u({\bf k}) +  \sum_{k > k_0} \mathcal{F}_u({\bf k}) +  \sum_{k > k_0} \mathcal{F}_\mathrm{LS}({\bf k}) - \sum_{k > k_0} D_u({\bf k}),
  \label{eq:ET:sum_Ek_gtr}  
  \eea
  one obtains  $ \Pi_u(k_0) =  \sum_{k > k_0} T_u({\bf k}) $~\cite{Lesieur:book:Turbulence,Verma:book:ET}.  Verma et al.~\cite{Verma:JGR1996DNS} employed Eq.~(\ref{eq:Pi_dEbydt_min_visc})  to compute the energy fluxes of magnetohydrodynamic turbulence.   
  
  Interestingly, $T_u(k)$-based energy flux formula can be  used to systems whose  nonlinear interactions differ from those of incompressible hydrodynamics.  For example, for dissipation-less Burgers equation,  $ \partial_t u = -   \partial_x u^2/2 $, the energy equation is  
  \be
  \frac{d}{d t} \frac{1}{2}  |u(k)|^2 = \sum_p  \Im[k u(k-p) u(p) u^*(k)] = T(k).
  \ee
Hence, the energy flux $ \Pi(k_0) = - \sum_{k<k_0 }T(k) $.    We can also use the above  procedure to compute the energy flux  for $\phi^4$ theory where the interaction is quartic and  $T(k) \sim \phi({\bf k}_1) \phi({\bf k}_2)  \phi({\bf k}_3)  \phi({\bf k}_4)$ with ${\bf k}_1 + {\bf k}_2 + {\bf k}_3 + {\bf k}_4 =0$~\cite{Amit:book:RG,Chaikin:book}.    Such schemes are useful for modelling energy transfers in binary fluids, and time-dependent Ginzburg-Landau and Cahn-Hilliard equations.  (see Sec.~\ref{sec:QT}).
  
 Kraichnan~\cite{Kraichnan:JFM1959},   Frisch~\cite{Frisch:book},  
 Alexakis et al.~\cite{Alexakis:PRL2005, Alexakis:PRE2005} and others have derived  formulas for the energy flux based on nonlinear interactions in Fourier space.  These formulas will not be discussed here due to lack of space and due to similarities with those discussed above.   Kolmogorov's formula for the energy flux~\cite{Kolmogorov:DANS1941Structure,Kolmogorov:DANS1941Dissipation}, which is based on the third-order structure function, will be discussed in Sec.~\ref{sec:struct_fn}.
 
   An important point to note:  the formulas of Eqs.~(\ref{eq:ET:flux_Dar}, \ref{eq:ET:Pi_Tk_less}) are applicable to isotropic as well as anisotropic flows. For anisotropic flows, the spectrum $ E_u(\textbf{k}) $ depends on the angle between \textbf{k} and  the anisotropic axis.  However, the energy flux, which involves a sum over the modes of a sphere,  is well-defined for such flows.   Due to these reasons, energy flux remains an important quantity for buoyant and MHD turbulence, even though such flows could be strongly anisotropic.

   The formalism of kinetic energy flux can be generalized to other energy transfers.  For example, the net energy transfer from wavenumber region $ A $ to  wavenumber region $ B$ is 
   \be
   T^{u,A}_{u,B} = \sum_{{\bf p} \in A} \sum_{{\bf k} \in B}
   \Im \left[ {\bf  \{  k \cdot u(q) \} \{ u(p) \cdot u^*(k) \} } \right] ,
   \ee
   which can be computed numerically using Fast Fourier Transform (FFT)~\cite{Dar:PD2001}.  In addition to the energy flux, the other two popular measures of energy transfers are \textit{shell-to-shell energy transfer} and \textit{ring-to-ring energy transfer}~\cite{Dar:PD2001, Teaca:PRE2009, Verma:book:ET,Sagaut:book}, but these topics are beyond the scope of this review.

In the next section, we will show how the energy flux varies with wavenumbers due to multiscale external force and dissipation.

\section{Variable energy flux}
\label{sec:VF}

The kinetic energy flux varies with wavenumbers in the presence of multiscale forcing and dissipation~\cite{Lesieur:book:Turbulence,Frisch:book,Davidson:book:Turbulence,Pope:book,Verma:book:BDF}.  In this review we summarise how  past works employed this  observation to deduce interesting properties of turbulence. We start  this section with an equation for the variable energy flux.  In this section we assume the wavenumber $k$ to be a continuous variable.


\subsection{Formalism}
\label{subsec:VF_formalism}

\begin{figure}[htbp]
\begin{center}
\includegraphics[scale=0.8]{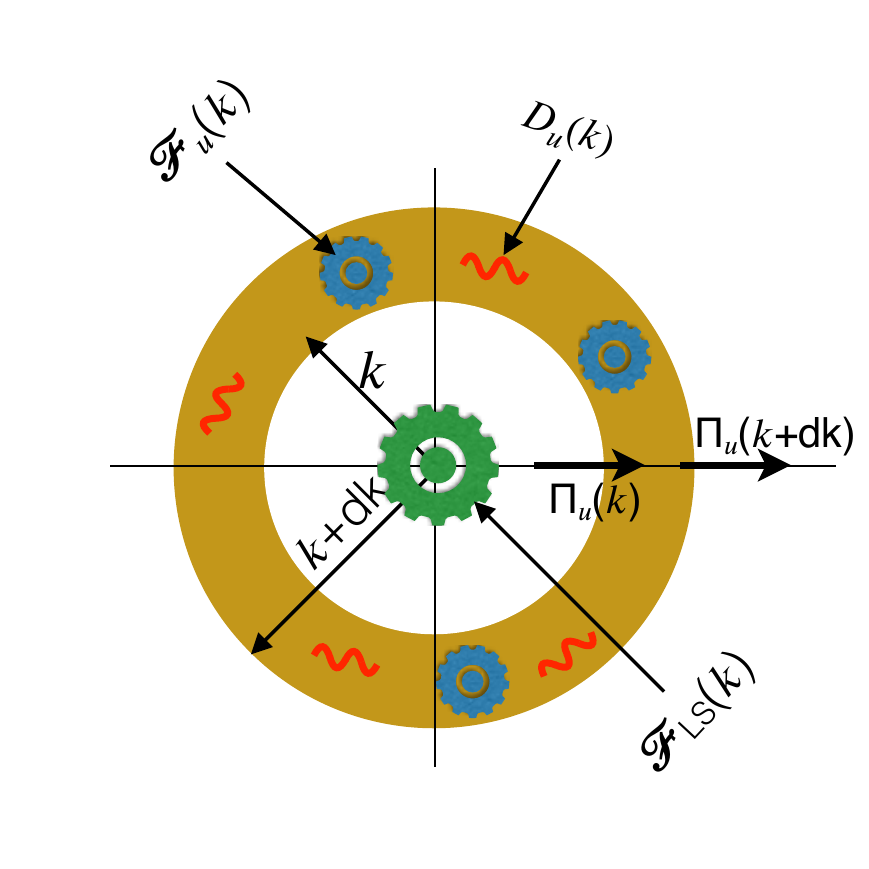}
\caption{ A figure illustrating the  energy flux difference, $\Pi_u(k+dk) - \Pi_u(k)$;   the energy injection rates $\mathcal{F}_u(k)dk$ and $\mathcal{F}_\mathrm{LS}(k)dk $ represented by wheels; and the viscous dissipation rate $D_u(k)dk$ represented by wavy lines.  Refer to Eq.~(\ref{eq:VF:Ek_energetics}).}
\label{fig:ET:energy_shell}
\end{center}
\end{figure}

Rewriting Eq.~(\ref{eq:ET:sum_Ek_less}) for  spheres of radii $k$ and $k+ dk$, and  then taking a difference between the two equations  yields
\bea
 \frac{d}{dt}  \sum_{k < k' \le k+dk} E_u({\bf k'}) & = &  \sum_{k < k' \le k+dk}  T_u({\bf k'}) +   \sum_{k < k' \le k+dk} \mathcal{F}_u({\bf k'})  \nonumber \\
 && +   \sum_{k < k' \le k+dk} \mathcal{F}_\mathrm{LS}({\bf k'}) -  \sum_{k < k' \le k+dk}   D_u({\bf k'}) .
 \label{eq:VF:Ek_energetics0}
   \eea
Using $ \sum_{k < k' \le k+dk}  T_u({\bf k'})  = [- \Pi_u(k+dk) + \Pi_u(k)]  $ and taking the limit $dk \rightarrow 0$, we obtain the following evolution equation for one-dimensional energy spectrum $E_u(k)$~\cite{Lesieur:book:Turbulence,Frisch:book,Davidson:book:Turbulence,Pope:book,Verma:book:BDF}:
\bea
 \frac{\partial}{\partial t}   E_u(k,t) & = & -\frac{\partial}{\partial k} \Pi_u(k,t) + \mathcal{F}_u(k,t)  
 + \mathcal{F}_\mathrm{LS}(k,t )-D_u(k,t),
 \label{eq:VF:Ek_energetics}
   \eea
   where
   \bea 
E_u(k)dk & = & \sum_{k < k' \le k+dk} E_u({\bf k'}), \\
\mathcal{F}_u(k) dk & = & \sum_{k < k' \le k+dk}  \Re[ {\bf F}_u({\bf k'}) \cdot  {\bf u^*(k')}],
\label{eq:VF:F_u_k_def}  \\
\mathcal{F}_\mathrm{LS}(k) dk & = & \sum_{k < k' \le k+dk}  \Re[ {\bf F}_\mathrm{LS}({\bf k'}) \cdot  {\bf u^*(k')}],
\label{eq:VF:F_LS_k_def}  \\
D_u(k) dk & = & 2 \nu \sum_{k < k' \le k+dk} k'^{2} E_u({\mathbf k}').
\eea  
Figure~\ref{fig:ET:energy_shell} illustrates the above quantities for a wavenumber shell whose inner and outer radii are $k$ and $k+dk$ respectively.   Also note that $T_u(k,t)$ is defined as~\cite{Leslie:book,Lesieur:book:Turbulence}
\bea
T_u(k,t) =  -\frac{\partial}{\partial k} \Pi_u(k,t).
\label{eq:VF:Tu_k}
\eea
Equation~(\ref{eq:VF:Ek_energetics}) describes how $\Pi_u(k)$ varies with $k$ in an unsteady flow.

 In this review we study the behaviour of the kinetic energy flux under a steady state.  Setting $\partial E_u(k,t) /\partial t = 0$ in Eq.~(\ref{eq:VF:Ek_energetics0}) yields the following equation for the wavenumber shell of radius $k$:
 \be
  \frac{d}{dk} \Pi_u(k) = \mathcal{F}_u(k)  +  \mathcal{F}_\mathrm{LS}(k) -D_u(k).
  \label{eq:VF:energetics_steady}
  \ee
  That is, the energy flux $ \Pi_u(k) $ varies with $k$ due to the energy injection rates  $\mathcal{F}_u(k)$ and $ \mathcal{F}_\mathrm{LS}(k) $, and the viscous dissipation rate $D_u(k)$.    We remark that Eq.~(\ref{eq:VF:energetics_steady}), derived using the energy conservation, is an exact relation in a statistical sense.   That is, the quantities $ \Pi_u(k) $,  $\mathcal{F}_u(k)$, $ \mathcal{F}_\mathrm{LS}(k) $, and $D_u(k)$ may fluctuate around their mean, but Eq.~(\ref{eq:VF:energetics_steady}) holds on an average.  This relation is analogous to Kolmogorov's four-fifth law, which is also related to the energy conservation~\cite{Frisch:book,Verma:book:ET}.  Note however that  Eq.~(\ref{eq:VF:energetics_steady}) is valid for anisotropic flows as well, which is the not the case for Kolmogorov's four-fifth law.  
    
\begin{figure}[htbp]
\begin{center}
\includegraphics[scale=0.6]{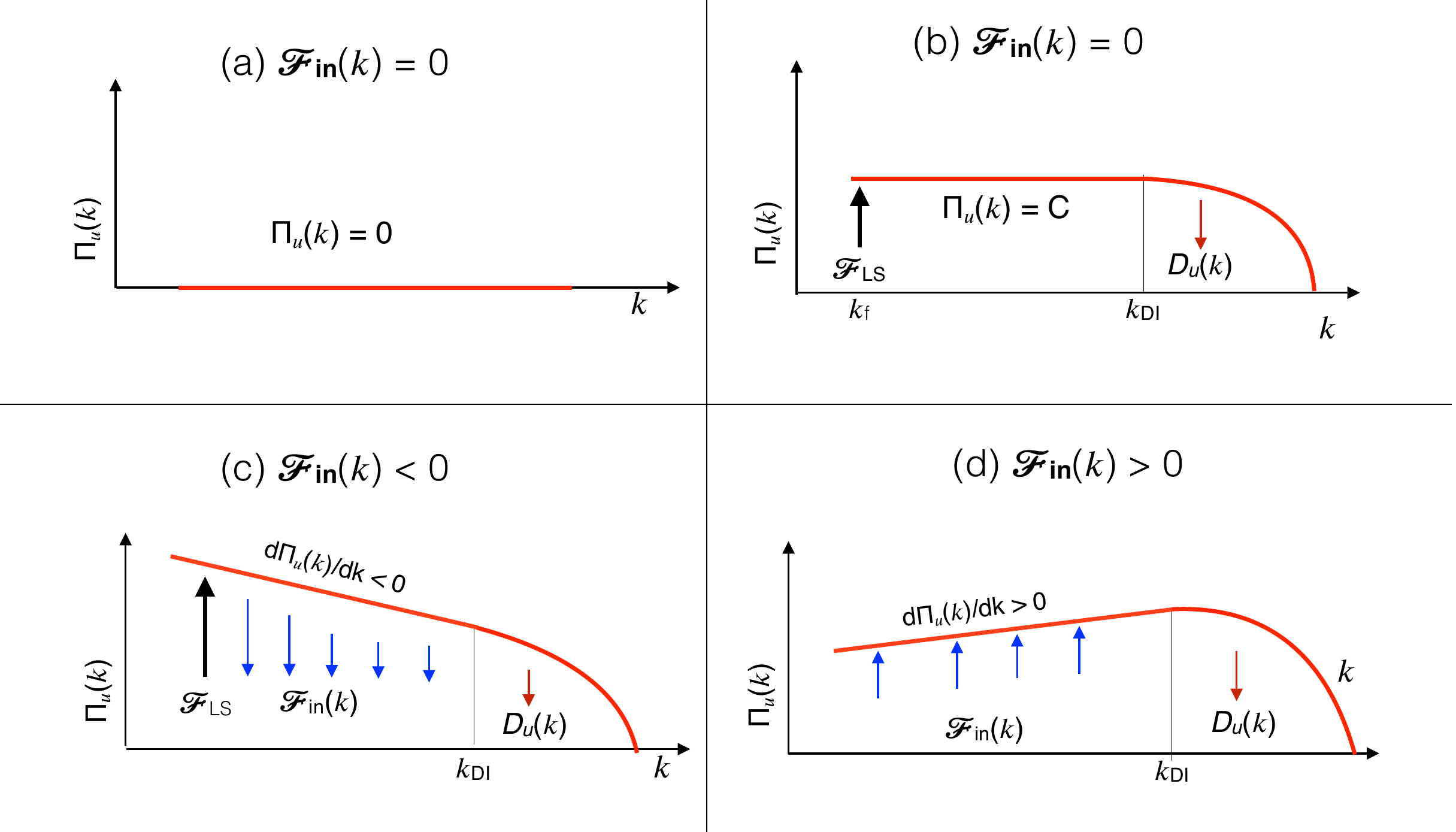}
\caption{Four categories of $\Pi_u(k)$ depending on the  inertial-range forcing $\mathcal{F}_\mathrm{in}(k)$: (a)  $\mathcal{F}_\mathrm{in}(k) = 0$ and $ \Pi_u(k) = 0$ (equilibrium);  (b)  $\mathcal{F}_\mathrm{in}(k) = 0 $ and $ \Pi_u(k) =C > 0$ (nonequilibrium); (c) $\mathcal{F}_\mathrm{in}(k) < 0  \Rightarrow d\Pi_u(k)/dk  < 0$; (d) $\mathcal{F}_\mathrm{in}(k) > 0  \Rightarrow d\Pi_u(k)/dk  > 0$.  Viscous dissipation $D_u(k)$ dominates the nonlinear term beyond wavenumber $k_\mathrm{DI}$. }
\label{fig:VF:energetics}
\end{center}
\end{figure}

In the intermediate wavenumber range, the nonlinear term dominates the viscous term, a reason for which this is called the \textit{inertial range}.  In addition,  $ \mathcal{F}_\mathrm{LS}(k) = 0$ and $D_u(k) \approx 0$ in this range.   Therefore,  
 \be
  \frac{d}{dk} \Pi_u(k) = \mathcal{F}_{u,\mathrm{in}}(k) ,
  \label{eq:VF:inertial_Pi}
  \ee
  where the subscript ``in" stands for the inertial range.  Thus,   the energy flux varies due to $\mathcal{F}_{u,\mathrm{in}}(k) $.  Behaviour of $\Pi_u(k)$  can be classified into the  following four categories:
\begin{enumerate}
\item \label{itm:absolute_eq}  $\mathcal{F}_{u,\mathrm{in}}(k) = 0$ and $\Pi_u(k) = 0$:  This case corresponds to the \textit{absolute equilibrium} scenario of Euler's equation (Navier Stokes equation with $\nu=0$)~\cite{Lee:QAM1952,Kraichnan:JFM1973}.  For this case, the average energy exchange between any two Fourier modes is zero.  That is, $\la S^{uu}({\bf k'|p|q}) \ra =0 $ for any triad, thus satisfying  detailed balance of kinetic energy transfer. Therefore, as in thermodynamics,  the modal kinetic energy spectrum $E_u({\bf k}) = $ constant, leading to $E_u(k) \sim k^2$  and $ \sim k $ for 3D and 2D hydrodynamic turbulence respectively~\cite{Lee:QAM1952,Kraichnan:JFM1973,Orszag:CP1973,Lesieur:book:Turbulence, Alexakis:PR2018,Verma:PR2004}.  We will discuss these cases in more detail in Sec.~\ref{sec:Euler}.  A related phenomenon to this cases is flux loop cascade~\cite{Boffetta:EPL2011,Falkovich:PRF2017}, which will be discussed in Sec.~\ref{sec:2D}.

\item  \label{itm:Pi>0} $\mathcal{F}_{u,\mathrm{in}}(k) = 0$ and $\Pi_u(k) =  C >0$:   This regime is described by Kolmogorov's theory of turbulence.  For this nonequilibrium case, $ \la S^{uu}({\bf k'|p|q}) \ra >0 $ for $k' > p$ and vice versa.  We describe this case in more detail in Sec.~\ref{subsec:Kolm_theory_turb_Fourier}.

\item $\mathcal{F}_{u,\mathrm{in}}(k) < 0$ and $d\Pi_u(k)/dk  < 0$: For  $\mathcal{F}_\mathrm{in}(k) < 0$, the kinetic energy flux decreases with $k$.  Notable examples in this category are Quasi-static MHD turbulence, flows with Ekman friction, and stably-stratified turbulence.  These flows are described in subsequent sections.

\item $\mathcal{F}_{u,\mathrm{in}}(k) >  0$ and $d\Pi_u(k)/dk > 0$:   The kinetic energy flux increases in this case. Leading examples for this case are turbulent thermal convection and  shear turbulence  that will be described  in later sections.
\end{enumerate}
The above four cases are illustrated in Fig.~\ref{fig:VF:energetics}.
 
In the following discussion, we provide several examples of variable energy flux in hydrodynamic turbulence.  However, first,  we begin with a brief description of Kolmogorov's theory of turbulence that forms a basis for many turbulence phenomenologies.

\subsection{Kolmogorov's theory of turbulence: constant energy flux}
\label{subsec:Kolm_theory_turb_Fourier}

In this section we briefly describe Kolmogorov's theory of turbulence, which corresponds to  case (ii) described above.  For a more detailed derivations, refer to \cite{Monin:book:v1,Monin:book:v2,Leslie:book,Lesieur:book:Turbulence,McComb:book:Turbulence,Frisch:book,Davidson:book:Turbulence,Sagaut:book,Pope:book, Orszag:CP1973,Mccomb:RPP1995,Verma:book:ET}.  Kolmogorov~\cite{Kolmogorov:DANS1941Structure, Kolmogorov:DANS1941Dissipation} considered a steady turbulent flow that is driven at large length scales or at a wavenumber band near $k_f \approx 1/L$, where $L$ is the system size.  The energy thus injected at the large scales cascades to the inertial   range and then to the  dissipative scales.  In the inertial  range, $\mathcal{F}_\mathrm{in}(k) = 0$ and $D_u(k)=0$. Therefore, under a steady state,  Eq.~(\ref{eq:VF:energetics_steady}) yields  $ \Pi_u(k) = \Pi_u =  \mathrm{const} $, as shown in Fig.~\ref{fig:VF:energetics}(b).  To be precise, for any $k>k_f$, an integration of Eq.~(\ref{eq:VF:energetics_steady})  yields
\bea
\Pi_u(k) =   \int_k^\infty D_u(k') dk'.
\label{eq:VF:Pi_u_eq_eps_u_k_infty}
\eea
The viscous dissipation  is negligible in the forcing band $(0,k_f)$.  Hence, for a wavenumber $ k $ in the inertial range,
\bea
\Pi_u(k) \approx \int_0^\infty D_u(k') dk' =\epsilon_u,
\label{eq:VF:Pi_u_eq_eps_u}
\eea
where $\epsilon_u$ is the total viscous dissipation rate.     Thus, under steady state,  $\epsilon_u$ is approximately equal to the energy injection rate by the large-scale forcing.   Note however that Eqs.~(\ref{eq:VF:Pi_u_eq_eps_u_k_infty}, \ref{eq:VF:Pi_u_eq_eps_u})  imply that   $\Pi_u(k) \lessapprox \epsilon_u$ in the inertial range. 

Another  important assumption of  Kolmogorov's theory of turbulence is that the physics in the inertial range is independent of the forcing and dissipative mechanisms~\cite{Leslie:book,Lesieur:book:Turbulence,Frisch:book}.  Hence, the energy spectrum $E_u(k)$ is isotropic, and  it depends only on $\epsilon_u$ and local wavenumber $k$.  Absence of any external length scale implies that $E_u(k)$  is a power law in $k$.  This assumption is related to the locality of interactions in hydrodynamic turbulence~\cite{Domaradzki:PF1990, Eyink:PD2005,Verma:Pramana2005S2S}, but this discussion will take us beyond the scope of this review.  

Using the above inputs and dimensional analysis, we derive that
\be
E_u(k) = K_\mathrm{Ko} \epsilon_u^{2/3} k^{-5/3},
\label{eq:VF:Kolm_Ek}
	\ee
where $  K_\mathrm{Ko} $ is Kolmogorov's constant.    The above theory of turbulence is  universal because  Eq.~(\ref{eq:VF:Kolm_Ek}) is independent of forcing and dissipative mechanisms, initial condition, and fluid properties such as viscosity and density.  The above law has been observed in many laboratory experiments, in natural flows such terrestrial atmosphere, and in numerical simulations~\cite{Lesieur:book:Turbulence,McComb:book:Turbulence,Frisch:book,Davidson:book:Turbulence,Sagaut:book,Pope:book}.   However, Kolmogorov's theory of turbulence  is inapplicable in the presence of multiscale external force and/or dissipation.  As described below, the formalism of variable energy flux is useful for modelling such flows.

\subsection{Variable energy flux  in the inertial-dissipation range and bottleneck effect}
\label{subsec:hydro_inertial_diss}

Kolmogorov's $k^{-5/3}$ energy spectrum is observed in the inertial range of hydrodynamic turbulence, but not in the dissipative range. Using the data from numerical simulations, Chen et al.~\cite{Chen:PRL1993} and   Martinez et al.~\cite{Martinez:JPP1997} proposed that in the far-dissipation range, the energy spectrum varies as $ k^\alpha \exp\left( -c k/k_d \right)$, where $ \alpha, c $ are constants, and   $ k_d = (\epsilon_u/\nu^3)^{1/4} $  is   Kolmogorov's wavenumber.    Pao~\cite{Pao:PF1965} argued in favour of $ k^{-5/3} \exp(- c (k/k_d)^{4/3}) $ spectrum.   Pope~\cite{Pope:book} showed that the following  energy spectrum is a good fit to many experimental observations (see e.g. \cite{Saddoughi:JFM1994}):
   \bea
E_u(k) = K_\mathrm{Ko} \epsilon^{2/3} k^{-5/3}  f_L(k L) f_\eta(k /k_d),
\label{eq:VF:Ek_Pope}
\eea
where $ f_L(kL),  f_\eta(\tilde{k})  $ represent the  large-scale and dissipative-scale components respectively.  
 
\begin{figure}[htbp]
 	\begin{center}
 		\includegraphics[scale = 0.7]{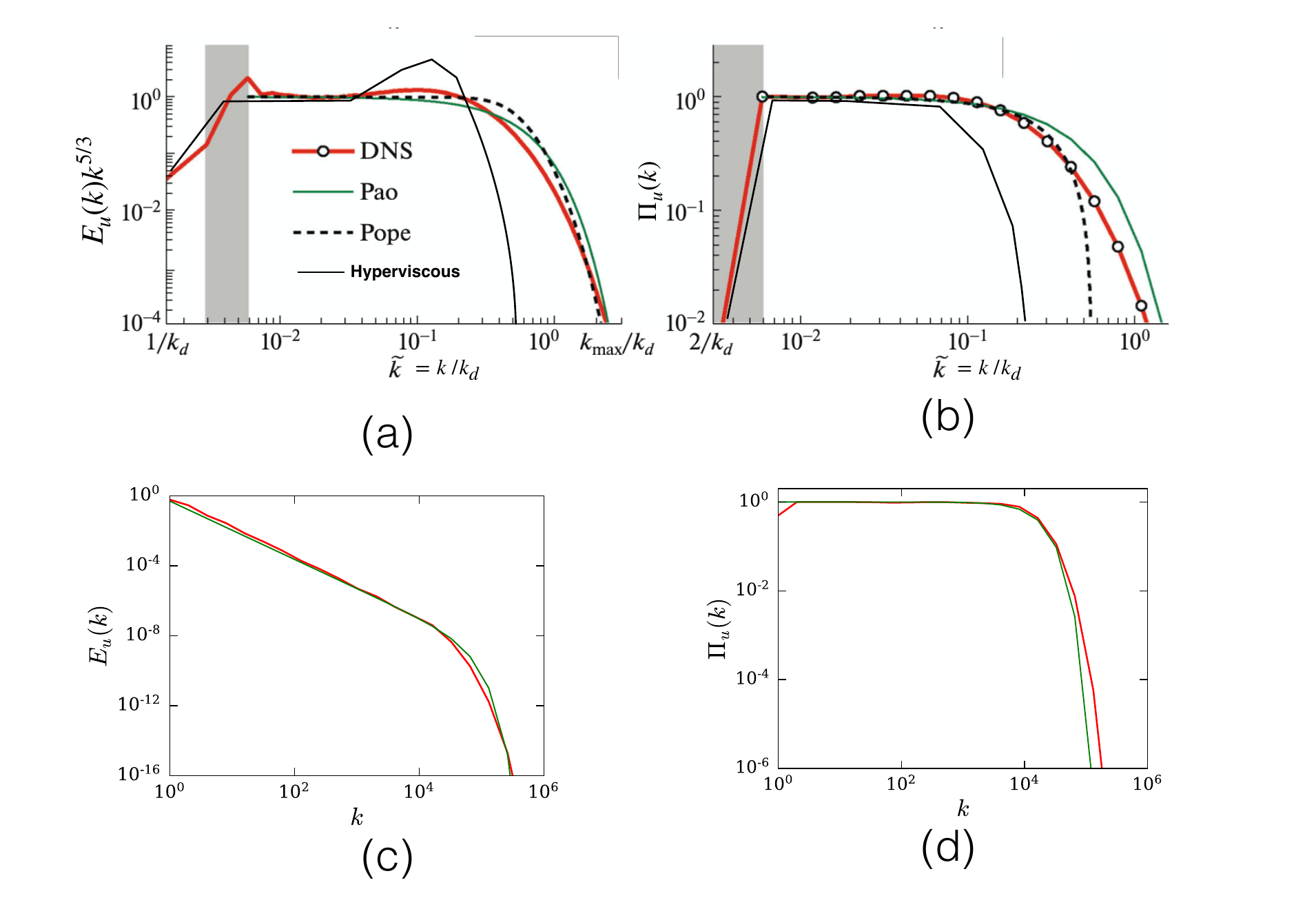}
 	\end{center}
 	\caption{ (a,b) The energy spectrum and flux of  hydrodynamic turbulence: for a $4096^3$ DNS (red curve), Pao's model (green curve), Pope's model (dashed curve), for a flow with hyperviscosity (schematic diagram only).   (c,d) The energy spectrum and flux of a shell model (red curves) are in good agreement with the predictions of Pao's model (green curves).  Figures (a,b) adopted from a figure of Verma et al.~\cite{Verma:FD2018}.  }
 	\label{fig:bottleneck}
 \end{figure}

Here we present Pao's model~\cite{Pao:PF1965} for the inertial-dissipation range  of hydrodynamic turbulence.  This formula is based on variable energy flux and the following ansatz. In the inertial-dissipation range, where  $\mathcal{F}_u = \mathcal{F}_\mathrm{LS} =0$, Eq.~(\ref{eq:VF:energetics_steady}) yields
 \bea
  \frac{d}{dk} \Pi_u(k) =   -2 \nu k^2 E_u(k).
  \label{eq:VF:3Dhydro_diss}
  \eea
 In addition,  Pao~\cite{Pao:PF1965}  assumed that in the inertial-dissipative range, $\Pi_u(k)/E_u(k)$ is independent of $\nu$ and  is  function only of $k$ and $\epsilon_u$ (also see \cite{Leslie:book}). Under these assumptions, dimensional analysis with an aid of Eq.~(\ref{eq:VF:Kolm_Ek}) yields 
\bea
\frac{E_u(k)}{\Pi_u(k)} = K_\mathrm{Ko}  \epsilon_u^{-1/3} k^{-5/3}.
\label{eq:VF:Pao_assumption}
\eea
The above conjecture  assumes locality of turbulent interactions because $\Pi_u(k)$ depends on local $k$ and $E_u(k)$.     Substitution of  Eq.~(\ref{eq:VF:Pao_assumption}) in Eq.~(\ref{eq:VF:3Dhydro_diss}) yields
   \bea
\Pi_u(k)  & =  & \epsilon_u \exp{\left(- \frac{3}{2} K_\mathrm{Ko}  (k/k_d)^{4/3}\right)}, \label{eq:VF:Pao_Pik}\\
E_u(k)  & = & K_\mathrm{Ko}  \epsilon_u^{2/3} k^{-5/3}  \exp{\left(- \frac{3}{2} K_\mathrm{Ko}  (k/k_d)^{4/3}\right)}.  \label{eq:VF:Pao_Ek}
\eea 
Pao~\cite{Pao:PF1965} showed that the above  energy spectrum describes many experimental observations~\cite{Grant:JFM1962} reasonably well\footnote{Using Eq.~(\ref{eq:VF:3Dhydro_diss}), Verma et al.~\cite{Verma:FD2018} argue that the energy flux and spectrum vary as $ \exp(-k) $ for laminar flows (small Reynolds number). Note that the energy flux is nonzero for flows with small but nonzero Reynolds number.}.  Recently, Verma et al.~\cite{Verma:FD2018}  showed that the above spectrum and flux describe  the results of high-resolution numerical simulations, except for a small discrepancy  in the dissipation range, where  Pao's model overpredicts the energy flux and spectrum (see Fig.~\ref{fig:bottleneck}).  This feature appears to be related to the  bottleneck effect, which will be discussed below.  Interestingly, Eqs.~(\ref{eq:VF:Pao_Pik}, \ref{eq:VF:Pao_Ek}) describe the energy spectrum and flux of the shell model of turbulence~\cite{Ditlevsen:book} quite well.  (see Fig.~\ref{fig:bottleneck}(c,d)).
 
In hydrodynamic turbulence, the normalised spectrum $E_u(k) k^{5/3}$ exhibits a hump between the inertial range and  the dissipation range~\cite{Falkovich:PF1994,Saddoughi:JFM1994,Verma:JPA2007,Kuchler:JSP2019,Donzis:JFM2010Bottleneck} (see Fig.~\ref{fig:bottleneck}(a)).    This phenomena, called {\em bottleneck effect}, is not yet fully understood. Falkovich~\cite{Falkovich:PF1994} related the bottleneck effect to the nonlinear energy transfer $T_u(k)$ of Eq.~(\ref{eq:VF:Tu_k}).  Verma and Donzis~\cite{Verma:JPA2007} argue that the observed bottleneck effect in most experiments and simulations is due to the insufficient inertial range available for facilitating the energy transfer from the large scales to small scales. Kuchler et al.~\cite{Kuchler:JSP2019} performed an experimental study of bottleneck effect and observed general agreement between the experimental results and model predictions~\cite{Falkovich:PF1994, Verma:JPA2007}.  Spyksma et al.~\cite{Spyksma:PF2012} showed that the bottleneck effect gets enhanced due to hyperviscosity, as illustrated schematically in Fig.~\ref{fig:bottleneck}(a).  Interestingly,  as shown Fig.~\ref{fig:bottleneck}(c), Pao's model and the shell model do not exhibit bottleneck effect.

\begin{figure}[htbp]
	\begin{center}
		\includegraphics[scale = 0.7]{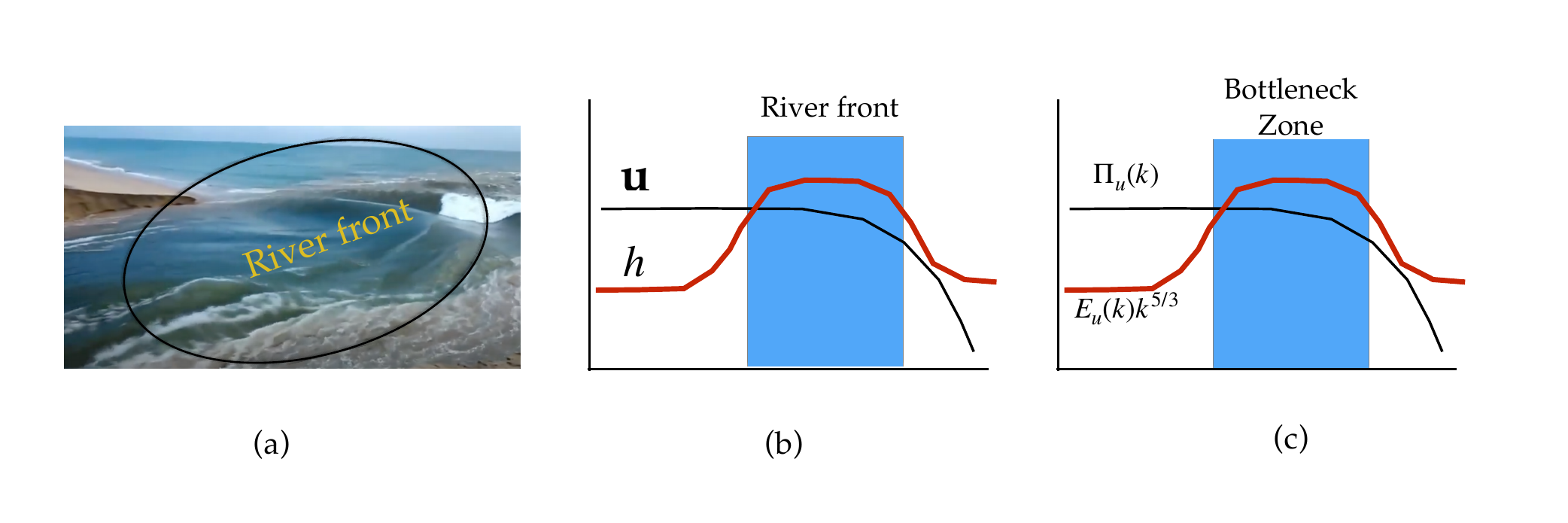}
	\end{center}
	\caption{ (a) River front meeting the ocean.  The height of the front increases when the river front is signficantly fast. (b) A schematic diagram of diminishing velocity \textbf{u} and increasing of the water level height \textit{h}. (c) In  the bottleneck zone of hydrodynamic turbulence, the normalized spectrum, $E_u(k)k^{5/3}$  increases due to the suppression of $\Pi_u(k)$ by viscous dissipation. 	Figure (a) is an adoption of a video   from https://www.youtube.com/watch?v=y0L5-owOr9U. }
	\label{fig:river_front}
\end{figure}

In the following discussion we  model the bottleneck effect by relating the kinetic energy flux  to  the mass flux in a river front when the front meets the ocean  (see Fig.~\ref{fig:river_front}(a)).  The mass flux of the water flow can be approximated by $ \rho h {\bf u} $, where $ \rho $ and $ {\bf u} $ are respectively the density and velocity of the water, and $ h $ is the  water level height of the river (we ignore the effects of boundary layers, waves, etc.).  A river moves with a certain velocity, thus carrying a mass flux.  However, the river  slows down  or $ {\bf u} $ decreases as the river approaches the ocean, as shown Fig.~\ref{fig:river_front}(b).  Note however that the mass flux should remain constant. Therefore, $ h $ tends to increase at the front in order to conserve the mass flux.  Note however that $ {\bf u} =0 $   when the river has attained equilibrium with the ocean (somewhat deep inside the ocean).  We also remark that the increase in $ h $ is small or insignificant  if the river has slowed down considerably before meeting the ocean.   The hump in $ h $ is observed only when the river front has significant mass flux when it meets the ocean.  Similar phenomena is observed in a traffic jam.  The traffic velocity  decreases considerably in a traffic jam leading to a pileup of vehicles at the bottleneck.  We also observe similar accumulation of people when a large crowd attempts to  exit  from a narrow gate. 

Now, let us connect the bottleneck effect to the aforementioned examples of river front.  The turbulent energy flux is analogous to the mass flux of the river or the traffic, while the energy spectrum to the   local  water volume.   The constant energy flux in the inertial range corresponds to the smooth flow of the river or traffic.   The turbulent energy flux  decreases in the dissipation range, similar to the mass flux in the river front during its merger with the ocean.   In Pao's model, the energy flux decreases  gradually, hence  it does not exhibit any bottleneck effect.  This feature is  akin  to a slow merger of a river front with the ocean, during which $ h $ does not increase.  However, as shown in Fig.~\ref{fig:bottleneck}(a), the energy flux in the DNS decreases more rapidly than Pao's model. Therefore,  $ E_u(k) k^{5/3}$ of the DNS exhibits a hump or bottleneck effect.  The hyperviscous dissipation suppresses the energy flux even further than the DNS, which leads to stronger  bottleneck effect.  Thus, the bottleneck effect is related to the extent of suppression of the turbulent energy flux in the dissipation range.  

Another interesting observation is that the numerical results of most shell models do not exhibit any bottleneck effect (see Fig.~\ref{fig:bottleneck}(c,d)).  These figures also illustrate that the energy spectrum and flux of the shell model are described reasonably well by Pao's model, which is  possibly because most shell models only incorporate local energy transfers (among neighbouring shells).  In contrast, the nonlocal shell models~\cite{Plunian:NJP2007,Plunian:PR2012} may exhibit bottleneck effect.  The aforementioned connections between locality, the  bottleneck effect, and Pao's model need to be explored further. 

The above discussion on Pao's model and the bottleneck effect illustrates the usefulness of variable energy flux.

\subsection{Quasi-static MHD turbulence}
\label{subsec:QSMHD}

In quasi-static (QS) MHD, which is applicable to liquid metals, the nonlinear term of the induction equation is ignored beause   $\mathrm{Re Pm } \rightarrow 0$, where $\mathrm{Pm}$ is the magnetic Prandtl number~\cite{Knaepen:ARFM2008, Verma:ROPP2017}.  For such flows, under a strong mean magnetic field ${\bf B}_0$, the Lorentz force is approximated by the following expression~\cite{Knaepen:ARFM2008, Verma:ROPP2017}:
\bea
{\bf F}_u({\bf k}) = - [N (U/L)  \cos^2 \theta] {\bf u}({\bf k}),
\label{eq:VF:QSMHD_force}
\eea
where $\theta$ is the angle between wavenumber ${\bf k}$ and ${\bf B}_0$;  $U, L$ are the large-scale velocity and length  respectively; and  $N = B_0^2 L/(U \kappa)$ is the interaction parameter with   $\kappa$ as the magnetic diffusivity.    

The effects of the Lorentz force in QS MHD turbulence is an important topic of research.  Laboratory experiments~\cite{Alemany:JdeM1979,Eckert:IJHFF2001,Kolesnikov:FD1974,Kit:MG1971,Sreenivasan:EJMB2000} and numerical simulations~\cite{Boeck:PRL2008,Potherat:JFM2014,Reddy:PF2014,Reddy:PP2014} of QS-MHD turbulence reveal that  for small interaction parameters ($N \lessapprox 1$), the Lorentz force has a weak effect on the flow, thus yielding a weak steepening of the energy spectrum compared to $ k^{-5/3} $ spectrum.   However, with the increase of $N$, $E_u(k)$ steepens significantly with the spectral index  approaching as low as $-5$ for $N \approx 30$.  For very large $N$, $E_u(k) \sim \exp(-b k)$ with $b$ as a positive constant. 

Several models have been constructed to explain the  steepening of the energy spectrum in QS MHD turbulence~\cite{Alemany:JdeM1979,Kolesnikov:FD1974,Kit:MG1971, Moffatt:JFM1967,Schumann:JFM1976,Moreau:book:MHD,Verma:PF2015QSMHD, Knaepen:ARFM2008,
	Anas:PRF2019}. In particular, many researchers invoked  two-dimensionalization of the flow to explain the steepened $ E_u(k)$~\cite{Kit:MG1971, Moffatt:JFM1967,Schumann:JFM1976,Moreau:book:MHD}.  However, in the following discussion we present how variable energy flux can explain the steepening of the spectrum~\cite{Verma:PF2015QSMHD,Verma:ROPP2017,Anas:PRF2019}.   In QS-MHD turbulence, the energy injected by the Lorentz force   gets converted to heat by the Joule dissipation $ D_J({\bf k}) $ (see Eq.~(\ref{eq:VF:QSMHD_force})) as shown below:
\bea
 \mathcal{F}_u({\bf k}) = \Re[ {\bf F}_u({\bf k}) \cdot  {\bf u^*(k)}] = -[2N (U/L)  \cos^2 \theta] E_u({\bf k}) = -D_J({\bf k}).
 \eea
In addition, the modal energy spectrum is not isotropic.  However,  the shell spectrum and energy flux, which is averaged over polar angle, obey the following equation~\cite{Verma:PF2015QSMHD, Verma:ROPP2017}:
\bea
\frac{d }{dk} \Pi_u(k) = -[2 \nu k^2 +2 c_2 N (U/L)  ] E_u(k),
\label{eq:VF:QSMHD_dPidk}
\eea
where $ c_2 $ is a constant.  

Anas and Verma~\cite{Anas:PRF2019} solved  Eq.~(\ref{eq:VF:QSMHD_dPidk})  by making an assumption similar to that by Pao~\cite{Pao:PF1965} for 3D hydrodynamic turbulence and derived the following formulas for the energy flux and spectrum.  For $ N \lessapprox 1 $,
\bea
\log \left( \frac{\Pi_u(k)}{\Pi_u(k_0)} \right)  & =  &   - \frac{3}{2} K_\mathrm{Ko}  
[(k/k_d)^{4/3} -(k_0/k_d)^{4/3}]  \nonumber \\
&&  +3  c_2 K_\mathrm{Ko} \left[ (k/k_{d2})^{-2/3} - (k_0/k_{d2})^{-2/3}  \right], \label{eq:VF:QSMHD_Pao_Pik} 
 \eea 
 \be
 E_u(k) = K_\mathrm{Ko} \Pi_u(k) \epsilon^{-1/3}  k^{-5/3},
 \ee
and for $N \gg 1$,
   \bea
E_u(k) & = & A \exp(-k/\bar{k}_d), \\
\Pi_u(k) & = & A \left[ 2 \nu \bar{k}_d ( k^2 + 2 k \bar{k}_d + 2 \bar{k}_d^2) + 2(NU/L) c_2 \bar{k}_d    \right] \exp(-k/\bar{k}_d),
\eea
where $k_{d2}, A, \bar{k}_d    $ are constants.  See Anas and Verma~\cite{Anas:PRF2019} for further details.  The above formulas provide a good description to past experimental and  numerical results, e.g.,~\cite{Alemany:JdeM1979,Eckert:IJHFF2001,Reddy:PF2014}.  Thus, variable energy flux helps describe the QS MHD turbulence quite well.   
 
\subsection{$k^{-1}$ spectrum in shear turbulence}
\label{subsec:k_minus_1_spectrum}

Many turbulent systems exhibit $k^{-1}$ energy spectrum at small wavenumbers or $1/f$ frequency spectrum at small frequencies.     For example, Tichen~\cite{Tchen:JRNBS1952} and Pereira et al.~\cite{Pereira:PRE2019} observed  $k^{-1}$ energy spectrum in shear-driven turbulence. In the solar wind, Matthaeus and Goldstein~\cite{Matthaeus:PRL1986} reported $1/f$ spectrum for  small frequencies  where solar wind jets may create shear turbulence.  Recently, Duguid et al.~\cite{Duguid:MNRAS2020}  reported $ 1/f $ spectrum for the total kinetic energy of  thermal convection at large time scales. 
To explain the $ k^{-1}$ spectrum at small $k$'s,  Tchen~\cite{Tchen:JRNBS1952} modelled  shear turbulence at large scales using Heisenberg's turbulence model~\cite{Heisenberg:PRSA1948};  he argued that the shear induces {\em strong resonance} in the flow, which in turn yields $u_k \sim \mathrm{constant}$.  Hence, $E_u(k) \sim u_k^2/k \sim k^{-1}$.

In the following discussion we derive $ k^{-1} $ spectrum using variable energy flux.   The velocity shear injects kinetic energy to the small wavenumber modes of the flow. Hence, we expect that $  \mathcal{F}_\mathrm{LS}(k) > C $ that leads to an increase in the energy flux $ \Pi_u(k) $ with $ k $, as in case (iv) discussed in Sec.~\ref{subsec:VF_formalism}.  Consequently, the energy spectrum $ E_u(k) \sim [\Pi_u(k)]^{2/3} k^{-5/3} $ is expected to be shallower than $ k^{-5/3} $.  In particular, for $  \mathcal{F}_\mathrm{LS}(k) = C $, Eq.~(\ref{eq:VF:inertial_Pi}) yields $ \Pi_u(k) = \int^k  \mathcal{F}_\mathrm{LS}(k')  dk' = C k $.   Now, using dimensional analysis we  derive that 
\bea
E_u(k) \sim   C^{2/3} k^{-1}.
\eea
As argued in Sec.~\ref{sec:flux}, the energy flux can  be defined for shear turbulence even though it is anisotropic.   The above wavenumber-dependent $  \mathcal{F}_\mathrm{LS}(k)$ is tune with the earlier works by Yakhot and Orszag~\cite{Yakhot:JSC1986} and Sain et al.~\cite{Sain:PRL1998}.

The above mechanism provides a plausible explanation for the $k^{-1}$ energy spectrum in shear turbulence~\cite{Tchen:JRNBS1952,Pereira:PRE2019,Matthaeus:PRL1986}.   The low-frequency $1/f$ spectrum in the solar wind may be due to the shear experienced by wind jets~\cite{Matthaeus:PRL1986}, while that in thermal convection may be due to shear among the large-scale thermal plumes~\cite{Duguid:MNRAS2020}.   Note that we convert the wavenumber spectrum to the frequency spectrum using  Taylor's hypothesis, which is applicable to the solar wind because it is much faster than   spacecrafts~\cite{Taylor:PRSA1954}. In thermal convection, Taylor's hypothesis is expected to work under certain conditions~\cite{Lohse:ARFM2010,Kumar:RSOS2018,Verma:INAE2020_sweeping}. 

The above theory for $k^{-1}$ spectrum in shear turbulence hinges on the assumption that $ \mathcal{F}_u({\bf k}) = C$, which needs to be tested using experiments and/or numerical simulations.   Interestingly, $1/f$ noise has been reported in a large number of physical systems---electric currents, ion-channel currents, music, earthquakes, etc.~(see \cite{Dutta:RMP1981,Banerjee:EPL2006,Verma:EPL2006}, and references therein).  It is possible that $1/f$ spectra in the electric and ion-channel currents are  connected to shear in the  electron flow.  This conjecture, however, needs to be tested.

In addition to the above, there are many more examples of variable energy flux.  In stably-stratified turbulence, $\Pi_u(k)$ decreases as $k^{-4/5}$ due to buoyancy~\cite{Bolgiano:JGR1959,Obukhov:DANS1959}.  On the contrary, $\Pi_u(k)$  increases marginally in turbulent thermal convection.  We will describe these fluxes in Sec.~\ref{sec:buoyant}.   In MHD turbulence, the kinetic energy flux varies in the inertial range itself due to the Lorentz force.  Similar variations are observed in solvents with polymers, quantum turbulence, binary-mixture turbulence, etc.  We will discuss these systems in Sections~\ref{sec:mhd} and \ref{sec:QT}.     In addition, the energy flux variations can be generalised to other quantities such as enstrophy and kinetic helicity; these issues will be discussed in Sec.~\ref{sec:enstrophyHk}.  

With this, we close our brief discussion on variable energy flux in hydrodynamic turbulence.  In the next section,  we describe such variations in flows with a secondary field, e.g., temperature,  magnetic field.

\section{Variable energy flux in flows with a secondary field: Formalism}
\label{sec:secondary}

Consider a secondary field  $\zeta$  advected by the velocity field ${\bf u}$.  This secondary field could be a scalar, a vector, or a tensor.  Leading examples of a scalar field are  density and temperature  of a fluid; that of a vector field are  magnetic field, dipolar field, and  flock velocity; and that of a tensor are the configuration tensor of a polymer and  stress tensor of an elastic fluid.  In this section, we present energy fluxes associated with a secondary field and those arising due to  interactions between the velocity and secondary fields.  As we describe below, many features of energy transfers  are common among the scalar, vector, and tensor secondary fields.  

 \begin{table}
	\begin{center}
		\caption{Energy ($E_\zeta$) and  modal energy ($E_\zeta({\bf k})$) of a secondary field}
		\label{tab:secondary:Ek}
		\begin{tabular}{lccc} 
			\hline\noalign{\smallskip}
			& scalar & vector & tensor \\  
			\noalign{\smallskip}\hline\noalign{\smallskip}
			$E_\zeta$ & $ \frac{1}{2} \la \zeta^2 \ra$ &  $\frac{1}{2}  \la  \bm{\zeta} \cdot  \bm{\zeta}  \ra $  &$ \frac{1}{2} \la \zeta_{ij} \zeta_{ij} \ra $  \\ 
			$E_\zeta({\bf k})$ & $ \frac{1}{2}  |\zeta({\bf k})|^2 $ &  $\frac{1}{2}  |  \bm{\zeta}({\bf k})|^2 $  &$ \frac{1}{2} \zeta_{ij}({\bf k}) \zeta^*_{ij}({\bf k})  $  \\ 
			\noalign{\smallskip}\hline
		\end{tabular}
	\end{center}
\end{table}

\subsection{Variable energy flux associated with a secondary field}
The equations for the velocity field ${\bf u}$ are same as those covered in Sec.~\ref{sec:flux}, except that the force field for the velocity field (${\bf F}_u$) could be a function of ${\bf u}$, $\zeta$, ${\bf r}$, and $t$.  The equations for the secondary field are
\bea
\mathrm{Scalar}: \frac{\partial{\zeta}}{\partial t} + ({\bf u}\cdot\nabla){\zeta}
& = &   F_\zeta({\bf u},\zeta) +  \kappa \nabla^2 {\zeta},   \label{eq:secondary:scalar} \\
\mathrm{Vector}:  \frac{\partial{ \bm{\zeta}}}{\partial t} + ({\bf u}\cdot\nabla){ \bm{\zeta}}
& = &    {\bf F}_\zeta({\bf u},  \bm{\zeta}) +  \kappa \nabla^2 { \bm{ \zeta}},   \label{eq:secondary:vector} \\
\mathrm{Tensor}:  \frac{\partial{{\zeta}_{ij}}}{\partial t} + ({\bf u}\cdot\nabla){{\zeta}_{ij}}
& = &   F_{{\zeta},ij}({\bf u},\zeta) + \kappa \nabla^2 {{\zeta}_{ij}}  ,   \label{eq:secondary:tensor} 
 \eea
 where $\kappa$ is the diffusion coefficient of the secondary field, and $F_\zeta$ is  the force field for the secondary field.    Two important nondimensional parameters are Prandtl number $\mathrm{Pr} = \nu/\kappa$ and P\'{e}let number $\mathrm{Pe}$, which is the ratio of the nonlinear term and the diffusion term in the equation for the secondary field, that is,
  \bea
 \mathrm{Pe} =  \frac{({\bf u}\cdot\nabla){\zeta}}{\kappa \nabla^2 {\zeta} } =   \frac{UL}{\kappa},
 \eea
 where $U,L$ are the large-scale velocity and length  respectively.

Similar to the kinetic energy, we define secondary energy and associated modal energy, as listed in Table~\ref{tab:secondary:Ek}.  For discrete wavenumbers,  one-dimensional secondary energy spectrum is defined as  $ E_\zeta(k) = \sum_{k-1 < k' \le k}  E_\zeta({\bf k'}) $, but for continuum wavenumbers,   $  E_\zeta(k) dk = \sum_{k < k' \le k+dk}  E_\zeta({\bf k'}) $.  The  evolution equation for the modal secondary energy  of a scalar is
 \bea
 \frac{d}{dt} E_\zeta(\mathbf{k}) & = &  T_\zeta({\bf k}) + \mathcal{F}_\zeta({\bf k}) - D_\zeta({\bf k}) \nonumber \\
 &=&  \sum_{\bf p} \Im \left[ {\bf \{  k \cdot u(q) \}  \{ \zeta(p) \zeta^*(k) \} }  \right] + \Re[ F_\zeta({\bf k}) \zeta^*({\bf k})] - 2 \kappa k^{2} E_\zeta({\mathbf k}).   \nonumber \\
 \label{eq:secondary:Ethetak}
  \eea  
  In the above equation, $T_\zeta({\bf k})$ is the nonlinear transfer of secondary energy to $\zeta({\bf k})$,   $\mathcal{F}_\zeta({\bf k})$ is the secondary energy transfer to  $\zeta({\bf k})$ by $F_\zeta({\bf k})$, and $ D_\zeta({\bf k}) $ is the diffusion or dissipation rate of $\zeta({\bf k})$.  The equations for the vector and tensor fields are very similar to the above equations, except that  the  field multiplication is performed appropriately  (scalar product or tensor product).   
  
  Several  important points regarding the secondary field are
 \begin{enumerate}
\item When $F_\zeta$ is a {\em linear function} of  ${\bf u} $,  the scalar energy injection rate $\mathcal{F}_\zeta({\bf k}) =  \Re[ F_\zeta({\bf k}) \zeta^*({\bf k})] $ is a function of ${\bf k}$,  ${\bf u(k)}$, and $\zeta({\bf k}) $.  We encounter such forms of $F_\zeta({\bf k}) $  in stably stratified turbulence and in thermal convection where $F_\zeta \propto u_z$.  Similar properties hold for $\mathcal{F}_u({\bf k})$ when $ {\bf F}_u $ is a linear function of $\zeta$.

\item When $F_\zeta$ is a {\em nonlinear function} of  ${\bf u} $ and/or $\zeta$,  the scalar energy injection rate   $\mathcal{F}_\zeta({\bf k})$ is a convolution.  Therefore, $\mathcal{F}_\zeta({\bf k}) $ involves wavenumbers other than ${\bf k}$.   For example,   in MHD turbulence, where ${\bf F}_{ \bm{\zeta}}  =  \bm{\zeta} \cdot \nabla {\bf u}$ with $  \bm{\zeta}$ as the magnetic field,\bea
\mathcal{F}_\zeta({\bf k}) & = & \sum_{\bf p} - \Im \left[ {\bf \{  k \cdot  \bm{\zeta}(k-p) \}  \{ u(p) \cdot  \bm{\zeta}^*(k)  \} } \right]
\eea
is a convolution.  The  nonlinear $F_u$ yields a similar convolution.
\end{enumerate}

The nonlinear term $ ({\bf u} \cdot \nabla) \zeta$ (and similar ones for vector and tensor) for the secondary field  facilitates scalar energy transfer.   For a wavenumber triad $ ({\bf k',p,q}) $, the {\em mode-to-mode secondary energy transfer} from wavenumber $ \textbf{p} $ to wavenumber $ \textbf{k} $ with the mediation of wavenumber $ \textbf{q} $ is~\cite{Verma:IJMPB2001,Verma:NJP2017,Verma:book:BDF}:
\bea
\mathrm{Scalar}:  S^{\zeta \zeta}({\bf k'|p|q}) & = & -\Im \left[  {\bf  \{  k' \cdot u(q) \} \{ {\zeta}({\bf p})  {\zeta}({\bf k'}) \} }  \right] ,\label{eq:secondary:M2M_scalar} \\
\mathrm{Vector}:  S^{\zeta \zeta}({\bf k'|p|q}) & = & -\Im \left[  {\bf  \{  k' \cdot u(q) \} \{ \bm{\zeta}({\bf p}) \cdot   \bm{\zeta}({\bf k'}) \} }  \right] , \label{eq:secondary:M2M_vector} \\
\mathrm{Tensor}:  S^{\zeta \zeta}({\bf k'|p|q}) & = & -\Im \left[  {\bf  \{  k' \cdot u(q) \}} \{ {\zeta}_{ij}({\bf p})  {\zeta}_{ij}({\bf k'}) \}  \right] .\label{eq:secondary:M2M_tensor} 
\eea
In the above equations, the giver and receiver modes are from the secondary field, while a velocity mode acts as a mediator for the secondary energy transfer. The aforementioned form of  energy transfer also follows from the structure of nonlinear term $ ({\bf u} \cdot \nabla) \zeta$ where $ {\bf u}$ advects the scalar field $ \zeta $.  Also, the superscript of $ S^{\zeta \zeta} $ refer to the receiver and giver fields, both being $ \zeta $.

Using incompressibility condition, we can show that the mode-to-mode  secondary energy transfer functions satisfy the following property:
\bea
S^{\zeta \zeta}({\bf k'|p|q}) = - S^{\zeta \zeta}({\bf p|k'|q}) .
\label{eq:secondary:Szetazeta_cancel}
\eea
 Using Eq.~(\ref{eq:secondary:Szetazeta_cancel}) we deduce that for a wavenumber region $A$ (including a triad),
\bea
\sum_{{\bf k'} \in A} \sum_{{\bf p} \in A}  S^{\zeta \zeta}({\bf k'|p|q}) =0.
\eea
 This  relation also implies that $E_\zeta$ is conserved when $F_\zeta =0$ and $\kappa=0$.   Using the formula for the mode-to-mode energy transfers, we  define the secondary energy flux for a wavenumber sphere of radius $k_0$ as~\cite{Leslie:book,Lesieur:book:Turbulence,Verma:IJMPB2001,Frisch:book,Verma:NJP2017,Verma:book:BDF}:
\bea
\Pi_{\zeta}(k_0) = \sum_{k'>k_0}   \sum_{p\le k_0} S^{\zeta \zeta}(\mathbf{k'|p|q}) .
\label{eq:secondary:secondary_flux}
\eea
Here, $\Pi_{\zeta}(k_0) $ is the net secondary energy transfer from all the modes inside the sphere to all the modes outside the sphere.   

Following similar lines of arguments as in   Sec.~\ref{sec:VF}, we derive the following evolution equation for the  scalar energy spectrum $E_\zeta(k)$~\cite{Lesieur:book:Turbulence,Verma:book:ET}:
\bea
\frac{\partial }{\partial t}   E_\zeta(k,t) =  -\frac{\partial }{\partial k} \Pi_\zeta(k,t) + \mathcal{F}_\zeta(k,t)  - D_\zeta(k,t),
\label{eq:secondary:Ek_energetics_Ezeta}
\eea
 where  
  \bea
 E_{\zeta}(k) dk & = & \sum_{k < k' \le k+dk} E_\zeta({\bf k'}), \\
\mathcal{F}_{\zeta}(k) dk & = & \sum_{k < k' \le k+dk}  \Re [F_\zeta({\bf k'}) \zeta^*({\bf k'})], \label{eq:secondary:energy_inj}  \\
D_{\zeta}(k) dk& = & \sum_{k < k' \le k+dk}   2 \kappa k'^2 E_\zeta({\bf k'}).
 \eea
Here, $\mathcal{F}_{\zeta}(k)$ represents  the scalar energy supply rate  by $F_\zeta$ to shell $k$, and $D_{\zeta}(k) $ represents the diffusion or dissipation rate of the scalar energy in  shell $k$.   For the vector and tensor secondary fields, Eq.~(\ref{eq:secondary:energy_inj}) involves vector and tensor products respectively. 
For a steady state ($\partial   E_\zeta(k)/\partial t = 0$), Eq.~(\ref{eq:secondary:Ek_energetics_Ezeta}) yields
 \bea
 \frac{d }{dk}  \Pi_\zeta(k) =  \mathcal{F}_\zeta(k)  -D_\zeta(k) .
\label{eq:secondary:dPi_dk_secondary}
\eea
 Thus, the energy flux of a secondary field is affected by $\mathcal{F}_\zeta(k) $ and $D_\zeta(k) $.   We may obtain a steady state  when $ \zeta $ field is forced at large scales. 
 
 In the inertial range, $D_\zeta(k) \approx 0$, hence
 \bea
 \frac{d}{dk} \Pi_\zeta = \mathcal{F}_{\zeta,\mathrm{in}}(k),
 \label{eq:secondary:dPi_z_dk_steady_inertial}
 \eea
 where the subscript ``in" of $\mathcal{F}_{\zeta,in}$ represents the inertial range.   Therefore, in the inertial range, similar to the description of  Sec.~\ref{sec:VF},  variations of $\Pi_\zeta(k) $  can be classified into four categories:
  \begin{enumerate}
\item $\mathcal{F}_{\zeta,\mathrm{in}}(k) = 0$ and $\Pi_\zeta(k) = 0$, 

 \item $\mathcal{F}_{\zeta,\mathrm{in}}(k) = 0$ and $\Pi_\zeta(k) = C > 0$ ,
 
 \item $\mathcal{F}_{\zeta,\mathrm{in}}(k) < 0$ and $d\Pi_\zeta(k)/dk < 0$ ,
 
 \item $\mathcal{F}_{\zeta,\mathrm{in}}(k) > 0$ and $d\Pi_\zeta(k)/dk > 0$ .
 
\end{enumerate}
The interpretation of the above four cases are very similar to those for the kinetic energy flux discussed in Sec.~\ref{sec:VF} and exhibited in Fig.~\ref{fig:VF:energetics}.  Note that the first case corresponds to the equilibrium configuration for the secondary field. We will describe these  cases in the subsequent sections. 

\subsection{Cross energy transfers  between the velocity and secondary fields}
\label{subsec:cross_energy}
 
\begin{figure}[htbp]
\begin{center}
\includegraphics[scale = 0.6]{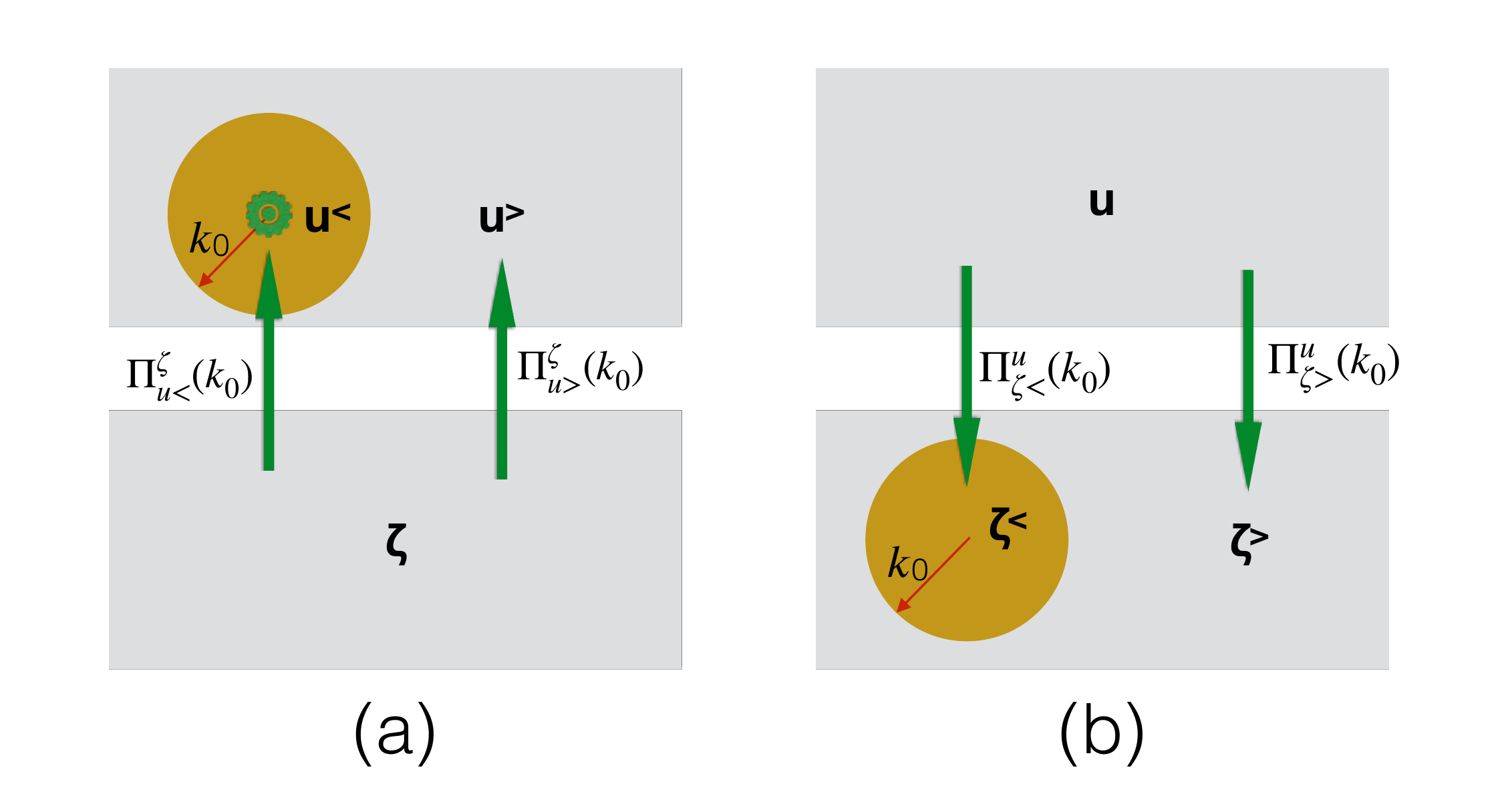}
\end{center}
\caption{(a) The flux $\Pi^\zeta_{u<}(k_0)$ ($\Pi^\zeta_{u>}(k_0)$) represents the energy transfer from $\zeta$ modes to the velocity modes inside (outside) the sphere. (b) The flux $\Pi^u_{\zeta<}(k_0) $ ($  \Pi^u_{\zeta>}(k_0) $) represents the reverse energy transfers, i.e., from all the velocity modes to $\zeta$ modes inside  (outside) the sphere. }
\label{fig:second:second_overall}
\end{figure}

 In this subsection we describe the energy transfers from the velocity field to the secondary field and vice versa.  We consider $k_0$ beyond $k_f$ and  rewrite Eq.~(\ref{eq:ET:sum_Ek_less}) as~\cite{Lesieur:book:Turbulence,Verma:book:ET}
 \bea
 \frac{d}{dt} \sum_{k \le k_0} E_u({\bf k}, t)  & = &  \sum_{k \le k_0} T_u +    \sum_{k \le k_0} \mathcal{F}_u({\bf k}) +    \sum_{k \le k_0} \mathcal{F}_\mathrm{LS}({\bf k}) - \sum_{k \le k_0} D_u({\bf k}) \nonumber \\
 & = &  -\Pi_u(k_0)  +  \Pi^\zeta_{u<}(k_0) + \varepsilon_\mathrm{inj} - \sum_{k \le k_0} D_u({\bf k}), 
 \label{eq:secondary:sum_Ek}
  \eea 
  where $\varepsilon_\mathrm{inj}$ is the kinetic energy injection rate by the large-scale force, and  $   \Pi^\zeta_{u<}(k_0)   $  is the net energy transfer from all the $ \bm{\zeta}$ modes to the velocity modes within the sphere of radius $k_0$.   In $  \Pi^\zeta_{u<}(k_0)$, the superscript and subscripts denote the giver and receiver field  variables respectively, while $ < $ denotes the modes within the sphere.  In the same vein, we define
  \bea
  \Pi^\zeta_{u>}(k_0)  = \sum_{k > k_0} \mathcal{F}_u({\bf k})
  \label{eq:secondary:Pi_zeta_u>}
  \eea
 as the net energy transfer from all the $ \bm{\zeta}$ modes to the velocity modes outside (represented by the symbol $>$) the sphere of radius $k_0$.  Using the evolution equation for $E_\zeta(\mathbf{k})$ we deduce that  
\bea
\Pi^u_{\zeta<}(k_0) =  \sum_{k \le k_0} \mathcal{F}_\zeta({\bf k});~~~\Pi^u_{\zeta>}(k_0)  = \sum_{k > k_0} \mathcal{F}_\zeta({\bf k})
\eea
are  the respective   energy transfers from all the velocity  modes to the  $\zeta$ modes inside and outside of the sphere of radius $k_0$.  Figure~\ref{fig:second:second_overall} illustrates these fluxes. 

The net energy transfer from $\zeta$ to ${\bf u}$ is
\bea
\mathcal{F}_u =  \sum_{\bf k} \mathcal{F}_u({\bf k}) =   \Pi^\zeta_{u<}(k) +   \Pi^\zeta_{u>}(k) .
  \label{eq:secondary:Pi_zeta_u_sum}
\eea
Since $\mathcal{F}_u $ is a fixed number, the sum $\Pi^\zeta_{u<}(k) +   \Pi^\zeta_{u>}(k)$ is constant in $k$.  Similarly, the net energy transfer from ${\bf u}$ to $\zeta$ is 
\bea
\mathcal{F}_\zeta =  \sum_{\bf k} \mathcal{F}_\zeta({\bf k}) =   \Pi^u_{\zeta<}(k) +   \Pi^u_{\zeta>}(k) .
  \label{eq:secondary:Pi_u_zeta_sum}
\eea
with the sum $  \Pi^u_{\zeta<}(k) +   \Pi^u_{\zeta>}(k) $ as a  constant in $k$.  Note however that the individual fluxes (e.g., $ \Pi^u_{\zeta<}(k) $) may vary with $k$.  For some systems,  $\int d{\bf r} (u^2+\zeta^2)/2$  represents the total energy and 
\bea
\mathcal{F}_u+\mathcal{F}_\zeta = 0~~~\mathrm{or}~~~\Pi^\zeta_{u<}(k) +   \Pi^\zeta_{u>}(k) +  \Pi^u_{\zeta<}(k) +   \Pi^u_{\zeta>}(k) =0.
\label{eq:secondary:F_u_F_zeta_sum_energy_cons}
\eea
 MHD turbulence and stably stratified turbulence are examples of such systems; they will be discussed later in this review.

\begin{figure}[htbp]
\begin{center}
\includegraphics[scale = 0.5]{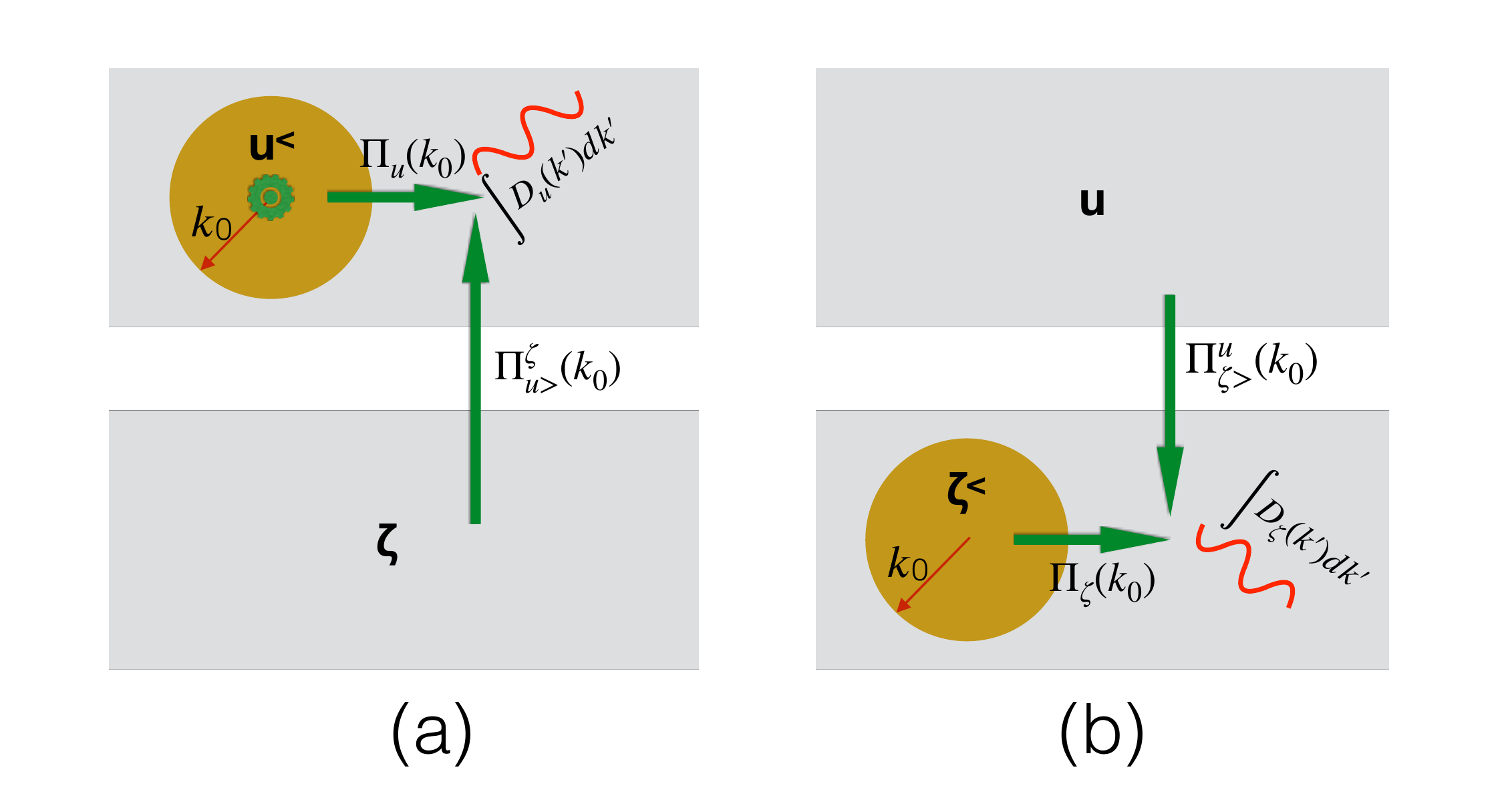}
\end{center}
\caption{Schematic diagram illustrating the identities (a) $\Pi_u(k_0)+ \Pi^\zeta_{u>}(k_0)  = \int_{k_0}^\infty dk' D_u(k')$; (b) $\Pi_\zeta(k_0) + \Pi^u_{\zeta>}(k_0) = \int_{k_0}^\infty dk' D_\zeta(k')$.   The wheel in the centre of (a) represents kinetic energy injection rate $\varepsilon_\mathrm{inj}$. }
\label{fig:second:identities}
\end{figure}

\begin{figure}[htbp]
\begin{center}
\includegraphics[scale = 0.6]{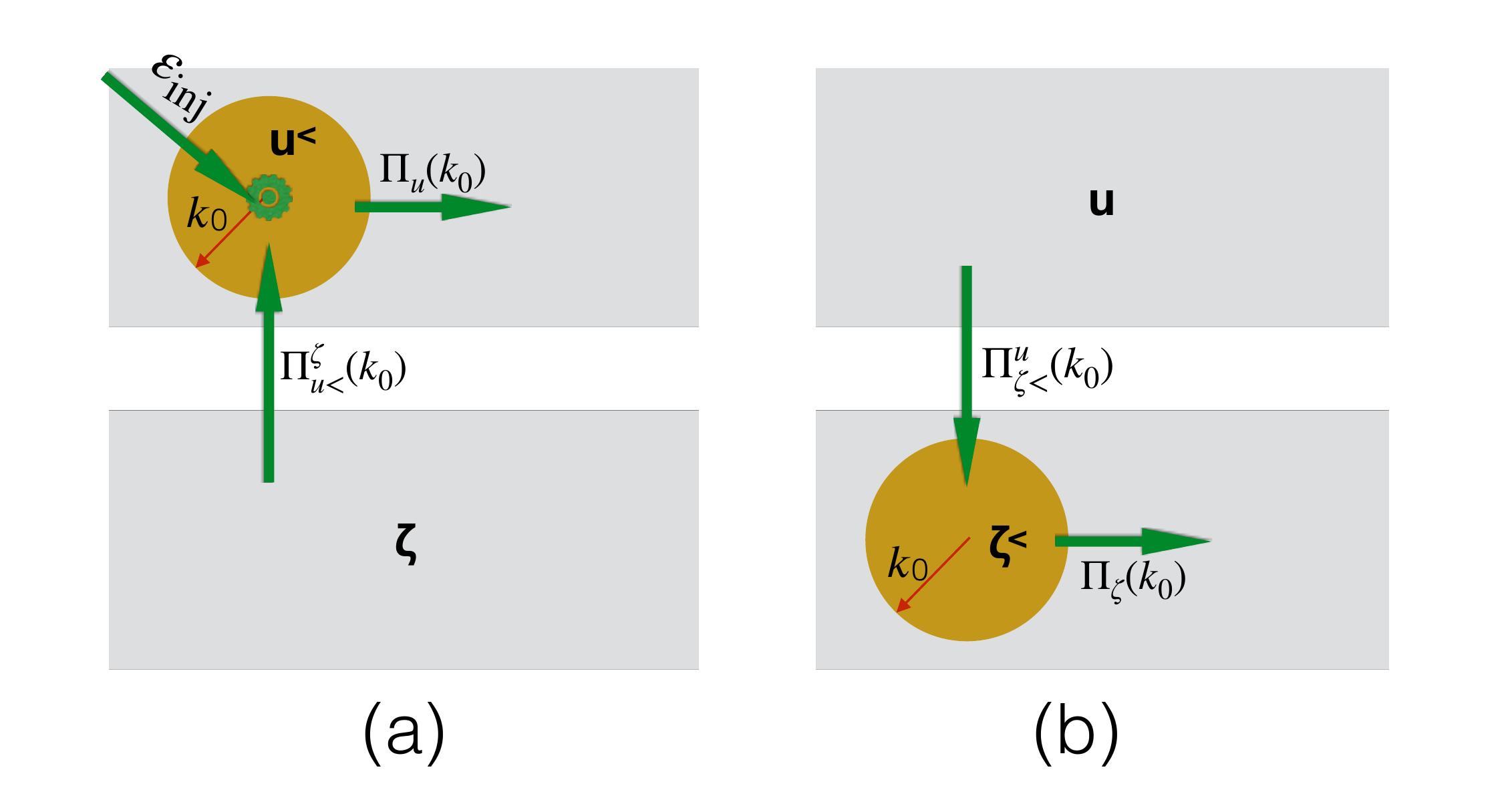}
\end{center}
\caption{ Schematic diagram illustrating identities for a steady state: (a) $\Pi_u(k_0) \approx  \Pi_{u<}^\zeta(k_0) + \varepsilon_\mathrm{inj}$; (b) $\Pi_\zeta(k_0)  \approx \Pi^u_{\zeta<}(k_0)$.  The wheel in the centre of (a) represents kinetic energy injection rate $\varepsilon_\mathrm{inj}$. }
\label{fig:second:identities2}
\end{figure}

Using the definitions of the above fluxes, we derive the following relations~\cite{Verma:book:ET}:
  \bea
   \frac{d}{dk} \Pi^\zeta_{u>}(k) = - \mathcal{F}_u(k);~~~
   \frac{d}{dk} \Pi^\zeta_{u<}(k) = \mathcal{F}_u(k); \nonumber \\
 \frac{d}{dk} \Pi^u_{\zeta>}(k) =  - \mathcal{F}_\zeta(k);~~~ 
  \frac{d}{dk} \Pi^u_{\zeta<}(k) =  \mathcal{F}_\zeta(k).
  \label{eq:secondary:Pi_u_zeta_F_zeta}
  \eea 
If we assume an absence of large scale forcing for $ \zeta $, then  the fluxes  obey the following properties under steady state:
   \bea
  \Pi_u(k_0)+ \Pi^\zeta_{u>}(k_0) =  \int_{k_0}^\infty dk D_u(k',t),  \label{eq:secondary:flux_1} \\
   \Pi_\zeta(k_0) + \Pi^u_{\zeta>}(k_0) =  \int_{k_0}^\infty dk D_\zeta(k), \\
    \Pi_u(k_0) =  \Pi_{u<}^\zeta(k_0) + \varepsilon_\mathrm{inj}, \\
     \Pi_\zeta(k_0)  = \Pi^u_{\zeta<}(k_0). \label{eq:secondary:flux_4} 
  \eea
For $ k_0 $ in the inertial range, 
$ \int_{k_0}^\infty dk D_{u,\zeta}(k',t) \approx \epsilon_{u,\zeta} $ respectively.   The above identities follow from the energy conservation.     See Figs.~\ref{fig:second:identities} and~\ref{fig:second:identities2} for  illustrations.
   
In  subsequent sections, we will describe the role of variable energy flux in several turbulent flows with secondary fields, namely, passive scalar flow, buoyancy driven flows,  magnetohydrodynamics, and polymer solution.   Note that in the turbulent limit, the nonlinear terms ${\bf (u \cdot \nabla) u}$ and ${\bf (u \cdot \nabla)  \bm{\zeta}}$ dominate the respective dissipation terms, and hence  $\mathrm{Re} \gg 1$ and $\mathrm{Pe} \gg 1$.  

In the next subsection we describe the scaling of passive secondary turbulence. 

\subsection{Turbulence with a passive secondary field}
\label{subsec:passive_secondary}
Consider a flow whose  $\mathcal{F}_{u}$ is independent of  the secondary field $\zeta$ and, hence, the velocity field is unaffected by the secondary field.  However, the secondary field $\zeta$ is affected by ${\bf u}$.  This is the reason why such a $\zeta$  is called a {\em passive secondary field}~ \cite{Monin:book:v1,Monin:book:v2,Lesieur:book:Turbulence, Leslie:book,Frisch:book, McComb:book:Turbulence}.  We will discuss the properties of such flows in this  subsection.

If the forces on the velocity and the passive secondary fields are active only at large scales, then $\mathcal{F}_{u,\mathrm{in}}(k)=0$ and  $\mathcal{F}_{\zeta,\mathrm{in}}(k) = 0$.   For such field configurations, using Eqs.~(\ref{eq:VF:inertial_Pi}, \ref{eq:secondary:dPi_z_dk_steady_inertial}) we deduce that $\Pi_u(k)$ and  $\Pi_\zeta(k)$ are constant in the inertial range.  Since the velocity field is unaffected by $\zeta$, the kinetic energy is described by Kolmogorov's theory of turbulence (see  Sec.~\ref{subsec:Kolm_theory_turb_Fourier}). Therefore, the kinetic energy spectrum is given by Eq.~(\ref{eq:VF:Kolm_Ek}).   

In the inertial range, $  \Pi_\zeta(k) = \mathrm{const} = \epsilon_\zeta $, 
where $\epsilon_\zeta$ is  the  energy dissipation rate of the secondary field.   Using dimension analysis and similar arguments as in Sec.~\ref{subsec:Kolm_theory_turb_Fourier}, one obtains~\cite{Lesieur:book:Turbulence,McComb:book:Turbulence,Frisch:book}
\bea
E_\zeta(k) = K_\mathrm{OC} \epsilon_\zeta (\epsilon_u)^{-1/3} k^{-5/3} ,
\label{eq:secondary:Etheta}
\eea
where $K_\mathrm{OC}$ is the {\em Obukhov-Corrsin constant}.  The above scaling has been verified using several numerical simulations and experiments~ \cite{Lesieur:book:Turbulence, McComb:book:Turbulence, Davidson:book:Turbulence,Frisch:book,Yeung:PF2005, Sreenivasan:PNAS2018}.   For the inertial-dissipation range, using arguments similar to those in Sec.~\ref{subsec:hydro_inertial_diss}, Pao~\cite{Pao:PF1968} derived that  
\bea
\Pi_\zeta(k)  & =  & \epsilon_\zeta  \exp{\left(- \frac{3}{2} K_\mathrm{OC} (k/k_c)^{4/3}\right)}, \label{eq:passive_scalar:Pao_Piscalar}\\
E_\zeta(k)  & = & K_\mathrm{OC}  \epsilon_\zeta \epsilon_u^{-1/3} k^{-5/3}  \exp{\left(- \frac{3}{2} K_\mathrm{OC}  (k/k_c)^{4/3}\right)}, \label{eq:passive_scalar:Pao_Escalar}
\eea 
where  $ k_c =   \left(\epsilon_u/\kappa^3\right)^{1/4} $ 
is  Kolmogorov's diffusion wavenumber.   Hence, $ k_c/k_d = \mathrm{Pr}^{3/4}$.

The above arguments are valid when  $\mathrm{Re} \gg 1$ and $\mathrm{Pe} \gg 1$.  The spectral properties are quite different for other regimes~\cite{Gotoh:book_chapter:passive_scalar,Donzis:FTC2010,Verma:book:BDF}, which are not discussed in this review.

 \subsection{Variable energy flux in anisotropic turbulence}
 \label{subsec:anisotropic_turb}
 
 Typically, a fluid flow becomes  anisotropic in the presence of a strong external field (e.g., mean magnetic field, buoyancy, external rotation field).  In such flows, we denote the velocity components perpendicular and parallel to the external field as ${\bf u}_\perp$ and $u_\parallel$, and the corresponding  energy spectra as 
 $  E_{u,\parallel}({\bf k}) = \frac{1}{2} |u_\parallel({\bf k})|^2 $ and $  E_{u,\perp}({\bf k}) =  \frac{1}{2} |{\bf u_\perp(k)}|^2 $ respectively.   The corresponding  energy fluxes are~ \cite{Verma:ROPP2017, Verma:book:BDF}:
 \bea
 \Pi_{u,\parallel}(k_0) & = & \sum_{k'>k_0}  \sum_{p\le k_0} -\Im\left\{ \left[{\bf k}'\cdot{\bf{u}}({\bf q})\right]\left[{u}_{\parallel}({\bf k}'){u}_{\parallel}({\bf p})\right]\right\}, \\
 \Pi_{u,\perp}(k_0) & = & \sum_{k'>k_0}  \sum_{p\le k_0} -\Im\left\{ \left[{\bf k}' \cdot{\bf{u}}({\bf q})\right]\left[{\bf{u}}_{\perp}({\bf k}')\cdot{\bf{u}}_{\perp}({\bf p})\right]\right\}, 
 \eea 
 where ${\bf k=p+q}$ and $ {\bf k'} = - {\bf k}$.
 
 The evolution equations for the one-dimensional spectra $E_{u,\perp}(k)$ and $E_{u,\parallel}(k) $ are 
 \bea
 \frac{\partial}{\partial t}   E_{u,\parallel}(k,t) & = & -\frac{\partial}{\partial k} \Pi_{u,\parallel}(k,t) +\mathcal{P}(k,t)  -2 \nu k^2 E_{u,\parallel}(k,t) + \mathcal{F}_{u,\parallel}(k) + \mathcal{F}_\mathrm{LS,\parallel}(k), \nonumber \\
 \label{eq:misc:Ek_pll}  \\
 \frac{\partial}{\partial t}   E_{u,\perp}(k,t) & = & -\frac{\partial}{\partial k} \Pi_{u,\perp}(k,t) -\mathcal{P}(k,t)  -2 \nu k^2 E_{u,\perp}(k,t) + \mathcal{F}_{u,\perp}(k) + \mathcal{F}_\mathrm{LS,\perp}(k), \nonumber \\ \label{eq:misc:Ek_perp} 
 \eea 
 where
 \bea
 \mathcal{P}(k)dk =\sum_{k < k' \le k+dk} \Im\left\{ k_{\parallel} {p}({\bf k}) u^*_\parallel({\bf k} ) \right\},  \label{eq:anisotropy:pperp}
 \eea
 and $\mathcal{F}_{u,\parallel}(k), \mathcal{F}_\mathrm{LS,\parallel}(k)$,  $\mathcal{F}_{u,\perp}(k)$, and $ \mathcal{F}_\mathrm{LS,\perp}(k)$ are the energy injection rates to $u_\parallel$ and ${\bf u}_\perp$ by the parallel and perpendicular components of  ${\bf F}_u$ and ${\bf F}_{\mathrm{LS}}$.   Under a steady state and in the inertial range with ${\bf F}_u=0$ and ${\bf F}_{\mathrm{LS}}=0$,
 \bea
 \frac{d}{d k} \Pi_{u,\parallel}(k)  = -  \frac{d}{d k} \Pi_{u,\perp}(k) = \mathcal{P}(k).   
 \eea 
 Thus, $\Pi_{u,\perp}(k)$ and $\Pi_{u,\parallel}(k)$ vary with $k$.  However, 
 \bea
 \frac{d}{d k} \Pi_u(k) =  \frac{d}{d k} [\Pi_\parallel(k) + \Pi_{u,\perp} (k) ] = 0.
 \eea
 That is, the energy flux $\Pi_u(k)$ is constant in inertial range,  as  expected from Kolmogorov's theory of turbulence.  Note that $  \mathcal{P}(k)  $ facilitates energy transfers between the perpendicular and parallel components of the velocity field.

 For QS MHD turbulence with a strong mean magnetic field, Reddy et al.~\cite{Reddy:PP2014} showed that $\mathcal{P}(k) $ is positive, and hence there is an energy transfer from $u_\perp$ to $u_\parallel$.  In such flows, $u_\perp$ is stronger than $u_\parallel$.  MHD turbulence  exhibits a similar behaviour~ \cite{Sundar:PP2017}.  The situation however is reversed in thermal convection where $\mathcal{P}(k) <0$ that leads to an energy transfer from  $u_\parallel$ to $u_\perp$.  
 
 The anisotropic energy transfers are  more complex in rotating turbulence.  Here, $\mathcal{F}_u({\bf k}) = 0$ because the velocity field is perpendicular to the Coriolis force ${\bf u} \times  \bm{\Omega}$.  However, the Coriolis force and the pressure field induce  asymmetry between $u_\parallel$ and ${\bf u}_\perp$.  The pressure field due by the Coriolis force  ($ p_\mathrm{Co} $) generates the following  $\mathcal{P}({\bf k}) $: 
 \bea
 \mathcal{P}({\bf k})  & = & k_\parallel \Im [ p_\mathrm{Co}({\bf k}) u^*_\parallel({\bf k}) ]
 \nonumber \\
 & = & 2 k_\parallel \Im [ i \frac{\Omega}{k} \sin \zeta u_1({\bf k}) u_2^*({\bf k}) \sin \zeta ] \nonumber \\
 & = &  \frac{ 2 \Omega k_\parallel}{k}  \sin^2 \zeta \Re[u_1({\bf k}) u_2^*({\bf k})],
 \eea
 where $u_1, u_2$ are components of the velocity field in Craya-Herring basis~\cite{Craya:thesis,Waleffe:PF1992,Verma:book:ET}.  The corresponding contribution by the nonlinear term  $( {\bf u \cdot \nabla}) {\bf u} $ is
 \bea
 \mathcal{P}({\bf k})  & = & k_\parallel \Im [ p_\mathrm{nlin}({\bf k}) u^*_\parallel({\bf k}) ]
 \nonumber \\
 & = &  -\frac{k_\parallel}{k^2} \sum_{\bf p} \Im [  u_\parallel^*({\bf k}) {\bf \{ k \cdot u(k-p) \}  \{ k \cdot u(p) \} } ]  .
 \eea
 The aforementioned quantities induce energy exchange   between $u_\parallel$ and ${\bf u}_\perp$, and they need to be quantified in future.   In Sec.~\ref{subsec:Q2D} we  show how the above energy transfers take an active part in quasi-2D turbulence generated by strong rotation, external magnetic field, or gravity.    We refer the reader to  earlier works~\cite{Sharma:PF2018, Sharma:PF2019, Waleffe:PF1992,Alexakis:PR2018} for further details.  We also remark that similar formulas need to be derived for the secondary fields as well.  For example, it will be interesting to investigate how the parallel and perpendicular components of the magnetic field exchange energy among themselves.

In the next section we describe how the ideas of variable energy flux yields interesting insights into the physics of buoyancy-driven turbulence.

\section{Variable energy flux in buoyancy-driven turbulence}
\label{sec:buoyant}
Buoyancy-driven flows can be broadly classified into two categories: stable and unstable~ \cite{Tritton:book,Davidson:book:TurbulenceRotating,Sagaut:book,Verma:book:BDF}.  We will show below that the properties of these two categories of flows are very different.   For brevity, our  focus would be on flows with linear stratification, which is a good approximation for a small region of planetary or stellar atmospheres.  

In the next two subsections we will describe turbulence phenomenologies of stably stratified and unstably stratified flows.  Since gravity affects the velocity field, the secondary fields in such flows are called {\em active fields}.  The buoyant flows are typically anisotropic due to external gravity.  Still, one-dimensional energy spectrum and flux are often employed to characterize such flows because they provide cumulative effects over the polar angles (angle between the buoyancy direction and wavenumber $ \textbf{k} $).

\subsection{Stably stratified turbulence}
\begin{figure}[htbp]
\begin{center}
\includegraphics[scale=0.4]{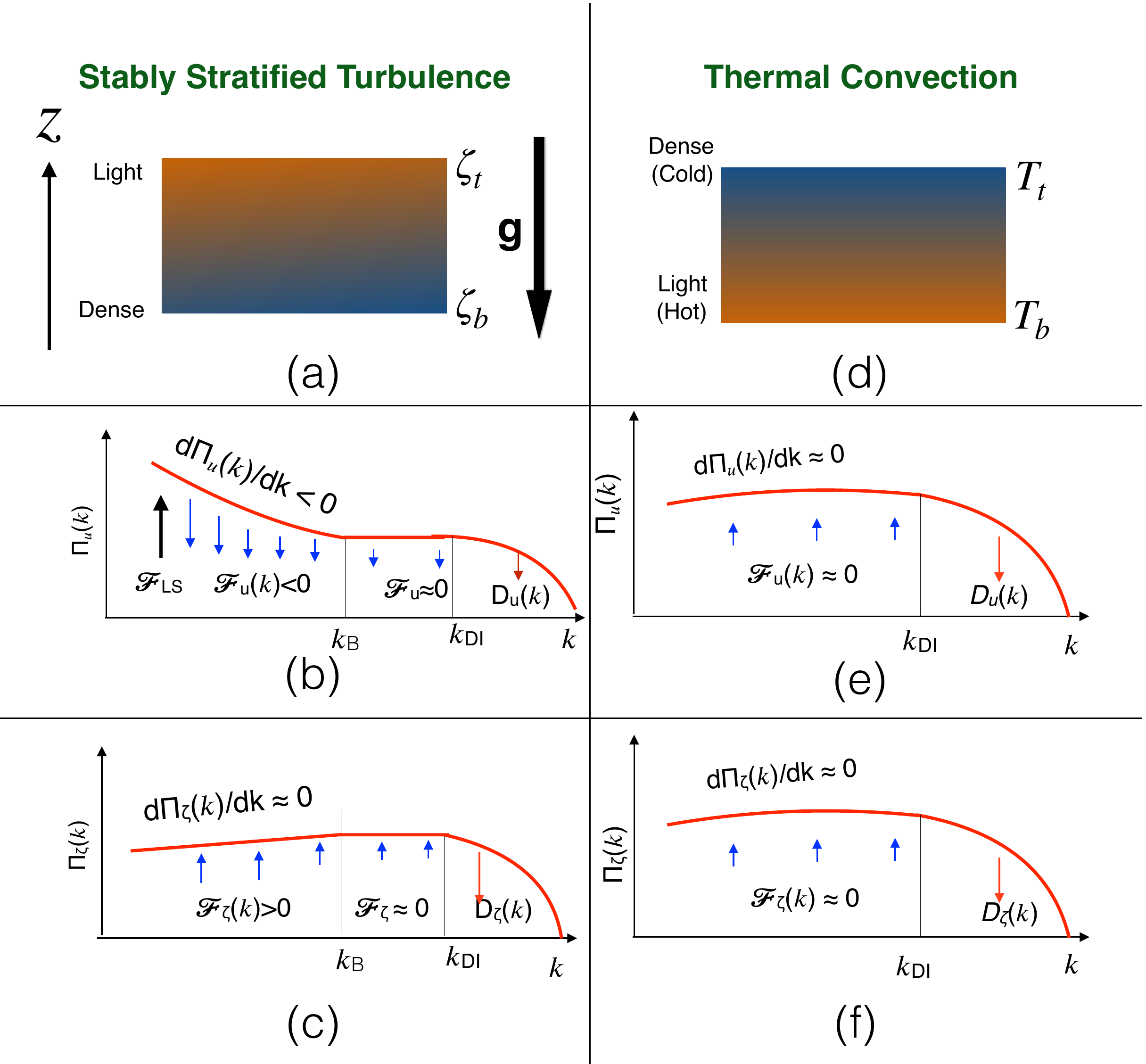}
\caption{For stably stratified turbulence: (a) a schematic diagram depicting the density $\zeta(z)$ that decreases with height;  (b) Kinetic energy flux, $\Pi_u(k)$, decreases for $k<k_B$, and is constant for $k_B < k< k_\mathrm{DI}$, where  $ k_B $ is the Bolgiano wavenumber, and $k_\mathrm{DI}$ is the wavenumber beyond which the dissipation range starts. (c) Secondary energy flux, $\Pi_\zeta(k)$, increases marginally for $k < k_B$, and is constant for $k_B < k< k_\mathrm{DI}$. For turbulent thermal convection: (d) a schematic diagram depicting a fluid between two plates whose temperatures are $T_b$ and $T_t$ ($T_b > T_t$); (e, f) $\Pi_u(k)$ and $\Pi_\zeta(k)$ are approximately  constant in the inertial range. }
\label{fig:secondary:BDF}
\end{center}
\end{figure}

A flow is said to be stably stratified when the density of a fluid under gravity decreases with height. See Fig.~\ref{fig:secondary:BDF}(a) for an illustration.  The background density profile is 
\bea
 \bar{\zeta}(z)  =  \zeta_b + \frac{d \bar{\zeta}}{d z} z =  \zeta_b + \frac{\zeta_t-\zeta_b}{d} z,
 \label{eq:linear_density}
 \eea
where $\bar{\zeta}(z)$ is the vertical density profile, which is assumed to be linear; gravity is along $-\hat{z}$; and $\zeta_b, \zeta_t$ are respective  densities at the bottom and top layers of the flow that is confined within a vertical distance $d$.  Stably stratified environment supports internal gravity waves with {\em Brunt-V\"{a}is\"{a}l\"{a} frequency}, which is given by~ \cite{Tritton:book, Davidson:book:TurbulenceRotating}
\bea
N = \sqrt{\frac{g}{\zeta_m} \left| \frac{d\bar{\zeta}}{dz} \right|},
\eea
where $\zeta_m$ is the  mean density of the whole fluid, and $g$ is the acceleration due to gravity.  Another useful nondimensional number is {\em Richardson number}, which is the ratio of  buoyancy and nonlinear term.

 The governing equations for the system are~ \cite{Tritton:book, Davidson:book:TurbulenceRotating,Sagaut:book,Verma:book:BDF} 
 \bea
\frac{\partial{\mathbf{u}}}{\partial{t}}+ (\mathbf{u}\cdot \nabla)\mathbf{u} & = &  - \nabla\sigma - N \zeta \hat{z} + \nu \nabla^{2}\mathbf{u} +{\bf F}_\mathrm{LS},  
\label{eq:u_SS} \\
\frac{\partial{\zeta}}{\partial{t}}+(\mathbf{u}\cdot\nabla)\zeta& = & N  u_{z} + \kappa \nabla^{2} \zeta, \label{eq:b_SS}
\eea 
where  $\sigma$ is the pressure,    $\zeta \rightarrow (g \zeta)/(N \zeta_m)$ is the density fluctuation in velocity units, and $ - N \zeta \hat{z} $ is  buoyancy.  For  periodic or vanishing boundary condition and in the absence of dissipative terms, the total energy,
\bea
E_u + E_\zeta = \int d{\bf r} \frac{1}{2} u^2  +   \int d{\bf r} \frac{1}{2} \zeta^2,   
\label{eq:buoyant:energy_conserv}
\eea
is conserved~ \cite{Tritton:book,Lindborg:JFM2006,	Davidson:book:TurbulenceRotating,Sagaut:book,Verma:book:BDF}.  In the above expression, $E_\zeta$ is the total potential anergy.

The forces related to buoyancy are
\bea
{\bf F}_u =  - N \zeta \hat{z};~~~F_\zeta = N  u_{z},
\eea
which are linear functions of the field variables.  Clearly, the energy injection rates by these two forces are
 \bea
\mathcal{F}_u({\bf k}) &= & -N \Re[ \zeta({\bf k}) u_z^*({\bf k}) ], \\
\mathcal{F}_\zeta({\bf k}) &= & N  \Re[ \zeta({\bf k}) u_z^*({\bf k}) ] .
\eea 
Hence, $ \mathcal{F}_\zeta({\bf k}) + \mathcal{F}_u({\bf k}) = 0 $.  Therefore, in the inertial range,
\bea
\frac{d}{dk} [ \Pi_u(k) + \Pi_\zeta(k)] = \mathcal{F}_u + \mathcal{F}_\zeta =0
\eea
leading to constancy of $\Pi_u(k) + \Pi_\zeta(k)$ in the inertial range.   The above two equations follow from the conservation of total energy in the inviscid limit (see Eq.~(\ref{eq:buoyant:energy_conserv})).  Due to the local interactions in $ \mathcal{F}_u({\bf k})$ and $ \mathcal{F}_\zeta({\bf k})$, the cross fluxes $ \Pi^{u<}_{\zeta>}(k)$,  $ \Pi^{u>}_{\zeta<}(k)$,  $ \Pi_{u>}^{\zeta<}(k)$, and $ \Pi_{u<}^{\zeta>}(k)$ are zeros.

Stably stratified turbulence has complex properties, which are discussed in detail in many books and  papers, for example, Lindborg~\cite{Lindborg:JFM2006}, Davidson~\cite{Davidson:book:TurbulenceRotating}, and references therein.  In this review, to present applications of variable energy flux in buoyant flows, we describe the spectral analysis for moderately stratified flows where  $ |{\bf u \cdot \nabla u}| \approx N \zeta$. For such flows, Richardson number is of the order of unity.

 For moderately stratified flows,   Bolgiano~\cite{Bolgiano:JGR1959} and  Obukhov~\cite{Obukhov:DANS1959} provided the first phenomenological model.  They assumed that for a wavenumber band in the inertial range,
 \bea
k u_k^2  = N \zeta_k;~~~
\Pi_\zeta =  k \zeta_k^2 u_k = \epsilon_\zeta \label{eq:SST_const_Pi_rho}
\eea 
that yield the following fluxes and spectra for the velocity and density fields:
  \bea
E_u(k) =  c_1 \epsilon_\zeta^{2/5}  N^{4/5} k^{-11/5},~~~
\Pi_u(k) & = & c_3  \epsilon_\zeta^{3/5} N^{6/5} k^{-4/5},  \label{eq:SST_pi_u} \\
E_\zeta(k) = c_2 \epsilon_\zeta^{4/5}  N^{-2/5} k^{-7/5},~~~
\Pi_\zeta(k) & = &  \epsilon_\zeta. \label{eq:SST_pi_zeta} 
\eea 
Clearly, {the kinetic energy flux $\Pi_u(k)$ decreases with $k$}, in contrast to  constant $\Pi_u$  in the inertial range of 3D hydrodynamic turbulence.  The reduction of $\Pi_u(k)$ occurs due to the conversion of kinetic energy to  potential energy, and it leads to a steepening of $E_u(k)$~ \cite{Bolgiano:JGR1959,Obukhov:DANS1959,Lindborg:JFM2006,Kumar:PRE2014,Davidson:book:TurbulenceRotating,Verma:book:BDF}.   See Fig.~\ref{fig:secondary:BDF}(b,c) for an illustration.

In addition,  Obukhov~\cite{Bolgiano:JGR1959} and Obukhov~\cite{Obukhov:DANS1959} predicted that buoyancy becomes weak for $k>k_B$, where
$k_B \approx N^{3/2} \epsilon_u^{-5/4} \epsilon_\zeta^{3/4}$
is  {\em Bolgiano wavenumber}. Due to weak buoyancy, Bolgiano and Obukhov predicted that  both kinetic and secondary energies exhibit $k^{-5/3}$ spectrum for $k_B < k < k_\mathrm{DI}$, where $k_\mathrm{DI}$ is the wavenumber beyond  which  dissipation becomes significant.     See Fig.~\ref{fig:secondary:BDF}(b) for an illustration. Using the constraint $k_B \ll k_d$,  where $ k_d $ is Kolmogorov's wavenumber,  Alam et al.~\cite{Alam:JFM2019}  showed that simultaneous presence of both the scaling regimes ($k^{-11/5}$ and $k^{-5/3}$) requires very large Reynolds number. Hence, the $k^{-5/3}$ regime of Bolgiano-Obukhov phenomenology is quite difficult to reproduce in numerical simulations.

  Kimura and Herring~\cite{Kimura:JFM1996}   reported Bolgiano-Obukhov scaling for a narrow wavenumber band in their decaying  simulation on a $128^3$ grid.   Kumar et al.~\cite{Kumar:PRE2014} performed a numerical simulation of stably stratified turbulence on a $ 1024^3 $ grid for Richardson number around unity and observed a good agreement between numerical results and the  predictions of  Eqs.~(\ref{eq:SST_pi_u}, \ref{eq:SST_pi_zeta}).       Rosenberg et al.~\cite{Rosenberg:PF2015}  reported Bolgiano-Obukhov scaling in their simulation of rotating stratified turbulence. 
  
 Strong buoyancy  (large Richardson number) makes the flow anisotropic, hence  Bolgiano-Obukhov scaling is inapplicable to such flows.    Lindborg~\cite{Lindborg:JFM2006} and  Davidson~\cite{Davidson:book:TurbulenceRotating} argued  that   the longitudinal and traverse velocity components  exhibit $k^{-3}$ and $k^{-5/3}$ spectra respectively.   Variable energy flux may provide interesting clues for this regime as well.  For two-dimensional stably stratified turbulence,     Kumar et al.~\cite{Kumar:JoT2017} derived several interesting relations using variable energy flux.   Also note that buoyancy is weak for flows with small Richardson number. Hence, such flows exhibit Kolmogorov's spectrum~ \cite{Kumar:PRE2014}.

 In the next subsection we will employ the ideas of variable energy flux to turbulent thermal convection.

\subsection{Turbulent thermal convection}

Thermal convection too is driven by buoyancy.  However, in contrast to the stably stratified flows, the fluid density increases with height that makes the flow unstable. A setup, exhibited in Fig.~\ref{fig:secondary:BDF}(d), consists of a thin fluid layer confined between two thermally conducting plates separated by a distance $d$.  The temperatures of the bottom and top plates are $T_b$ and $T_t$ respectively. 

In thermal convection, the  temperature is a sum of externally-imposed linearly varying   temperature  $\bar{T}(z)$ and fluctuation $\zeta(x,y,z)$:
\bea
T(x,y,z) = \bar{T}(z) + \zeta(x,y,z),
\label{eq:buoyant:T}
\eea
where
\bea
\bar{T}(z) = T_b + \frac{d\bar{T}}{dz} z = T_b - \frac{T_b-T_t}{d} z.
\eea
The equations for thermal convection under Boussinesq approximation are~ \cite{Chandrasekhar:book:Instability}
 \bea
 \frac{\partial {\bf u}}{\partial t} + ({\bf u} \cdot \nabla) {\bf u} & = &  -\frac{1}{\zeta_m} \nabla \sigma   + \alpha g \zeta \hat{z} + \nu \nabla^2 {\bf u}, \label{eq:buoyant:RBC1}
 \\
\frac{\partial \zeta}{\partial t} +  ({\bf u} \cdot \nabla)  \zeta & = &  \frac{\Delta}{d} u_z +\kappa \nabla^2 \zeta, \label{eq:buoyant:RBC2} \\
\nabla \cdot {\bf u} = 0, \label{eq:buoyant:RBC3}
\eea 
where $\alpha, \kappa$ are respectively  the thermal expansion coefficient and thermal diffusivity  of the fluid,  $g$ is the acceleration due to gravity, and $\Delta = T_b-T_t$.   The two important parameters of turbulent thermal convection are   Prandtl number, $\mathrm{Pr}=\nu/\kappa$, and  Rayleigh number, 
\bea
\mathrm{Ra}  =  \frac{\alpha g d^3 \Delta }{\nu \kappa}.
\eea
Note that the forces
\bea
{\bf F}_u = \alpha g \zeta \hat{z},~~~ 
F_\zeta = \frac{\Delta}{d} u_z
\eea
are linear functions of the field variables.

In thermal convection, hot plumes ascend and cold ones descend.  Hence, $ \langle \zeta({\bf r}) u_z({\bf r}) \rangle > 0 $.  Therefore, using Parceval's theorem we deduce that 
\bea
\sum_{\bf k} \Re \left[ \la \zeta({\bf k}) u_z^*({\bf k}) \ra \right] >0.
\eea
Further, numerical simulations  reveals that $\Re \left[ \la \zeta({\bf k}) u_z^*({\bf k}) \ra \right] > 0$ for most Fourier modes of thermal convection. Hence,
 \bea
\mathcal{F}_u({\bf k}) &= & \alpha g \Re[ \zeta({\bf k}) u_z^*({\bf k}) ] > 0
\label{eq:buoyant:mathcal_F_u} \\
\mathcal{F}_\zeta({\bf k}) &= & \frac{\Delta}{d} \Re[ \zeta({\bf k}) u_z^*({\bf k}) ] > 0 
\label{eq:buoyant:mathcal_F_zeta} 
\eea 
 Therefore, in the inertial range,
\bea
\frac{d}{dk} [ \Pi_u(k) - \frac{\alpha g d}{\Delta} \Pi_\zeta(k)]  = 0
\eea
leading to
\bea
\Pi_u(k) - \frac{\alpha g d}{\Delta} \Pi_\zeta(k) = \mathrm{const}.
\label{eq:buoyant:flux_diff_RBC}
\eea
In the dissipationless limit, $\int d{\bf r} \frac{1}{2} [u^2 - (\alpha g d/\Delta) \zeta^2]$, a sum of kinetic energy and potential energy, is conserved.  Here, the potential energy, $-\frac{1}{2} (\alpha g d/\Delta) \zeta^2$,  is negative because the system is unstable.  Since this potential energy can be converted to kinetic energy, it is also called \textit{available potential energy}.  Contrast this potential energy with that  for stably stratified turbulence.

Equation~(\ref{eq:buoyant:mathcal_F_u}) implies that $\Pi_u(k)$ should increase with $k$.   However,  for $\mathrm{Pr} \sim 1$, Verma et al.~\cite{Verma:NJP2017} showed that $\mathcal{F}_u(k) \sim (kL)^{-5/3}$, hence, $\mathcal{F}_u(k)$ is small  in the inertial range because $kL \gg 1$.  The function $\mathcal{F}_u(k) $ is even steeper for small Pr~\cite{Verma:book:BDF}.  These observations indicate that large-scale thermal plumes drive the flow, similar to the forcing in Kolmogorov's theory for hydrodynamic turbulence. These results 
 falsify the popular statements that thin thermal plumes drive thermal convection.
 
 Based on similarities between the forcing in thermal convection and Kolmogorov's model of turbulence, Kumar et al.~\cite{Kumar:PRE2014} and Verma et al.~\cite{Verma:NJP2017} argued that   for $ \mathrm{Pr} \lessapprox 1 $,   $\Pi_u(k)$ remains an approximate constant in the inertial range  and $E_u(k) \sim k^{-5/3}$.   Verma et al.~\cite{Kumar:PRE2014, Verma:NJP2017}  verified the above phenomenology using high-resolution numerical simulations.    The above flux-based arguments resolve the long impasse in the field regarding the energy spectrum.  Note that several past works~ \cite{Procaccia:PRA1990, Lvov:PRL1991,Lvov:PD1992,Rubinstein:NASA1994} projected that  Bolgiano-Obukhov's scaling  for  stably stratified turbulence ($E_u(k) \sim k^{-11/5}$)  holds for turbulent thermal convection as well, while  experimental and numerical works were inconclusive. For large Prandtl numbers, the flow is dissipative and $E_u(k) \sim k^{-13/3}$~\cite{Pandey:PRE2014,Verma:book:BDF}.

However, there is a major difference between hydrodynamic turbulence and turbulent convection.  In turbulent convection, inertial-range kinetic energy flux is a fraction of the total viscous dissipation rate.  This is because thermal plumes drive the flow at all scales.  In particular, under steady state, for a $k$ in the inertial range, 
   \bea
    \Pi_u(k) & \approx & \int_0^{k} dk' \mathcal{F}_u(k') =  \int_0^{\infty} dk' \mathcal{F}_u(k') -  \int_{k}^\infty dk' \mathcal{F}_u(k')  \nonumber \\
    & = & \epsilon_u -  \int_{k}^\infty dk' \mathcal{F}_u(k') ,  \label{eq:buoyant:Pi_u_RBC_decrease} 
    \eea 
    where $ \epsilon_u $ is the total viscous dissipation rate.     Using Eq.~(\ref{eq:buoyant:Pi_u_RBC_decrease}) we deduce that the inertial-range kinetic energy flux $   \Pi_u(k)< \epsilon_u$  due to the presence of buoyancy  at all scales.  Using numerical simulations,  Bhattacharya et al.~\cite{Bhattacharya:PF2019:SF} showed that for $\mathrm{Pr}=1$, the inertial-range $     \Pi_u(k) $ is around  one-third of $ \epsilon_u$.

There are many turbulent flows with unstable stratification, notably Rayleigh-Taylor turbulence~\cite{Boffetta:ARFM2016}, bubbly turbulence~\cite{Lakkaraju:PNAS2013}, Taylor-Couette turbulence~\cite{Lewis:PRE1999,Grossmann:ARFM2016}, etc.    Based on the arguments of this subsection,  we expect that the  turbulence properties of unstable stable stratification are similar to those of hydrodynamic turbulence, e.g., $E_u(k) \sim k^{-5/3}$~ \cite{Verma:NJP2017, Verma:book:BDF}.    The results on Rayleigh-Taylor turbulence~ \cite{Banerjee:JFM2010,Akula:JFM2016,Boffetta:ARFM2016} and  Taylor-Couette turbulence~ \cite{Lewis:PRE1999} are consistent with the above observations.

In the aforementioned buoyant flows,   ${\bf F}_u$ and ${\bf F}_\zeta$ are linear functions of $\zeta$ and ${\bf u}$ respectively.  Hence the cross energy transfer occurs among the modes  with same wavenumbers.  Therefore, $\mathcal{F}_u({\bf k})$ and $ \mathcal{F}_\zeta({\bf k})$ are functions only of local wavenumber ${\bf k}$. However, when ${\bf F}_u$ and ${\bf F}_\zeta$ are  nonlinear functions of ${\bf u}$ and/or $\zeta$,  the cross energy transfers are convolutions of  ${\bf u}$ and $\zeta$ Fourier modes.  In the following section, we illustrate such energy transfers in MHD turbulence.


\section{Variable energy fluxes in magnetohydrodynamic turbulence}
\label{sec:mhd}

Magnetohydrodynamic (MHD) turbulence is a vast area of research with many astrophysical and engineering applications. It is covered in several books, e.g.,  \cite{Cowling:book, Davidson:book:MHD,Biskamp:book:MHDTurbulence}, review articles, e.g.,  \cite{Verma:PR2004,Brandenburg:PR2005,Goldstein:ARAA1995,Alexakis:PR2018}, and references therein.  The present section does not attempt to summarise  vast number of results of MHD turbulence, but focusses on variable energy fluxes of MHD turbulence.

A magnetofluid, which is a quasi-neutral and electrically-conducting collisional plasma, is described   by  velocity field and magnetic field, which is denoted by $ \bm{\zeta}$.    In the following subsection, we describe the governing equations  for MHD turbulence.

\subsection{MHD turbulence: Governing equations and cross transfers}

The dynamical equations for the velocity and magnetic fields of MHD are~ \cite{Cowling:book, Davidson:book:MHD,Biskamp:book:MHDTurbulence}
 \bea
\frac{\partial{\mathbf{u}}}{\partial{t}}  +  ({\bf u} \cdot \nabla) {\bf u}  & = &  -  \nabla p  +   ( \bm{\zeta}\cdot \nabla)  \bm{\zeta} +  {\bf F}_\mathrm{LS} +  \nu \nabla^2 {\bf u}, \label{eq:mhd:MHDu_fluct} \\
\frac{\partial{ \bm{\zeta}}}{\partial{t}} + ({\bf u} \cdot \nabla)  \bm{\zeta} & = &   ( \bm{\zeta}\cdot \nabla) {\bf u}  + \kappa  \nabla^2  \bm{\zeta},  \label{eq:mhd:MHDb_fluct} \\
\nabla \cdot {\bf u} =  \nabla \cdot  \bm{\zeta} &= & 0, \label{eq:MHD_formalism:MHDdel_u_zero_fluct}  
\eea 
where $p$ is the total (thermodynamic + magnetic) pressure, ${\bf F}_\mathrm{LS}$ is the large-scale external force, and  $\nu, \kappa$ are the kinematic viscosity and magnetic diffusivity respectively.  In the above equation, the magnetic field is in Alfv\'{e}nic units, which has same dimension as the velocity field.  In the above equations, the forces on  ${\bf u}$ and $  \bm{\zeta}$ fields are 
 \bea
{\bf F}_u = ( \bm{\zeta}\cdot \nabla)  \bm{\zeta};~~~
{\bf F}_\zeta = ( \bm{\zeta}\cdot \nabla) {\bf u}.
\label{eq:MHD:forces}
\eea 
For  inviscid flow ($\nu = \kappa =0$), under  periodic or vanishing boundary condition,  the total energy
\bea
E_u + E_\zeta = \int d{\bf r} \frac{1}{2} u^2  +   \int d{\bf r} \frac{1}{2} \zeta^2   
\label{eq:mhd:energy_conserv}
\eea
is conserved~ \cite{Cowling:book, Davidson:book:MHD,Biskamp:book:MHDTurbulence}.  Here, $E_\zeta$ is the total magnetic anergy.  The other conserved quantities of inviscid MHD are cross helicity, $  \int d{\bf r}  \frac{1}{2} ({\bf u} \cdot \bm{\zeta})$, and magnetic helicity, $  \int d{\bf r}  \frac{1}{2} ({\bf A \cdot \bm{\zeta}})$, where $ {\bf A} $ is the vector potential. In this review, we focus on kinetic and magnetic energies only.

In Fourier space, the forces of  Eq.~(\ref{eq:MHD:forces}) are
 \bea
{\bf F}_u({\bf k}) =i \sum_{\bf p} \{ {\bf k} \cdot  \bm{\zeta}({\bf q})  \}   \bm{\zeta}({\bf p});~~~
{\bf F}_\zeta({\bf k}) = i \sum_{\bf p} \{ {\bf k} \cdot  \bm{\zeta}({\bf q})  \}  {\bf u}({\bf p}), 
\eea 
where ${\bf q=k-p}$.  The equation for the kinetic and magnetic modal energies are 
  \bea 
 \frac{d}{dt} E_u(\mathbf{k}) & = &  T_u({\bf k}) + \mathcal{F}_u({\bf k})  + {\bf F}_\mathrm{LS}({\bf k})- D_u({\bf k}) \nonumber \\
 &=&  \sum_{\bf p} \Im \left[ {\bf \{  k \cdot u(q) \}  \{ {\bf u(p) \cdot u^*(k)} \} }  \right] + \mathcal{F}_u({\bf k}) + {\bf F}_\mathrm{LS}({\bf k}) - 2 \nu k^{2} E_u({\mathbf k}), 
 \label{eq:mhd:Euk}  \nonumber \\  \\
 \frac{d}{dt} E_\zeta(\mathbf{k}) & = &  T_\zeta({\bf k}) + \mathcal{F}_\zeta({\bf k}) - D_\zeta({\bf k}) \nonumber \\
 &=&  \sum_{\bf p} \Im \left[ {\bf \{  k \cdot u(q) \}  \{  \bm{\zeta}(p) \cdot  \bm{\zeta}^*(k) \} }  \right] + \mathcal{F}_\zeta({\bf k}) - 2 \kappa k^{2} E_\zeta({\mathbf k}), 
 \label{eq:mhd:Ethetak}
  \eea   
  where the kinetic and magnetic energy injection rates by the forces ${\bf F}_u({\bf k})$ and ${\bf F}_\zeta({\bf k})$  are
 \bea
\mathcal{F}_u({\bf k}) & = & \sum_{\bf p} -\Im \left[ {\bf \{  k \cdot  \bm{\zeta}(q) \}  \{  \bm{\zeta} (p) \cdot  u^*(k) \} } \right], 
\label{eq:mhd:Fu} \\
\mathcal{F}_\zeta({\bf k}) & = & \sum_{\bf p} -\Im \left[ {\bf \{  k \cdot  \bm{\zeta}(q) \}  \{ u(p) \cdot  \bm{\zeta}^*(k)   \} } \right].
\label{eq:mhd:Fzeta}
\eea 
Note that ${\bf F}_u({\bf k})$ facilitates energy transfer from $ \bm{\zeta}$ to ${\bf u}$, while ${\bf F}_\zeta({\bf k})$  yields energy transfer from ${\bf u}$ to  $ \bm{\zeta}$.  As remarked earlier,  $\mathcal{F}_u({\bf k}), \mathcal{F}_\zeta({\bf k}) $ are convolutions because ${\bf F}_u, {\bf F}_\zeta$ are nonlinear functions of ${\bf u}$ and $ \bm{\zeta}$.  In contrast,  the corresponding transfers for the buoyant flows are functions of fields at  local wavenumber (see Sec.~\ref{sec:buoyant}). 

The nonlinear structure of the cross transfers between ${\bf u}$ and $ \bm{\zeta}$ can be formulated in terms of mode-to-mode energy transfers.   For a triad ${\bf (k',p,q)}$ satisfying ${\bf k'+p+q}=0$,  Dar et al.~\cite{Dar:PD2001} and  Verma~\cite{Verma:PR2004}  derived the following formulas for the mode-to-mode energy transfers from ${\bf u}$ to  $ \bm{\zeta}$ and vice versa:
\bea
S^{u \zeta}({\bf k'|p|q}) & = & \Im \left[ {\bf \{  k' \cdot  \bm{\zeta}(q) \}  \{  \bm{\zeta}(p) \cdot  u(k') \} } \right]
= -\Im \left[ {\bf \{  k \cdot  \bm{\zeta}(q) \}  \{  \bm{\zeta}(p) \cdot  u^*(k) \} } \right], \nonumber
\label{eq:mhd:Suzeta} \\
\eea
\bea
S^{\zeta u}({\bf k'|p|q}) & = &  \Im \left[ {\bf \{  k' \cdot  \bm{\zeta}(q) \}  \{ u (p) \cdot  \zeta(k') \} } \right] 
 = - \Im \left[ {\bf \{  k \cdot  \bm{\zeta}(q) \}  \{ u (p) \cdot  \zeta^*(k) \} } \right]. \nonumber 
\label{eq:mhd:Szetau} \\
\eea
The former  is the mode-to-mode energy transfer from $ \bm{\zeta}({\bf p})$ to ${\bf u}({\bf k'})$ with the mediation of $ \bm{\zeta}({\bf q})$, while the latter provides the energy transfer from ${\bf u}({\bf p})$  to $ \bm{\zeta}({\bf k'})$  with the mediation of $ \bm{\zeta}({\bf q})$. In $S^{ab}({\bf k'|p|q}) $, the superscript $a b$ refer to receiver  field $ a $ and giver field $ b $.

  These transfers  satisfy the property:
\bea
S^{u \zeta}({\bf k'|p|q}) = -S^{\zeta u}({\bf p|k'|q}),
\label{eq:mhd:Suzeta_zetau}
\eea
that is, the energy gained  by ${\bf u}({\bf k'})$ from $ \bm{\zeta}({\bf p})$ is negative of the energy gained by $ \bm{\zeta}({\bf p})$ from   ${\bf u}({\bf k'})$.  This is a property of energy exchange.  Using the above property, we can show that  for any region $A$ of Fourier space, including a triad,
 \bea
\sum_{{\bf k'} \in A} \sum_{{\bf p} \in A} [S^{u \zeta}({\bf k'|p|q}) +S^{\zeta u}({\bf p|k'|q})] = 0.
\label{eq:mhd:Suzeta_zetau_sum}
\eea  
By summing over all the Fourier modes, we deduce that  $\mathcal{F}_u + \mathcal{F}_\zeta=0$,  where $\mathcal{F}_u$ and $\mathcal{F}_\zeta$ are respectively the total energy gained by the velocity and secondary fields via cross energy transfers.   Note that stably-stratified turbulence too has $\mathcal{F}_u + \mathcal{F}_\zeta=0$.

In the next subsection we define the energy fluxes of MHD turbulence.

\subsection{Various energy fluxes of MHD turbulence}

\begin{figure}[htbp]
\begin{center}
\includegraphics[scale = 0.7]{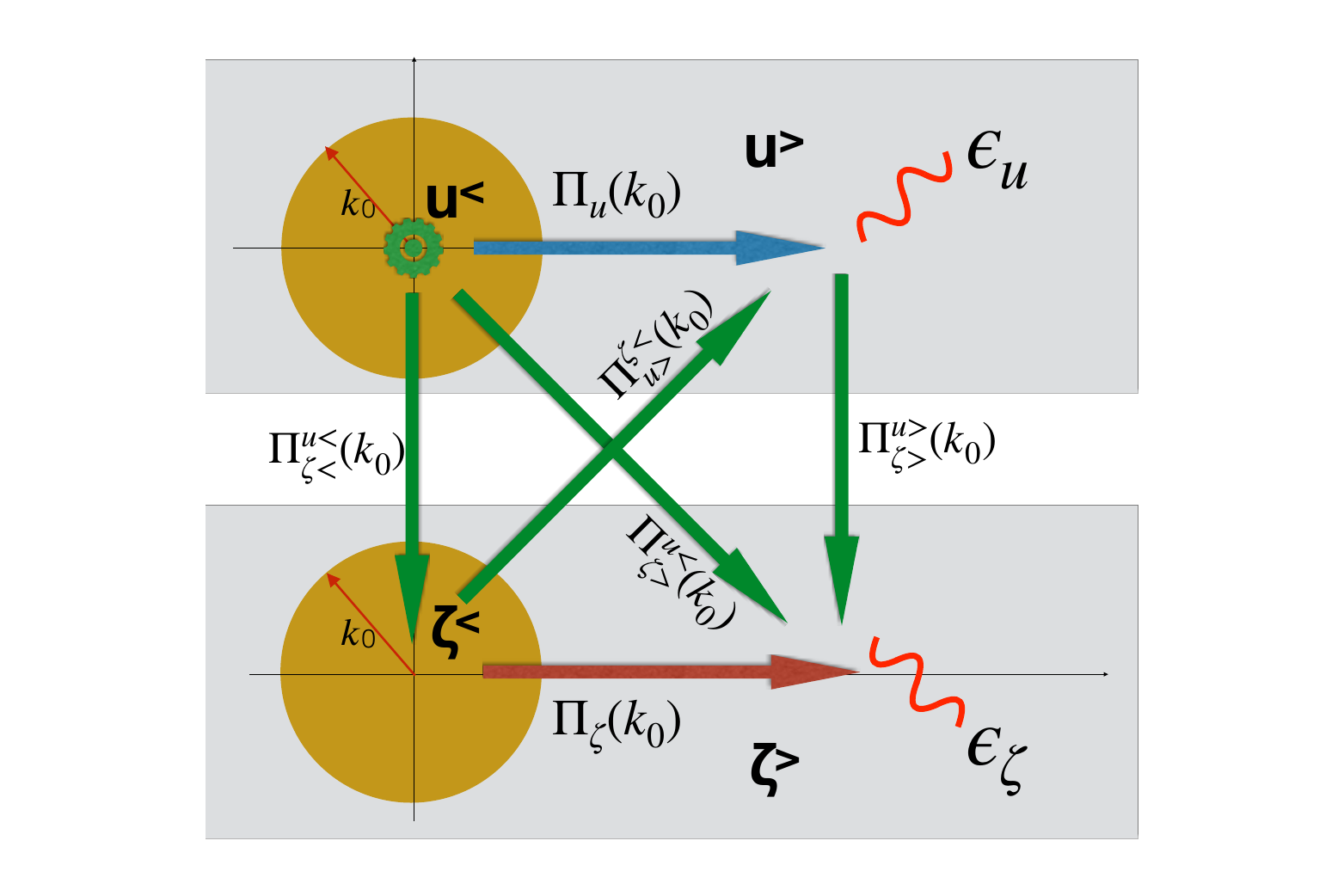}
\end{center}
\caption{The six energy fluxes of MHD turbulence. $\epsilon_u,  \epsilon_\zeta$ are the dissipation rates of the velocity and magnetic fields respectively.  The wheel in the centre of the velocity sphere represents the  external forcing at  large scales. }
\label{fig:second:second_flux}
\end{figure}

Using the mode-to-mode energy transfers of Eqs.~(\ref{eq:mhd:Suzeta}, \ref{eq:mhd:Szetau}),  Dar et al.~\cite{Dar:PD2001} and  Verma~\cite{Verma:PR2004} derived   formulas for the energy fluxes of MHD turbulence for a wavenumber sphere of radius $k_0$.  Note that  $\Pi_u(k_0)$ and $\Pi_\zeta(k_0)$ are respective fluxes for the kinetic  and magnetic energies, while the energy flux  $\Pi^{u<}_{\zeta>}(k_{0})$ represents the net energy transfer from all the velocity modes  inside the sphere to all the magnetic modes outside the sphere, that is,
\bea
\Pi^{u<}_{\zeta>}(k_{0})= \sum_{p\le k_{0}} \sum_{k'>k_{0}} S^{\zeta u}({\bf k'|p|q}).
\label{eq:mhd:MHD_flux_u<_b>}
\eea
The other fluxes, $\Pi^{\zeta<}_{u>}(k_{0})$, $\Pi^{u<}_{\zeta <}(k_{0})$, and  $\Pi^{u>}_{\zeta>}(k_{0})$  are defined similarly.  Note that an application of Eq.~(\ref{eq:mhd:Suzeta_zetau}) yields
\bea
 \Pi^{u a}_{\zeta b}(k_0) = -  \Pi^{\zeta b}_{u a}(k_0),
 \eea
 where $a,b$ represent $<$ or $>$.     The above energy fluxes of MHD turbulence, depicted in Fig.~\ref{fig:second:second_flux}, have been studied in great detail~\cite{Dar:PD2001,Debliquy:PP2005,Alexakis:PRE2005}.  Interestingly, $\Pi^{u<}_{\zeta>}(k)$ and $\Pi^{\zeta<}_{u>}(k)$ are absent in buoyant flows due to the  products of  field variables with the same wavenumber~(see Sec.~\ref{sec:buoyant}).

Now we relate the above quantities to the fluxes $\Pi^u_\zeta$ and $\Pi^\zeta_u$ discussed in Sec.~\ref{subsec:cross_energy}.  Clearly,
 \bea
\Pi^\zeta_{u<}(k_0) = \sum_{k \le k_0}    \mathcal{F}_u({\bf k}) 
& = &  \Pi^{\zeta<}_{u<}(k_0) +  \Pi^{\zeta>}_{u<}(k_0),  \label{eq:mhd:Pi_zeta_u<} \\
\Pi^\zeta_{u>}(k_0) = \sum_{k >  k_0}    \mathcal{F}_u({\bf k}) 
& = &  \Pi^{\zeta<}_{u>}(k_0) +  \Pi^{\zeta>}_{u>}(k_0),  \label{eq:mhd:Pi_zeta_u>} \\
\Pi^u_{\zeta<}(k_0) = \sum_{k \le k_0}    \mathcal{F}_\zeta({\bf k}) 
& = &  \Pi^{u<}_{\zeta<}(k_0) +  \Pi^{u>}_{\zeta<}(k_0),  \label{eq:mhd:Pi_u_zeta<} \\
\Pi^u_{\zeta>}(k_0) = \sum_{k \le k_0}    \mathcal{F}_\zeta({\bf k}) 
& = &  \Pi^{u<}_{\zeta>}(k_0) +  \Pi^{u>}_{\zeta>}(k_0).  \label{eq:mhd:Pi_u_zeta>} 
\eea 
The identities of Eqs.~(\ref{eq:secondary:Pi_zeta_u_sum}, \ref{eq:secondary:Pi_u_zeta_sum}) are translated to the following identities for MHD turbulence:
\bea
\mathcal{F}_u & = &  \Pi^{\zeta<}_{u<}(k) + \Pi^{\zeta<}_{u>}(k)  + \Pi^{\zeta>}_{u<}(k) + \Pi^{\zeta>}_{u>}(k)  = C_1, \label{eq:mhd:C1} \\
\mathcal{F}_\zeta  & = & \Pi^{u<}_{\zeta<}(k) + \Pi^{u<}_{\zeta>}(k)  + \Pi^{u>}_{\zeta<}(k) + \Pi^{u>}_{\zeta>}(k)  = C_2,  \label{eq:mhd:C2}
\eea
where $C_1$ and $C_2$ are constants.  The above two sums are constant in $k$ even  though the individual flux in the sum can vary with $k$.   Note that $ C_1 $ is the net energy transfer from   $ \bm{\zeta} $  to  \textbf{u}, while  $ C_2 $ is the net energy transfer from   \textbf{u} to   $ \bm{\zeta} $.   Interestingly, the vorticity field has similar properties as the magnetic field (with some important deviations)~\cite{Batchelor:PRSA1950}.  In Sec.~\ref{subsec:3D_enstrophy_flux} we will describe various fluxes associated with the vorticity field.


Under a steady state with \textit{no} external forcing for $ \bm{\zeta}$,  the net nonlinear energy transfer to the magnetic field,  $\mathcal{F}_\zeta$, balances the magnetic diffusion or dissipation rate, $\epsilon_\zeta$. That is,
\bea
\mathcal{F}_\zeta   = \sum_{\bf k} 2 \eta k^2 E_\zeta(k) = \epsilon_\zeta > 0.
\label{eq:mhd:F_zeta}
\eea
Note however that an imbalance in $\mathcal{F}_\zeta $ and $\epsilon_\zeta$ makes the flow unsteady.   For the velocity field, $\mathcal{F}_u = -\mathcal{F}_\zeta <0$. In addition, the viscous force  dissipates the kinetic energy.  Hence,  for the velocity field, a large-scale force, ${\bf F}_\mathrm{LS}$, is required to maintain a  steady  state. Using energy balance we obtain
\bea
\varepsilon_\mathrm{inj} + \mathcal{F}_u =  \sum_{\bf k} 2 \nu k^2 E_u(k) = \epsilon_u,
\label{eq:mhd:F_u_eps_int}
\eea
where $\varepsilon_\mathrm{inj}$ is the total energy injection rate by ${\bf F}_\mathrm{LS}$.  In astrophysics, a supernova is an example of one such energy  source.   Note that under a steady state, the net energy injection rate by  ${\bf F}_\mathrm{LS}$ equals the sum of viscous dissipation and Joule dissipation, i.e., $ \varepsilon_\mathrm{inj} = \epsilon_u + \epsilon_\zeta $.

%
  
Equations~(\ref{eq:secondary:flux_1}-\ref{eq:secondary:flux_4}) are applicable to MHD turbulence as well.   When $k$ is in the inertial range, the  identities of Eqs.~(\ref{eq:secondary:flux_1}-\ref{eq:secondary:flux_4}) get simplified to
 \bea
   \Pi_u(k)+ \Pi^{\zeta}_{u>}(k)  = \epsilon_u;~~~ 
      \Pi_\zeta(k) + \Pi^u_{\zeta>}(k) = \epsilon_\zeta,  \label{eq:mhd:identity_1_2} \\
      \Pi_u(k) + \Pi^{u<}_\zeta(k) =  \varepsilon_\mathrm{inj},~~~
       \Pi_\zeta(k) -\Pi^u_{\zeta<}(k) = 0.  \label{eq:mhd:identity_3_4}
  \eea 
 The above equations follow from the energetics.  For example, a physical interpretation of the equation, $  \Pi_u(k) + \Pi^{u<}_\zeta(k) =  \varepsilon_\mathrm{inj} $, is as follows. Here,  the velocity field is  forced at large scales by $ {\bf F}_\mathrm{LS} $.    A part of the injected kinetic energy by $ {\bf F}_\mathrm{LS} $ cascades to the inertial range as  $\Pi_u$, but a fraction of it is transferred to the magnetic energy as $\Pi^{u<}_\zeta(k)>0$~ \cite{Dar:PD2001,Verma:PR2004,Brandenburg:PR2005,Alexakis:PRE2005}.   This cross transfer from ${\bf u}$ to  $ \bm{\zeta}$ amplifies the magnetic field; this mechanism is responsible for the generation of  large-scale magnetic field in planets, stars, and galaxies~ \cite{Moffatt:book,Zeldovich:book:Magneticfield, Brandenburg:PR2005,Verma:PR2004}.     Note that this cross energy transfer in MHD turbulence makes  both    $\Pi_u(k)$ and $ \Pi^{u<}_\zeta(k)$ functions of $k$, in contrast to constant $\Pi_u(k)$ in the inertial range of hydrodynamic turbulence.  Using similar arguments we can  show that the other MHD fluxes also vary with $k$. However, the total energy flux is constant under a steady state.  That is,
\bea
\varepsilon_\mathrm{inj}  &=  & \Pi_u(k) + \Pi^{u<}_\zeta(k) = \Pi_u(k) 
+ \Pi^{u<}_{\zeta<}(k) +  \Pi^{u<}_{\zeta>}(k)  \nonumber \\
& = &  \Pi_u(k) + \Pi_\zeta(k) + \Pi^{\zeta<}_{u>}(k) + \Pi^{u<}_{\zeta>}(k)  = \Pi_\mathrm{tot}(k) = \epsilon_u + \epsilon_\zeta.
\label{eq:mhd:total_flux}
\eea
Here we employ the identity $\Pi^{u<}_{\zeta<} = \Pi_\zeta + \Pi^{\zeta<}_{u>}(k)$.  These relations do not hold for unsteady flows.  For example, in decaying MHD turbulence, magnetic field feeds energy to the velocity field when $ b^2 \gg u^2 $ and vice versa~\cite{Debliquy:PP2005}.    

The cross energy transfer  $ \Pi^{u<}_\zeta(k)$ is also responsible for the drag reduction in MHD turbulence~\cite{Verma:PP2020}. Since  $ \Pi^{u<}_\zeta(k)>0$, using Eq.~(\ref{eq:mhd:identity_3_4}) we deduce that $  \Pi_u(k) < \varepsilon_\mathrm{inj} $.   Note that in hydrodynamic turbulence, $\Pi_u(k) \approx  \varepsilon_\mathrm{inj}$. Therefore,  for  the same $ \varepsilon_\mathrm{inj}$, $ \Pi_u(k)$ in MHD turbulence is lower than that for hydrodynamic turbulence.  Therefore, the turbulent drag, $F_D \approx \Pi_u/U$, will be lower for MHD turbulence compared to its  hydrodynamic counterpart~\cite{Verma:PP2020}.  In the next section we will show that a similar dynamics is a work in turbulent flows with dilute polymer.

Variable energy flux also provides valuable insights into the dynamics and spectral laws of MHD turbulence and dynamo.  For example, the kinetic and magnetic energy spectra are often modelled as~ \cite{Kraichnan:PF1965MHD,Iroshnikov:SA1964,Biskamp:book:MHDTurbulence,Goldstein:ARAA1995,Verma:PP1999,Verma:PRE2001,Verma:PR2004}  
\bea
E_{u,\zeta}(k) \sim \epsilon_{u,\zeta}^{2/3} k^{-5/3},~~\mathrm{or}~~~(B_0 \epsilon_{u,\zeta})^{1/2} k^{-3/2},
\eea
where $B_0$ is the magnitude of the mean magnetic field.  The above equations need proper interpretations because $ \Pi_u(k)$ and $ \Pi_\zeta(k)$, which are variables of $k$, cannot be simply replaced by $ \epsilon_u $ and $ \epsilon_\zeta $ respectively.  For the  spectral studies, it is more appropriate to employ  the total energy flux or the fluxes of Els\"{a}ser variables (topic of the next subsection) because they are constant in $k$.

\subsection{Energy fluxes associated with Els\"{a}ser variables}

An  alternative formulation of MHD turbulence is in terms of Els\"{a}ser variables, ${\bf z}^\pm = {\bf u} \pm  \bm{\zeta}$.   The MHD equations in terms   ${\bf z}^\pm$  and a mean magnetic field ${\bf B}_0$ are~ \cite{Cowling:book, Biskamp:book:MHDTurbulence}
 \bea
\frac{\partial{\mathbf{z}^\pm}}{\partial{t}}  \mp ({\bf B}_0 \cdot \nabla) {\bf z}^\pm + ( {\bf z}^\mp \cdot \nabla) {\bf z}^\pm   =  -  \nabla p  + \nu_+\nabla^2 {\bf z}^\pm + \nu_-\nabla^2 {\bf z}^\mp, \label{eq:MHD_formalism:MHDzpm}    \\
\nabla \cdot {\bf z}^\pm = 0, \label{eq:MHD_formalism:MHDdel_zpm_zero}
\eea 
where $ \nu_\pm = \frac{1}{2} (\nu \pm \eta)$.  For ${\bf z}^+$, in a triad $({\bf k',p,q})$, the mode-to-mode energy transfer from ${\bf z^+ (p)}$ to ${\bf z^+ (k')}$ with the mediation of ${\bf z^- (q)}$ is~ \cite{Verma:PR2004}
 \bea
S^{z^+ z^+}({\bf k'|p|q}) & = & -\Im \left[ {\bf \{  k' \cdot z^-(q) \}  \{ z^+ (p) \cdot  z^{+}(k') \}}  \right]. \ \eea
For ${\bf z}^-$, the mode-to-mode energy transfer from ${\bf z^- (p)}$ to ${\bf z^- (k')}$ with the mediation of ${\bf z^+(q)}$ is
\bea
S^{z^- z^-}({\bf k'|p|q}) & = & -\Im \left[ {\bf \{  k' \cdot z^+(q) \}  \{ z^- (p) \cdot  z^{-}(k') \}}  \right].
\eea
Note that there is no cross  transfer from ${\bf z^+}$ to ${\bf z^-}$ and vice versa.  
The corresponding energy fluxes are
\bea
\Pi_{z^+}(k_0)=  \sum_{p \le k_{0}} \sum_{k>k_{0}} S^{z^+ z^+}({\bf k|p|q});~
\Pi_{z^-}(k_0)=  \sum_{p \le k_{0}} \sum_{k>k_{0}} S^{z^- z^-}({\bf k|p|q}) .
\eea
	
Due to the absence of cross  transfers between  ${\bf z}^+$ and  ${\bf z}^-$, $\Pi_{z^\pm}(k_0)$ are constant in the inertial range of MHD turbulence.  In addition, in the inertial range,   $  \Pi_{z^\pm}(k) = \epsilon_{z^\pm} $, where $\epsilon_{z^\pm}$ are the total dissipation rates of ${\bf z}^\pm$.  Constancy of $\Pi_{z^\pm}(k) $ in the inertial range makes them  suitable candidates for modelling the energy spectrum of MHD turbulence.  For example,  Marsch~\cite{Marsch:RMA1991} argued that
\bea
E_{z^\pm}(k) = K_{z^\pm}(k)  \epsilon_{z^\pm}^{4/3} \epsilon_{z^\mp}^{-2/3} k^{-5/3},
\eea
where $ K_{z^\pm}(k)$ are nondimensional constants.   Refer to  Bismamp~\cite{Biskamp:book:MHDTurbulence} and   Verma~\cite{Verma:PR2004} for further details.

In the next section we will briefly describe the energy fluxes for a solvent with dilute  polymers.

\section{Variable energy fluxes in a turbulent flow with dilute polymers}
\label{sec:polymer}
In this section, we discuss the energy transfers and drag reduction  in a solution of dilute polymers~ \cite{deGennes:book:Polymer,Tabor:EPL1986, Sreenivasan:JFM2000, Benzi:PRE2008, Benzi:PD2010, Benzi:ARCMP2018}.     In such flows, the polymer is often described by finitely extensible nonlinear elastic-Peterlin (FENE-P) model.   The equations for the  velocity field and  polymer-conformation tensor $\zeta$ of FENE-P model are  \cite{Fouxon:PF2003, Perlekar:PRL2006}:
\bea
\frac{\partial{u_i}}{\partial t} +u_j \partial_j u_i
= -\frac{1}{\rho} \partial_i p +  \nu \partial_{jj} u_i + \frac{\mu}{\tau_p} \partial_j (f   \zeta_{ij}) + F_{\mathrm{LS},i} ,\label{eq:tensor:FENE-u} \\
\frac{\partial{\zeta_{ij}}}{\partial t} +  u_l \partial_l \zeta_{ij}
=    \zeta_{il} \partial_l u_j +  \zeta_{il} \partial_j u_l +  \frac{1}{\tau_p}   [f \zeta_{ij} -\delta_{ij} ],
\label{eq:tensor:FENE-C} \\
\partial_i u_i  = 0, 
\eea
where $p$ is the pressure, $\rho$ is the mean density of the solvent, $\nu$ is the kinematic viscosity, $\mu$ is an additional viscosity parameter, $\tau_p$ is the polymer relaxation time, and $f$ is the renormalized Peterlin's function. 
  We also remark that energetics of polymer turbulence has many similarities with those for MHD turbulence.  As we describe below, there is a preferential energy transfer from the velocity field to the polymer, just like the energy transfer from the velocity field to the magnetic field in dynamos. Refer to  deGennes~\cite{deGennes:book:Intro},  Fouxon and Lebdev~ \cite{Fouxon:PF2003}, and references therein for  details.

 In the above equations, the following  forces   (apart from constants) associated with ${\bf u}$ and $\zeta$  induce cross energy transfers:
 \bea
 F_{u,i} =   \partial_j (f   \zeta_{ij}),~~~
 F_{\zeta,ij} = \zeta_{il} \partial_l u_j +  \zeta_{il} \partial_j u_l.
 \eea
In Fourier space, the respective energy feed by these forces to the kinetic energy and the tensor energy  are 
  \bea
 \mathcal{F}_u({\bf k}) & = &  - \sum_{\bf p} \Im \left[  k_j f({\bf q}) \zeta_{ij}({\bf p}) u^*_i({\bf k}) \right], \\
  \mathcal{F}_{\zeta}({\bf k}) & = &  -\sum_{\bf p} \Im \left[  \zeta_{il}({\bf q}) p_l u_j({\bf p})  \zeta^*_{ij}({\bf k})  +  \zeta_{il}({\bf q}) p_j u_l({\bf p})  \zeta^*_{ij}({\bf k})  \right],
 \eea
where ${\bf q = k-p}$.   Both, $ \mathcal{F}_u({\bf k}) $ and $\mathcal{F}_{\zeta}({\bf k})$ are convolutions similar to those in MHD turbulence.   However, the structure of the nonlinear terms for the polymers is  more complex than that for MHD turbulence.  Till date, there are no formulas for the mode-to-mode energy transfers from ${\bf u}$ to $\zeta$ and vice versa. Yet, the following equations can be used to describe the energy fluxes from the velocity field to the polymer field.
\bea
\Pi^u_{\zeta<}(k_0) = \sum_{k \le k_0}    \mathcal{F}_\zeta({\bf k});~~~\Pi^u_{\zeta>}(k_0) = \sum_{k > k_0}    \mathcal{F}_\zeta({\bf k}) .
\eea 
  In fact, Eqs.~(\ref{eq:mhd:identity_1_2}, \ref{eq:mhd:identity_3_4}) too are applicable to turbulent flows with dilute polymers.

From Eq.~(\ref{eq:mhd:identity_3_4}), under a steady state and in the inertial range,
 \bea
 \Pi_u(k) =  \varepsilon_\mathrm{inj} - \Pi^{u<}_\zeta(k) 
  \label{eq:mhd:Pi_u_polymer_eps_inj}
 \eea
 where $\varepsilon_\mathrm{inj}$ is the energy injection rate by  large-scale forcing, and  $ \Pi^{u<}_\zeta(k)$ is the energy transfer from the large-scale velocity field to $ \zeta $.    See Fig.~\ref{fig:second:identities2}(a) for an illustration.  In  polymeric flows, it has been shown that the velocity field stretches the polymers~ \cite{deGennes:book:Polymer,Tabor:EPL1986}, similar to the stretching of the magnetic field  in MHD turbulence and  dynamo.  Therefore,  we expect that 
 \bea
 \Pi^{u<}_\zeta(k) > 0  \Rightarrow \Pi_u(k) <  \varepsilon_\mathrm{inj}.
 \eea
 That is, for the same $ \varepsilon_\mathrm{inj}$,  the kinetic energy flux in the polymer solution  will be reduced compared to  hydrodynamic turbulence.  Therefore, the turbulent drag, $F_D \approx \Pi_u/U$, for a polymeric flow will be lower than its  hydrodynamic counterpart~\cite{Verma:PP2020}.  Several numerical simulations exhibit the aforementioned reduction in $\Pi_u(k)$~\cite{Valente:JFM2014, Valente:PF2016}.    Thus,  variable energy flux provides valuable insights into the mechanism of turbulent drag reduction in polymer solution and in  MHD turbulence. 


In the next section we describe the flux associated with enstrophy and  kinetic helicity.

\section{Variable enstrophy and helicity fluxes}
\label{sec:enstrophyHk}

In this section we focus on the enstrophy and kinetic-helicity fluxes of 3D  hydrodynamics.  Note that $ \textbf{F}_u =0 $ for pure hydrodynamics.  However, we will retain the large-scale forcing  ($ {\bf F}_{\mathrm{LS}}$).  

The vorticity field,  $\bm{\omega} = \nabla  \times {\bf u}$, plays an important role in hydrodynamic turbulence.   The dynamical equation of $\bm{\omega}$ is~\cite{Lesieur:book:Turbulence, Frisch:book, Davidson:book:Turbulence}
\bea 
\frac{\partial \bm{\omega} }{\partial t}    = \nabla \times ({\bf u}\times \bm{\omega})  +{\bf F}_{\omega,\mathrm{LS}}   +  \nu \nabla^2  \bm{\omega}, 
\label{eq:gov:omega_dot0}
\eea
or
\bea 
\frac{\partial \bm{\omega} }{\partial t} + ({\bf u} \cdot \nabla)\bm{\omega} = ( \bm{\omega}\cdot \nabla)  {\bf u}   + {\bf F}_{\omega,\mathrm{LS}}  + \nu \nabla^2  \bm{\omega},
\label{eq:gov:omega_dot}
\eea
where   ${\bf F}_{\omega,\mathrm{LS}}   =  \nabla \times {\bf F}_\mathrm{LS}$. The total  enstrophy, $E_\omega = \frac{1}{2} \int d{\bf r} \omega^{2}$, and the modal enstrophy, $E_\omega({\bf k}) = \frac{1}{2}  | \bm{\omega}({\bf k})|^2$, are   important quantities of hydrodynamics.   The evolution equation for the latter is
\bea
\frac{d}{dt} E_\omega(\mathbf{k}) & = &\sum_{\bf p} \left\{ \Im \left[ {\bf \{ k \cdot u(q) \}  \{  \bm{\omega}(p) \cdot  \bm{\omega}^* (k) \} } \right]  - \Im \left[ \{ {\bf k \cdot  \bm{\omega}(q) \}   \{ u(p) \cdot  \bm{\omega}^*(k) \} } \right] \right\} \nonumber \\
&&+ \mathcal{F}_{\omega,\mathrm{LS}}(\mathbf{k})  - 2 \nu k^{2} E_\omega({\mathbf k}),
\label{eq:gov:Eomega_dot}
\eea  
where
\bea  
 \mathcal{F}_{\omega,\mathrm{LS}}(\mathbf{k}) & = & \Re[ i {\bf k \times F}_{u,\mathrm{LS}}({\bf k}) 
\cdot  \bm{\omega}^*({\bf k}) ] = k^2 \Re[  {\bf u}^*({\bf k}) \cdot {\bf F}_\mathrm{LS}({\bf k})]  
\label{eq:enstrophy_injectoon_rate}
\eea
is the enstrophy injection rate by $ {\bf F}_{\mathrm{LS}} $. In Eq.~(\ref{eq:gov:Eomega_dot}),  the first term in the right-hand-side represents the advection of the vorticity field by the velocity field, while  the second term represents  vortex stretching.  For inviscid and force-free 3D hydrodynamics, $E_\omega $ is  \textit{not} conserved  due to the vortex stretching by the velocity field~\cite{Lesieur:book:Turbulence, Frisch:book, Davidson:book:Turbulence}.    Note however that $E_\omega $ is conserved in 2D hydrodynamics; this issue will be discussed in Sec.~\ref{sec:2D}.

\subsection{Variable enstrophy flux}
\label{subsec:3D_enstrophy_flux}
When we compare Eq.~(\ref{eq:gov:omega_dot}) with Eq.~(\ref{eq:secondary:vector}) for the secondary vector $ \bm{\zeta}$, we obtain the following correspondence 
\bea
\bm{\zeta} \rightarrow  \bm{\omega};~~~{\bf F}_\zeta \rightarrow  {\bf F}_\omega = (  \bm{\omega}\cdot \nabla)  {\bf u},
\eea
and
\bea
\mathcal{F}_\omega({\bf k}) & = & \sum_{\bf p} - \Im \left[ {\bf \{  k \cdot  \bm{\omega}(q) \}  \{ u(p) \cdot  \bm{\omega}^*(k)  \} } \right];~~~
\mathcal{F}_u({\bf k}) = 0.
\eea 
Here, $\mathcal{F}_\omega({\bf k}) $ induces   enstrophy enhancement via vortex stretching  by the velocity field.  This process is similar to the magnetic field stretching by the velocity field in MHD turbulence (see Sec.~\ref{sec:mhd}), first proposed by Batchelor~\cite{Batchelor:PRSA1950}.   Note however that vorticity field does not back-react on the velocity field  because $\mathcal{F}_u({\bf k})=0$.  

By making an analogy with MHD turbulence, we  define the following enstrophy fluxes for a  wavenumber sphere of radius $k_0$ (see Sec.~\ref{sec:mhd}):
\bea
\Pi^u_{\omega<}(k_0) =  \sum_{k \le k_0} \mathcal{F}_\omega({\bf k});~~~
\Pi^u_{\omega>}(k_0) =   \sum_{k > k_0} \mathcal{F}_\omega({\bf k}).
\eea 
Here, $\Pi^u_{\omega<}(k_0)$ ($\Pi^u_{\omega>}(k_0)$) represents the net enstrophy gain by the vorticity modes within (outside) the sphere due to the nonlinear interactions with  the velocity modes.  Following Eq.~(\ref{eq:secondary:Pi_u_zeta_sum}), we deduce that the net enstrophy enhancement rate due to the vortex stretching is
\bea
\Pi^u_{\omega<}(k) + \Pi^u_{\omega>}(k) = \mathcal{F}_\omega 
= \mathrm{const}.
\label{eq:Pi_u_omega_sum_F_omega}
\eea
The sum in the above equation is independent of $k$ even though its constituents, $\Pi^u_{\omega<}(k)$ and $\Pi^u_{\omega>}(k) $, may vary with $k$.   

Since $\mathcal{F}_u({\bf k})=0$ (no back reaction from the vorticity to the velocity field), both $ \Pi^\omega_{u<}(k)$ and $ \Pi^\omega_{u>}(k)$ are zeros.  Hence, the vortex dynamics is similar to that of kinematic dynamo where the magnetic field does not back-react on the velocity field~\cite{Moffatt:book}.  Note however that in MHD turbulence, the magnetic field back-reacts on the velocity field.  Also note that the  enstrophy and  kinetic energy have different dimensions.

The term $({\bf u} \cdot \nabla) \bm{\omega} $ provides advection to the vorticity field,  analogous to the advection of the secondary vector by the term $({\bf u} \cdot \nabla) \bm{\zeta} $ (see Sec.~\ref{sec:secondary}).  Consequently, following Eq.~(\ref{eq:secondary:M2M_vector}), we define the mode-to-mode enstrophy transfer from $ \bm{\omega}({\bf p})$ to  $ \bm{\omega}({\bf k'})$ with the mediation of ${\bf u(q)}$ as~\cite{Verma:book:ET,Sadhukhan:PRF2019}
\bea
S^{\omega \omega}({\bf k'|p|q}) & = & -\Im \left[  {\bf  \{  k' \cdot u(q) \} \{ \bm{\omega}({\bf p}) \cdot   \bm{\omega}({\bf k'}) \} }  \right] .
\eea
Hence, the  enstrophy flux is (see Eq.~(\ref{eq:secondary:secondary_flux})) 
\bea
\Pi_{\omega}(k_0) = \sum_{k'>k_0}   \sum_{p\le k_0} S^{\omega \omega}(\mathbf{k'|p|q}) .
\label{eq:enstrophy_flux}
\eea
Now, following Eqs.~(\ref{eq:secondary:dPi_dk_secondary}, \ref{eq:secondary:Pi_u_zeta_F_zeta}), we deduce that during a steady state, in the inertial range, 
\bea
\frac{d }{dk}  \Pi_\omega(k) =  \mathcal{F}_\omega(k)  = - \frac{d}{dk} \Pi^u_{\omega>}(k)  =   \frac{d}{dk} \Pi^u_{\omega<}(k) .
\label{eq:enstrophy:dPi_dk_scalar}
\eea
Therefore, 
\bea
\Pi_\omega(k)  +  \Pi^u_{\omega>}(k) = C_3;~~~
\Pi_\omega(k)  -  \Pi^u_{\omega<}(k) = C_4,
\label{eq:enstrophy_flux_C1_C2}
\eea
\label{eq:enstrophy_identitities}
where $C_3, C_4$ are constants.  Since $  \Pi^u_{\omega>}(k)$ and $ \Pi^u_{\omega<}(k)  $ are nonzero, we conclude that $ \Pi_\omega(k)  $ varies with $k$, unlike $\Pi_u(k)$, which is constant  in the inertial range.  The above arguments and Eq.~(\ref{eq:Pi_u_omega_sum_F_omega}) yields $C_3-C_4 = \mathcal{F}_\omega$.  

We derive  several new formulas by making an  analogy between the vorticity field and the magnetic field.  For example, a comparison of vorticity dynamics with Eqs.~(\ref{eq:mhd:Suzeta}, \ref{eq:mhd:Szetau})  yields the following  formulas for the mode-to-mode enstrophy transfers:
\bea
S^{u \omega}({\bf k'|p|q}) &= & 0, \\
S^{\omega u}({\bf k'|p|q}) &= & \Im \left[ {\bf \{  k' \cdot  \bm{\omega}(q) \}  \{  u(p) \cdot  \bm{\omega}(k')   \} } \right]. 
\eea
Consequently, analogous to MHD fluxes, we can define  four enstrophy fluxes: $\Pi^{u<}_{\omega<}$, $\Pi^{u<}_{\omega>}$, $\Pi^{u>}_{\omega<}$, $\Pi^{u>}_{\omega>}$.  See Fig.~\ref{fig:second:second_flux} of Sec.~\ref{sec:mhd} for an illustration.
Note that 
\bea
\Pi^u_{\omega<}(k) = \Pi^{u<}_{\omega<}(k) + \Pi^{u>}_{\omega<}(k);~~~
\Pi^u_{\omega>}(k) = \Pi^{u<}_{\omega>}(k) + \Pi^{u>}_{\omega>}(k). 
\eea 
Substitutions of these relations in the identity of Eqs.~(\ref{eq:Pi_u_omega_sum_F_omega}) yields the following relation~\cite{Sadhukhan:PRF2019}:
\bea
\Pi^{u<}_{\omega<}(k) + \Pi^{u>}_{\omega<}(k) +  \Pi^{u<}_{\omega>}(k) + \Pi^{u>}_{\omega>}(k)
= \mathcal{F}_\omega.
\label{eq:misc:enstrophy_flux_total}
\eea
The above sum is constant in $k$ (both in inertial and dissipation range), akin to Eq.~(\ref{eq:mhd:C2}) for MHD turbulence.

\begin{figure}[htbp]
	\begin{center}
		\includegraphics[scale = 0.25]{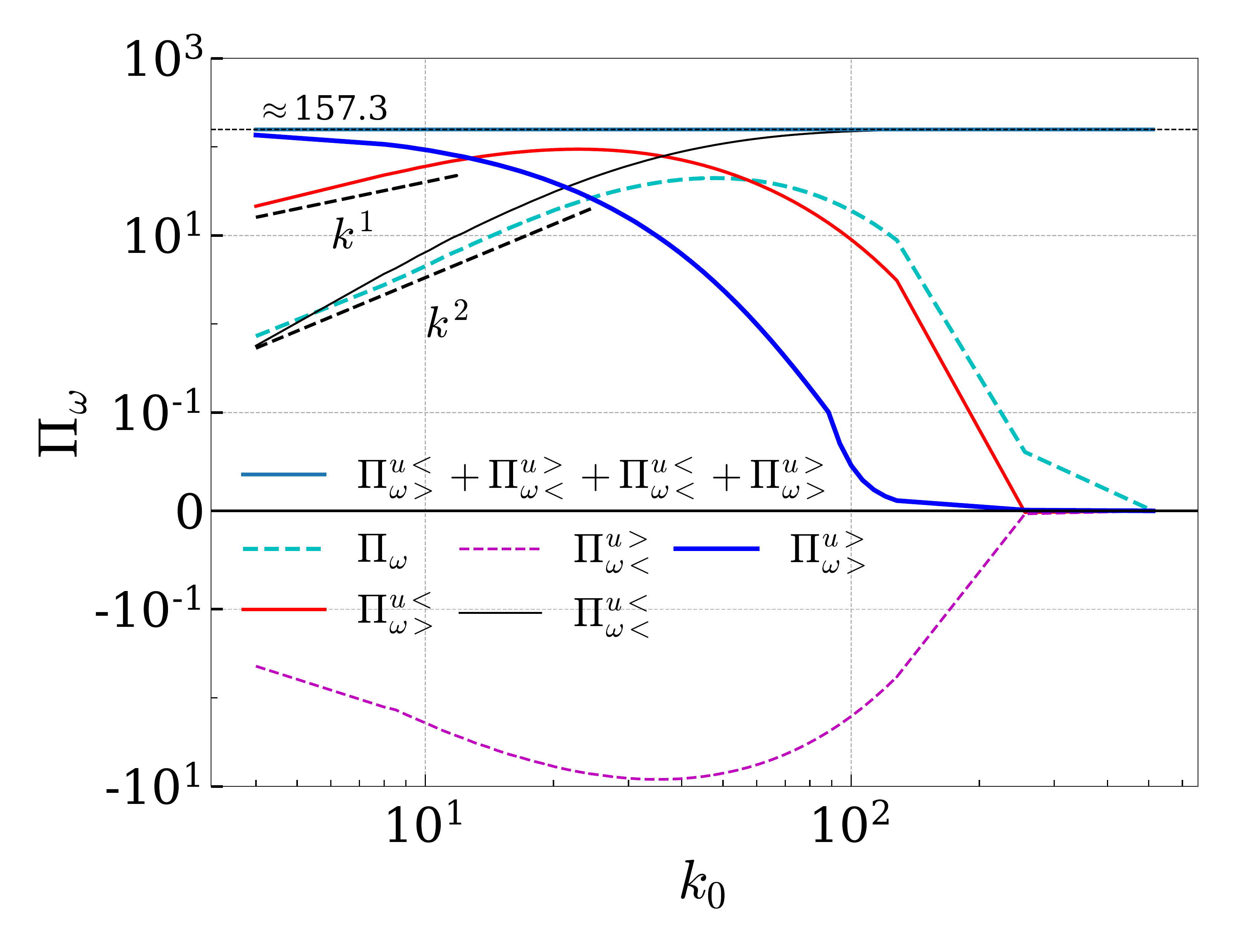}
	\end{center}
	\caption{ For the numerical simulations by Sadhukhan et al.~\cite{Sadhukhan:PRF2019}, plots of enstrophy fluxes and the conserved flux of Eq.~(\ref{eq:misc:enstrophy_flux_total}). From  Sadhukhan et al.~\cite{Sadhukhan:PRF2019}.  Reprinted with permission from APS. }
	\label{fig:misc:enstrophy_flux}
\end{figure}

Sadhukhan et al.~\cite{Sadhukhan:PRF2019} performed numerical simulations of hydrodynamic turbulence and computed the aforementioned enstrophy fluxes, as well as the conserved quantity of Eq.~(\ref{eq:misc:enstrophy_flux_total}).  These quantities are exhibited in Fig.~\ref{fig:misc:enstrophy_flux}.  Note that the individual fluxes may vary with $k$ due to cross transfers, but the sum of Eq.~(\ref{eq:misc:enstrophy_flux_total}) is a constant.  Interestingly, $ \Pi_\omega(k) \sim k^2 $ and $ \Pi^{u<}_{\omega >}(k)  \sim k$.  Verma~\cite{Verma:book:ET} argued that $ \Pi_\omega(k) \sim k^2 $  due to the term  $   \bm{\omega}({\bf p}) \cdot   \bm{\omega}({\bf k'})  $ of $ S^{\omega \omega}({\bf k'|p|q}) $. At small wavenumbers, $ \Pi^{u>}_{\omega>}(k) $ is the most dominant among all the fluxes implying  that the  intermediate and small-scale vortices are stretched most significantly.

We summarise the  similarities and  dissimilarities between the  vorticity field and magnetic field in Table~\ref{tab:misc:vort_vs_B}.  A major difference between the two fields is related to the back-reaction on the velocity field---vorticity does not back-react, but magnetic field does.  Hence, enstrophy dynamics  is similar to  that of  kinematic dynamo where the magnetic field does not affect the velocity field.   Note that enstrophy fluxes do not have  a relation equivalent to the conservation of total energy flux in MHD turbulence (Eq.~(\ref{eq:mhd:total_flux})).  
\begin{table}[h]
	\centering
	\caption{ Similarities (first nine rows) and dissimilarities (last two rows) between the magnetic and vorticity fields. }
	{\begin{tabular}{c|c} \hline
			Magnetic field & Vorticity \\ \hline
			$S^{\zeta \zeta}$ & $S^{\omega \omega}$ \\
			$S^{\zeta u}$ & $S^{\omega u}$ \\
			$\Pi_\zeta$ & $\Pi_\omega$ \\
			$\Pi^u_\zeta$ & $\Pi^u_\omega$ \\
			Stretching of magnetic  field lines & Vortex stretching \\ 
			$u$-to-$\zeta$ transfer &  $u$-to-$\omega$ transfer  \\
			Growth of magnetic field & Enhancement of enstrophy \\
			$\zeta$-to-$\zeta$ transfer (forward for nonhelical) & $\omega$-to-$\omega$ transfer (forward) \\
			No dynamo in 2D & No vortex stretching in 2D  \\ \hline 
			$S^{u \zeta} \ne 0$ & $S^{u \omega} = 0$ \\ 
			$\zeta$-to-$u$ transfer (back-reaction), &  No $\omega$-to-$u$ transfer (no back-reaction) \\ 
			except for kinematic dynamo where $S^{u \zeta} = 0$ & \\
			 \hline
	\end{tabular}}
	\label{tab:misc:vort_vs_B}
\end{table}

In the next subsection, we will the describe variable kinetic-helicity flux.

\subsection{Flux of kinetic helicity}
\label{subsec:HK}

Kinetic helicity, which is defined as $H_K = \frac{1}{2}  \int d{\bf r} ( {\bf u} \cdot \bm{\omega})$,   plays a major role in the growth of vorticity and magnetic field~ \cite{Lesieur:book:Turbulence,Davidson:book:Turbulence,Moffatt:book}.   Using Eqs.~(\ref{eq:gov:NS_ur}, \ref{eq:gov:omega_dot0}) we can show that for $\nu=0$, $ {\bf F}_u =0,  {\bf F}_\mathrm{LS} = 0$, and  periodic or vanishing  boundary condition, the total kinetic helicity  is conserved in 3D hydrodynamics~\cite{Lesieur:book:Turbulence, Frisch:book, Davidson:book:Turbulence}; this is in addition to the conservation of  total kinetic energy.

 The evolution equation for the modal kinetic helicity, $H_K({\bf k})  = \frac{1}{2}  \Re[{\bf u}({\bf k}) \cdot  \bm{\omega}^*({\bf k}) ]$, is~\cite{Lesieur:book:Turbulence, Frisch:book, Davidson:book:Turbulence,Verma:book:ET}
\be
\frac{d}{dt}H_K({\bf k}) =  \sum_{\bf p} \Re[{\bf u(q)} \cdot \{  \bm{\omega}({\bf p}) \times   \bm{\omega}^*({\bf k}) \}]  +  \mathcal{F}_{H_K,\mathrm{LS}}({\bf k})  -2 \nu k^2 H_K({\bf k}) ,
\label{eq:fourier:Hk_dot}  
\ee
where $ \mathcal{F}_{H_K,\mathrm{LS}}(\mathbf{k}) =  \Re[  \bm{\omega}^*({\bf k}) \cdot {\bf F}_{\mathrm{LS}}({\bf k})]$ is the kinetic helicity injection rate by  $ {\bf F}_{\mathrm{LS}} $.   Based on Eq.~(\ref{eq:fourier:Hk_dot}), researchers have derived various formulas for the kinetic helicity flux.  For example, M\"{u}ller et al.~\cite{Muller:PRE2012} argued that the kinetic flux is given by 
\bea
\Pi_{H_K}(k_0) = \sum_{k>k_0}   \{ i {\bf k \times [ u \times  \bm{\omega}} ]({\bf k}) \} \cdot   {\bf u^*(k)} +c.c. ,
\label{eq:Muller_HKflux}
\eea
where $c.c.$ stands for the complex conjugate.   In the following discussion we will describe more  flux formulas for the kinetic helicity.

Following the structure of the nonlinear term of Eq.~(\ref{eq:fourier:Hk_dot}),     Verma~\cite{Verma:book:ET}, Sadhukhan et al.~\cite{Sadhukhan:PRF2019}, and Plunian et al.~\cite{Plunian:JPP2019}  showed that the mode-to-mode kinetic helicity from wavenumber ${\bf p}$ to wavenumber ${\bf k'}$ with the mediation of wavenumber ${\bf q}$ is
\bea
S^{H_K}(\mathbf{k'|p|q}) =  \Re[{\bf u(q)} \cdot \{  \bm{\omega}({\bf p}) \times   \bm{\omega}({\bf k'}) \}]  .
\eea
In terms of $S^{H_K}(\mathbf{k'|p|q})$, the kinetic helicity flux for a wavenumber sphere of radius $k_0$ is
\bea
\Pi_{H_K}(k_0) = \sum_{p \le k_0}   \sum_{k'>k_0} S^{H_K}(\mathbf{k'|p|q}) .
\label{eq:helical_turb:HK_flux}
\eea
Following the same lines of arguments as in Sec.~\ref{sec:VF}, we obtain
\bea
\frac{d}{d k} \Pi_{H_K}(k) =   \mathcal{F}_{H_K,\mathrm{LS}}(k)    - D_{H_K}(k),
\label{eq:dPi_HK_dk_FHK}
\eea
where $D_{H_K}(k) = 2 \nu k^2 H_K(k)$ is the dissipation rate of kinetic helicity in shell $k$, and $ \mathcal{F}_{H_K,\mathrm{LS}}(k) $ is    the kinetic helicity injection rate  in the shell due to the  force ${\bf F}_\mathrm{LS}$.        In the inertial range,  $\mathcal{F}_{H_K,\mathrm{LS}}(k)  =0$ and $D_{H_K}(k) =0 $,  hence $ \Pi_{H_K}(k)  = \epsilon_{H_K} = \mathrm{constant}$,  where $ \epsilon_{H_K} $ is the total dissipation rate of kinetic helicity.   Using the constancy of $\Pi_{H_K}(k) $, dimensional analysis and field-theoretic arguments, the following inertial-range spectrum for  kinetic helicity has been derived~\cite{Lesieur:book:Turbulence, Zhou:PRE1993RG, Avinash:Pramana2006}:
\bea
H_K(k) = K_H \epsilon_{H_K} (\epsilon_u)^{-1/3} k^{-5/3} ,
\label{eq:helical_turb:HK(k)}
\eea
where $K_H$ is a nondimensional constant, whose value has been estimated to be of the order of unity.   The above scaling has been verified in many numerical simulations~\cite{Lesieur:book:Turbulence,Teimurazov:CCM2017,Sadhukhan:PRF2019}.   Interestingly, Sadhukhan et al.~\cite{Sadhukhan:PRF2019} also modelled the kinetic helicity spectrum and flux in the inerital-dissipation range using a  generalized Pao's model~\cite{Pao:PF1965}.

The helical turbulence is also described using Craya-Herring and helical basis~\cite{Waleffe:PF1992, Kessar:PRE2015,Alexakis:PR2018,Sahoo:FDR2018,Plunian:JPP2019}.   In Craya-Herring basis~\cite{Craya:thesis,Herring:PF1974,Lesieur:book:Turbulence,Sagaut:book,Waleffe:PF1992, Verma:book:BDF}, the unit vectors for a wavenumber \textbf{k} are
\bea
\hat{e}_3({\bf k}) = \hat{k};~~
\hat{e}_1({\bf k}) = \frac{ \hat{k} \times \hat{n}}{|\hat{k} \times \hat{n}|};~~
\hat{e}_2({\bf k}) = \hat{e}_3({\bf k})  \times \hat{e}_1({\bf k}) ,  
\eea 
where $\hat{k}$ is the unit vector along  wavenumber ${\bf k}$, and $\hat{n}$ could be along any  direction, but it is typically chosen along the anisotropy direction.  We denote the velocity components along $\hat{e}_1({\bf k})$, $\hat{e}_2({\bf k})$,  $\hat{e}_3({\bf k})$ as $ u_1({\bf k}), u_2({\bf k}), u_3({\bf k}) $ respectively.  Among them $u_3({\bf k}) = 0$  due to  incompressibility,  hence
${\bf u}({\bf k}) =u_1({\bf k})  \hat{\textbf{e}}_1(\textbf{k}) + u_2({\bf k})   \hat{\textbf{e}}_2(\textbf{k})$.  In this  basis,  the mode-to-mode kinetic energy transfer  from wavenumber $ \textbf{p} $ to wavenumber $ \textbf{k} $ with the mediation of wavenumber $ \textbf{q} $ is 
\bea
S^{uu}({\bf k'|p|q}) =S^{u_1 u_1}({\bf k'|p|q}) + S^{u_2 u_2}({\bf k'|p|q}),
\label{eq:helical_basis:Suu_kpq_CH}
\eea
where
\bea
S^{u_1 u_1}({\bf k'|p|q})& = &k' \sin\bar{\beta} \cos \bar{\gamma}  \Im \{u_1({\bf q}) u_1({\bf p}) u_1({\bf k'}) \} ,  \label{eq:helical_basis:Su1u1} \\
S^{u_2 u_2}({\bf k'|p|q}) & = &- k' \sin\bar{\beta} \Im \{u_1({\bf q}) u_2({\bf p}) u_2({\bf k'}) \} ,\label{eq:helical_basis:Su2u2}
\eea 
with $\bar{\alpha}, \bar{\beta}, \bar{\gamma}$ as the internal angles across $ k,p,q $ of the triangle formed by the wavenumbers ${\bf (k',p,q)}$~ \cite{Leslie:book,Verma:book:ET}.   

Another useful basis called {\em helical basis}~\cite{Waleffe:PF1992,Lesieur:book:Turbulence,Sagaut:book}  is constructed using the Craya-Herring vectors.  In this basis, the unit vectors  are
\bea
\hat{e}_{s_k}({\bf k}) = \frac{1}{\sqrt{2}} [\hat{e}_2({\bf k}) - i s_k \hat{e}_1({\bf k})],
\eea
where $s_k $ takes values $+1$ or $-1$.  In terms of these unit vectors, the velocity field is $
{\bf u}({\bf k}) =u_+({\bf k}) \hat{e}_+({\bf k})  + u_{-} ({\bf k}) \hat{e}_-({\bf k})$, 
where
\bea
u_{s_k}({\bf k}) & = & \frac{1}{\sqrt{2}}  [u_2({\bf k}) + i s_k u_1({\bf k})].  \label{eq:helical_basis:upm} 
\eea
In the helical basis, the mode-to-mode kinetic energy transfer from ${\bf u(p)}$ to ${\bf u(k')}$ with the mediation of ${\bf u(q)}$ is 
\bea
S^{uu}({\bf k'|p|q}) & = & \sum_{s_p, s_{k'}}  S^{uu}_{s_{k'} s_p}({\bf k'|p|q}),
\eea
where $S^{uu}_{s_{k'} s_p}({\bf k'|p|q})$,  the kinetic energy transfer from mode $u_{s_p}({\bf p})$ to $u_{s_{k'}}({\bf k'})$ with the mediation of ${\bf u(q)}$, is~\cite{Verma:book:ET}
\bea
S^{uu}_{s_{k'} s_p}({\bf k'|p|q}) & = & -\Im \left[ {\bf  \{  k' \cdot u(q) \}}  
u_{s_p}({\bf p}) u_{s_{k'}}({\bf k'}) 
\{ \hat{e}_{s_p}({\bf p}) \cdot \hat{e}_{s_{k'}}({\bf k'}) \} \right]  \nonumber \\
& = & -\frac{k'}{2} \sin\bar{\beta} (1+s_p s_{k'} \cos \bar{\gamma})
\Im \{u_1({\bf q}) u_{s_p}({\bf p}) u_{s_{k'}}({\bf k'}) \}.  \label{eq:helical_basis:Suu_helical_modes}
\eea
Similarly, the mode-to-mode kinetic helicity transfer from wavenumber ${\bf p}$ to wavenumber ${\bf k'}$ with the mediation of wavenumber ${\bf q}$ is~\cite{Verma:book:ET}
\bea
S^{H_K}({\bf k'|p|q}) & = &\sum_{s_p, s_{k'}}  S^{H_K}_{s_{k'} s_p}({\bf k'|p,q}) ,
\eea
where $S^{H_K}_{s_{k'} s_p}({\bf k'|p,q}) $, the elemental helicity transfer from $u_{s_p}({\bf p})$  to $u_{s_{k'}}({\bf k'})$ with the mediation of ${\bf u(q)}$,  is
\bea
S^{H_K}_{s_{k'} s_p}({\bf k'|p,q}) & = & -\frac{1}{2} pk' [s_{k'}  \sin \bar{\beta} + s_{p}\sin \bar{\alpha}]
\Im \{u_1({\bf q}) u_{s_p}({\bf p}) u_{s_{k'}}({\bf k'}) \}  \nonumber \\
&& +\frac{1}{2} pk' \sin \bar{\gamma} \Re \{u_2({\bf q}) u_{s_p}({\bf p}) u_{s_{k'}}({\bf k'}) \}.
\eea
The kinetic energy flux from helical mode $u_{s_g}$ to mode $u_{s_r}$, where $s_g$ and $s_r$ are the signs of giver and receiver modes respectively, can be written as
\bea
\Pi^{u_{s_g} <}_{u_{s_r} >}(k_{0}) & = & \sum_{p \le k_{0}} \sum_{k' >k_{0}} S^{uu}_{s_g s_r} ({\bf k'|p|q}) .
\label{eq:helical_basis:fluid_flux_helical}
\eea
The corresponding kinetic helicity flux is
\bea
\Pi^{H_K s_g}_{H_K s_r}(k_{0}) =  \sum_{p \le k_{0}} \sum_{k' >k_{0}}   S^{H_K}_{s_r, s_g} ({\bf k'|p|q}).
\eea
Alexakis and Biferale~\cite{Alexakis:PR2018} and  Sahoo et al.~\cite{Sahoo:FDR2018} constructed another set of formulas for the energy flux, which is
\bea
\Pi^{s_1 s_2 s_3}_u  = \la {\bf u}^{< k} \cdot \mathcal{P}^{s_1} [{\bf u}^{s_2} \times  \bm{\omega}^{s_3} ] \ra ,
\label{eq:misc:Pi_HK_Biferale}
\eea
where $s_i$ takes values $\pm 1$ depending on the sign of kinetic helicity, and $\mathcal{P}^{s_1}$ represents projection along $\hat{e}_{s_1}$.   The above flux formula is related to Eq.~(\ref{eq:helical_basis:Suu_helical_modes}).

For  large-scale external force, the energy and helicity spectra exhibit $k^{-5/3}$ spectra and constant fluxes~\cite{Lesieur:book:Turbulence,Frisch:book}.  However, the energy transfers become more complex when the external force is employed at intermediate scales, or when only homochiral modes (modes with same sign of kinetic helicity)  are present.     Biferale et al.~\cite{Biferale:PRL2012} showed that the nonlinear interactions among homochiral modes yield an inverse cascade of kinetic energy in 3D hydrodynamic turbulence.  Sahoo et al.~\cite{Sahoo:PRL2017} varied the  strengths of different triadic interactions involving helical modes and observed a discontinuous transition from inverse energy cascade to forward energy cascade.    Plunian et al.~\cite{Plunian:JFM2020} performed numerical simulations with realistic helical modes (rather than homochiral modes)  and obtained similar results as Sahoo et al.~\cite{Sahoo:PRL2017}.   Detailed discussions on these  issues are beyond the scope of this review. For  details, refer to references \cite{Kessar:PRE2015,Alexakis:PR2018,Sahoo:FDR2018,Buzzicotti:PRF2018,Biferale:JFM2019,Plunian:JFM2020}.

We close this subsection with a comment that the secondary fields also induces kinetic helicity.   For example, bouyancy and Lorentz force induce kinetic helicity. These topics, however, are too complex to be discussed here.  Refer to   the references~\cite{Lesieur:book:Turbulence,Sagaut:book,Waleffe:PF1992,Verma:book:ET}  for further details.

\subsection{Variable kinetic energy flux in decaying  hydrodynamic turbulence}

So far, we focussed on the steady-state behaviour of turbulence.  In this section, we will briefly discuss how variable energy flux plays an important role in the decay of   hydrodynamic turbulence for which  ${\bf F}_u= {\bf F}_\mathrm{LS}=0$~\cite{Batchelor:PRSA1948_decay_init,Batchelor:PRSA1948_decay_final,Saffman:PF1967,Kolmogorov:DANS1941Degeneration,Davidson:book:Turbulence, Lesieur:book:Turbulence, Frisch:book}. 
  For a viscous   flow (flow with zero nonlinearity), the kinetic energy flux  is zero, while the total kinetic energy decays as
\bea
E_u(t) = \sum_{\bf k} E_u({\bf k},0) \exp(-\nu k^2 t),
\eea
where $E_u({\bf k},0)$ is the initial modal kinetic energy of wavenumber ${\bf k}$.  The decay law for a turbulent flow is more complex, which will be discussed below.

 To analyse the evolution of decaying turbulence, we start with Eq.~(\ref{eq:VF:Ek_energetics}) and integrate over small wavenumbers up to integral wavenumber $k_\ell \sim 1/l $  ($l$ is the integral length scale). The wavenumber region $ [0,k_l] $ contains most of the kinetic energyof the  turbulent flows.  The above  operation yields~\cite{Kolmogorov:DANS1941Degeneration, Davidson:book:Turbulence, Lesieur:book:Turbulence, Frisch:book}
\bea
\frac{dE}{dt} \approx  \frac{\partial}{\partial t}  \int_0^{k_\ell} dk E_u(k,t) \approx
-\Pi(k_\ell) -\int_0^{k_\ell} dk D_u(k,t).
\eea
Since the viscous dissipation is negligible in the wavenumber band $[0,k_\ell]$, we obtain
\bea
\frac{dE}{dt} = -\Pi(k_\ell) = -\frac{U^3}{\ell} ,
\label{eq:misc:decay0}
\eea
where $U$ is the large-scale velocity. Thus, the kinetic energy decays due to the kinetic energy flux  $\Pi(k_\ell)$. In decaying turbulence, $U$ decreases and $\ell$ increases, hence, $\Pi(k_\ell) $ is a time-dependent quantity.

Equation~(\ref{eq:misc:decay0}) can be solved if we know the relationship between $U$ and $\ell$.   To derive this relation,  Kolmogorov~\cite{Kolmogorov:DANS1941Degeneration} employed conservation of Loitsyansky integral, $\int r^2 \la {\bf u \cdot u'} \ra  d{\bf r}$ and obtained $ U^2 \ell^5 = \mathrm{constant}$~\cite{Monin:book:v1,Monin:book:v2}.  This conservation law  is related to the spectrum $E_u(k) \sim k^4$ at small wavenumbers:
\bea
E_u = U^2/2 \sim \int_0^{k_0} E_u(k) dk \sim k_0^5 \sim \ell^{-5}.
\eea
Using  $ U^2 \ell^5 = \mathrm{constant}$ we transform Eq.~(\ref{eq:misc:decay0}) to
\bea
\frac{dE}{dt} = -  \frac{U^3}{\ell}  = - c U^{17/5} = -c' E^{17/10},
\label{eq:misc:decay_law}
\eea
where $c$ and $c'$ are constants.  The above equation yields the following decaying solution:
\bea
E(t) \sim t^{-10/7};~~\ell(t) \sim t^{2/7}.
\eea

Saffman~\cite{Saffman:PF1967} derived another decay law based on the assumption that $E_u(k) \sim k^2$  at small wavenumbers.     This infrared spectrum yields  $ U^2 \ell^3 = \mathrm{constant} $,   substitution of which in  Eq.~(\ref{eq:misc:decay0}) yields
\bea
E(t) \sim t^{-6/5};~~\ell(t) \sim t^{2/5}.
\eea

Experiments on grid turbulence~\cite{Krogstad:JFM2010,Comte-Bellot:JFM1966,Mohamed:JFM1990} tend to favour 6/5 decay law of Saffman.  However,  numerical simulations exhibit one of the two laws depending on initial $E_u(k)$ at  small wavenumbers~\cite{Ishida:JFM2006, Davidson:JFM2010}:  10/7 law for $k^4$ spectrum and 6/5 law for $k^2$ spectrum.  The above experimental results tend to suggest that small wavenumber spectrum in grid turbulence may follow  $E_u(k) \sim k^2$  for $k < k_f$, similar to that observed in Kolmogorov flow~\cite{Dallas:PRL2015} (see Sec.~\ref{sec:Euler}).

For MHD turbulence, Biskamp~\cite{Biskamp:book:MHDTurbulence} derived decay laws based on conservation of magnetic helicity.  Verma~\cite{Verma:PR2004} employed variations in the energy flux to derive decay laws for MHD turbulence.  Refer to the original papers for details.

\section{Brief review of variable energy fluxes in 2D  and quasi-2D turbulence}
\label{sec:2D}

In this section, we present a brief review of variable energy fluxes in 2D and quasi-two-dimensional (quasi-2D)  hydrodynamic,   buoyancy-driven, and  MHD turbulence.  We start with a description of the energy and enstrophy fluxes of 2D hydrodynamic turbulence.

\subsection{Fluxes of 2D hydrodynamic turbulence}

For a 2D flow field ${\bf u} = u_x \hat{x}+u_y \hat{y}$, the vorticity field ${\bm \omega} = \omega \hat{z} = (\partial_x u_y - \partial_y u_x) \hat{z}$ is perpendicular to the plane of the flow.   Hence, Eq.~(\ref{eq:gov:omega_dot}) yields the following   evolution equation  for 2D  vorticity:
\bea 
\frac{\partial {\omega} }{\partial t} + ({\bf u} \cdot \nabla){\omega}    = 
 F_{\omega,\mathrm{LS}}  + \nu \nabla^2 {\omega}.
\label{eq:gov:omega2D}
\eea
Note the absence of vortex stretching term in the above equation.   For a 2D field, the total enstrophy $E_\omega = \int d{\bf r} \frac{1}{2} \omega^{2}$ is conserved~\cite{Lesieur:book:Turbulence, Frisch:book, Davidson:book:Turbulence}, which  is in addition to the conservation of total energy.   Note that in 2D hydrodynamics, the total kinetic helicity vanishes identically because ${\bf u}$ and $\bm{\omega}$ are perpendicular to each other.

 In 2D hydrodynamics, $ \Pi^u_{\omega>}(k) =   \Pi^u_{\omega<}(k) = 0 $ due to the absence of ${\bm \omega} \cdot \nabla  {\bf u}  $ term in Eq.~(\ref{eq:gov:omega2D}) (see  Sec.~\ref{subsec:3D_enstrophy_flux}).  Here, we  assume that  the external force supplies kinetic energy and enstrophy at large scales (see Eq.~(\ref{eq:enstrophy_injectoon_rate})). Following Eq.~(\ref{eq:enstrophy_flux_C1_C2}), we deduce that  $\Pi_\omega(k)$  is constant in the inertial range and it equals the enstrophy dissipation rate ($\epsilon_\omega$). This is in contrast  to 3D hydrodynamic turbulence where  $ \Pi_{\omega}(k)  $ varies with $k$ due to the  variability of $ \Pi^u_{\omega<}(k)  $ and $ \Pi^u_{\omega>}(k)  $.  Kraichnan~\cite{Kraichnan:PF1967_2D} showed that the above  enstrophy flux yields  $E_\omega(k) \sim \epsilon_\omega^{2/3} k^{-1}$  in the inertial-range.  Note that $ \Pi_u(k) $ is small in the inertial range. The above picture is a part of  Kraichnan~\cite{Kraichnan:PF1967_2D}'s framework for 2D turbulence that will be described below. 

\begin{figure}[htbp]
\begin{center}
\includegraphics[scale = 0.7]{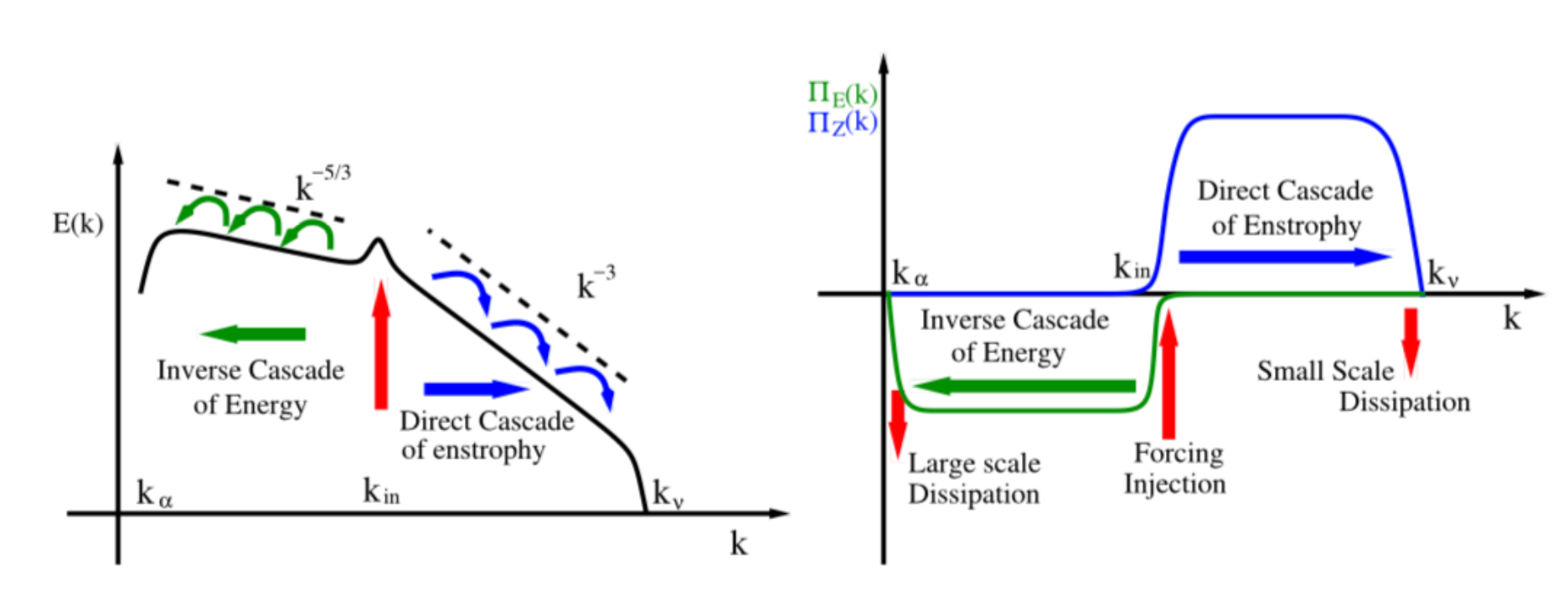}
\end{center}
\caption{For 2D turbulence, (left panel) schematic illustrations of energy and entropy spectra; (right panel) energy and enstrophy fluxes.  Note that $k_\mathrm{in} \equiv k_f$ and $k_\nu \equiv k_d$.  From  Alexakis and Biferale~\cite{Alexakis:PR2018}.  Reprinted with permission from Elsevier. }
\label{fig:2d:twoD_schematic}
\end{figure}

Two-dimensional turbulence gets more complex when it is forced at an intermediate wavenumber $ k_f$.    Using conservation laws and field-theoretic  tools,  Kraichnan~\cite{Kraichnan:PF1967_2D} constructed a phenomenology for such a scenario.   In the inertial range of $k < k_f$ region, $ \Pi_u(k)  =  -\epsilon_u  $ and  $ E_u(k) = K_{2D} \epsilon_u^{2/3} k^{-5/3} $, where $\epsilon_u$ is the magnitude of the kinetic energy flux, and  $K_{2D}$ is a constant whose numerical value is between 5.5 and 7.0.  However, in the inertial range of  $k>k_f$ regime, $ \Pi_\omega(k) = \epsilon_\omega > 0 $ and $  E_u(k) = K'_{2D} \epsilon_\omega^{2/3} k^{-3}  $, where  $K'_{2D}$ is another constant whose numerical value is between 1.3 and 1.7.   See Fig.~\ref{fig:2d:twoD_schematic} for an illustration.  The results of many experiments~(see \cite{Belmonte:PF1999,Rutgers:PRL1998,Rivera:PRL1998,Tabeling:PR2002} and references therein) and numerical simulations~(see  \cite{Boffetta:ARFM2012, Tabeling:PR2002}, and references therein) are consistent with the above predictions.  We do not detail these results here.

Sharma et al.~\cite{Sharma:PF2018} and  Gupta et al.~\cite{Gupta:PRE2019} extended   Kraichnan's  phenomenology~\cite{Kraichnan:PF1967_2D} beyond the inertial range by extrapolating Pao's model~\cite{Pao:PF1965} to 2D turbulence. In the following we extend their calculations by including Ekman friction.   For steady state, away from the forcing band, the kinetic energy and enstrophy fluxes obey the following equations:
\bea
\frac{d}{dk} \Pi_u(k)& = &  -2(\nu k^2+\alpha)  E_u(k), \label{eq:2d:Pi_u_steady_eq} \\
\frac{d}{dk} \Pi_\omega(k)  & = &  -2( \nu k^2 + \alpha) E_\omega(k) = -2 (\nu k^2+\alpha) k^2 E_u(k). \label{eq:2d:Pi_w_steady_eq}
\eea 
For $k<k_f$, it is assumed that $E_u(k)/\Pi_u(k)$   depends only  on the kinetic  energy dissipation rate ($\epsilon_u$) and $k$, but not on $\nu$.  Hence, 
\bea
\frac{E_u(k)}{\Pi_u(k)} = - K_{2D}  \epsilon_u^{-1/3} k^{-5/3},
\label{eq:2d:Pao_assumption}
\eea
where $\epsilon_u = -\Pi_u(k_0)$ with $k_0$ being a wavenumber near $k_f$.   Substitution of the above in  Eq.~(\ref{eq:2d:Pi_u_steady_eq}) and integration from $k$ to $k_0$ yields
\bea
\Pi_u(k)  & =  &  -\epsilon_u \exp{\left( \frac{3}{2} K_{2D}  [(k/k_d)^{4/3}- (k_0/k_d)^{4/3}] \right)} \nonumber \\
 && 	\times \exp{\left(-\frac{3\alpha K_{2D}}{\nu k_d^2}  [(k_d/k)^{2/3}- (k_d/k_0)^{2/3}] \right)}, \label{eq:2d:Pao_Pik}  \\
E_u(k)  & = &- K_{2D}  \Pi_u(k) \epsilon_u^{-1/3} k^{-5/3}, \label{eq:2d:Pao_Ek}
\eea  
where $k_d =  (\epsilon_u/ \nu^3)^{1/4}$ is Kolmogorov's wavenumber for 2D turbulence.   Interestingly, $ \Pi_u(k)  \rightarrow 0 $ as $k \rightarrow 0$ due to the Ekman friction.  Substitution of the above $E_u(k)$ in Eq.~(\ref{eq:2d:Pi_w_steady_eq}) yields the following enstrophy flux:
\bea
\Pi_\omega(k)   & = &   \Pi_\omega(k_0)  + 2  K_{2D}  \epsilon_u^{1/3}  
\int_{k_0}^{k} (\nu k'^2 + \alpha) \Pi_u(k') k'^{1/3} dk' .
\label{eq:2d:Piw_k<kf}
\eea

For $k>k_f$, following  Pao~\cite{Pao:PF1965}, it has been argued that $E_\omega(k)/\Pi_\omega(k)$  depends only on the enstrophy dissipation rate  ($\epsilon_\omega$) and $k$, but not on $ \nu $.  Hence,
\bea
\frac{E_\omega(k)}{\Pi_\omega(k)} & = & K'_{2D} \epsilon_\omega^{-1/3} k^{-1}.
\label{eq:2d:Pao_enstrophy}
\eea
Under this assumption, using Eq.~(\ref{eq:2d:Pi_w_steady_eq}) we obtain
\bea
\Pi_\omega(k) =   \Pi_\omega(k_0) \left( \frac{k}{k_0} \right)^{-2 \alpha K'_{2D} \epsilon_\omega^{-1/3} } 
\exp\left( - \frac{K'_{2D}}{k_{d2D}^2} (k^2-k_0^2) \right),
\label{eq:2d:Piomegak_k>k_f}
\eea
where $ k_{d2D} =\epsilon_\omega^{1/6}/\sqrt{\nu}$, and    $ \Pi_\omega(k_0) $ is the reference value of the enstrophy flux at $k=k_0$~\cite{Anas:PRF2019,Verma:EPL2012}.  We choose $k_0 \approx k_f$ and $\Pi_\omega(k_f)  \approx \epsilon_\omega$, where $\epsilon_\omega$ is the enstrophy injection rate.    Substitution of the above  in the equations for variable flux  yields
\bea
E_u(k) & = &  K'_{2D} \epsilon_\omega^{2/3} k^{-3} \left( \frac{k}{k_0} \right)^{- 2\alpha K'_{2D} \epsilon_\omega^{-1/3} }   \exp\left( - \frac{K'_{2D}}{k_{d2D}^2} (k^2-k_0^2) \right),
\label{eq:2d:Euk_k>k_f}  \\
\Pi_u(k) & =  & K'_{2D} \exp(x_0) \frac{\epsilon_\omega }{k_{d2D}^2}  \int_{ x}^\infty 
\frac{1}{x'} \left(1+ \frac{\beta}{x'} \right)  \left(\frac{x'}{x_0} \right)^{- \beta}   \exp{(-x')} dx' , 
\label{eq:2d:Piuk>kf}  
\eea
where $x=K'_{2D} (k/k_{d2D})^2$, $x_0=K'_{2D} (k_0/k_{d2D})^2$, and $\beta = K'_{2D} \alpha/\nu k_{d2D}^2$.  Note that $   E_u(k) $ is steeper than $ k^{-3} $ due to the Ekman friction and the viscous dissipation.  Asymptotically, 
\bea
\frac{\Pi_u(k)}{\epsilon_\omega }  \approx   \frac{K'_{2D} }{k_{d2D}^2}  E_1(K'_{2D} (k/k_{d2D})^2) \ll 1.
\label{eq:2d:Piuk>kf_limiting}
\eea
That is, $\Pi_u(k) \ll \epsilon_\omega$ for $k \gg k_f$, and $ \Pi_u(k) \sim \log k $ for $ \alpha = 0 $.   The aforementioned scaling relations are consistent with the analytical results of Gotoh~\cite{Gotoh:PRE1998}, numerical results of Gupta et al.~\cite{Gupta:PRE2019} and Anas and Verma~\cite{Anas:PRF2019}, as well as the experimental results of Boffetta et al.~\cite{Boffetta:EPL2005}. 

Thus, the above phenomenology describes how the viscous and Ekman dissipation steepens the 2D energy spectrum beyond the inertial range scaling.

 \subsection{Fluxes in quasi-2D turbulence}
 \label{subsec:Q2D}
 
Strong  rotation, gravity,  magnetic field, and shear tend to make the flows quasi-two-dimensional with dominant velocity field perpendicular to the direction of the external field.   Note that the above flows are stably stratified.   However, unstable stratification such as thermal convection tends to strengthen the parallel component of the velocity field.  In such flows, pressure facilitates energy exchange  between the parallel and perpendicular components of the velocity fields (see Sec.~\ref{subsec:anisotropic_turb}).  In this subsection, we sketch  some past work in this area   without getting into  complex details.

For fast rotation, researchers have reported strong two-dimensionalization of the flow~\cite{Godeferd:AMR2015,Alexakis:JFM2015,Sharma:PF2018}.  Such flows contain strong vortical structures. They exhibit  inverse energy cascade at small wavenumbers and forward energy cascade  at large wavenumbers~\cite{Sharma:PF2018,Sharma:PF2018_forced}.  Similar energy transfers and fluxes are observed in liquid-metal MHD turbulence under strong external magnetic field~\cite{Sommeria:JFM1982,Moreau:book:MHD,Knaepen:ARFM2008,Reddy:PP2014,Verma:ROPP2017}.  Under strong external magnetic field, MHD turbulence too exhibits similar anisotropic behaviour~\cite{Shebalin:JPP1983,Sommeria:JFM1982,Sundar:PP2017,Bandyopadhyay:JFM2019}.  Biferale et al.~\cite{Biferale:PF2017} showed how  flows transform from 2D to quasi-2D turbulence when the helical modes are truncated.    In contrast,  buoyancy destabilizes  convective flows and generates  plume structures; here, the velocity field parallel to buoyancy is stronger than the perpendicular component.

Boffetta et al.~\cite{Boffetta:EPL2011} and Bartello~\cite{Bartello:JAS1995} studied the energy fluxes of stably-stratified turbulence; they observed an inverse cascade of kinetic energy and a forward cascade of potential energy.  They termed these forward and backward fluxes as a  flux-loop.  Kumar et al.~\cite{Kumar:JoT2017} simulated 2D stably-stratified turbulence and observed variety of flow patterns.   Falkovich and Kristsuk~\cite{Falkovich:PRF2017} studied compressible turbulence and observed that planar structures and  wave turbulence exhibit inverse and  forward energy cascades respectively. 

These applications reveal importance of variable energy flux in the description of quasi-2D turbulence.  In all such fiows, pressure plays an important role in the energy transfer between the parallel and perpendicular components.   In addition, there are interesting works on  2D turbulence with buoyancy and MHD turbulence.  However, these topics have been skipped in this review.

\section{Variable energy flux in dissipationless turbulence}
\label{sec:Euler}

\subsection{Hydrodynamic systems}

An inviscid or dissipationless fluid flow  is described using \textit{Euler equation}, which is Navier-Stokes equation with $\nu=0$.   Depending on the choice of initial  condition, Euler turbulence exhibits either zero kinetic energy flux or a combination of positive and zero kinetic energy fluxes.   Passive scalars turbulence and MHD turbulence without dissipation  too exhibit similar properties.  We describe these systems in the present section.

In the following discussion, we will focus on the  properties of truncated Euler turbulence where   large-wavenumber Fourier modes   are  absent.  Kraichnan~\cite{Kraichnan:JFM1973} and Lee~\cite{Lee:QAM1952}  showed that  the Liouville  theorem is applicable to the phase space formed by  the Fourier modes of truncated Euler turbulence.   Further, they argued that the system is in  equilibrium and ergodic in the available phase space.    Based on these observations, which are similar to those of equilibrium thermodynamics, and using conservation of kinetic energy and kinetic helicity,  Kraichnan~\cite{Kraichnan:JFM1973} and Lee~\cite{Lee:QAM1952}  showed that $ E_u(\textbf{k}) $ and $ H_K(\textbf{k}) $ are random variables with the following probability distribution: 
\be
P(E_u(\textbf{k}), H_K(\textbf{k})) =  \frac{1}{Z}  \exp[-\beta E_u(\textbf{k}) - \gamma H_K(\textbf{k})],
\ee
where $ E_u(\textbf{k})$ and $ H_K(\textbf{k}) $ are the modal kinetic energy and kinetic helicity respectively, and $Z$ is a prefactor. 

It is convenient to write down  distribution functions for the helical energy spectra, $E_{u\pm}(\textbf{k})  $,  where $ u_\pm $ are the  helical variables defined in Sec.~\ref{subsec:HK}~\cite{Waleffe:PF1992,Verma:book:ET}.  These distribution functions are
\be
P(E_{u_+}(\textbf{k}), E_{u_-}(\textbf{k})) =   \beta_+ \beta_-  \exp[-\beta_+  E_{u_+}(\textbf{k}) - \beta_-  E_{u_-}(\textbf{k}) ]
\ee
that lead to $ \la E_{u_\pm}(\textbf{k}) \ra = 1/\beta_\pm$ (because $ u_\pm $ are independent variables).  Using a change of variable, $ \beta_\pm = \beta \mp \gamma k $, we obtain~\cite{Lee:QAM1952,Kraichnan:JFM1973},
\bea
\la E_{u}(\textbf{k}) \ra  = \la E_{u_+} \ra + \la E_{u_-} \ra = \frac{2 \beta}{\beta^2- \gamma^2 k^2 }, \\
\la H_{K}(\textbf{k}) \ra =k (\la E_{u_+} \ra - \la E_{u_-} \ra)   =\frac{2 \gamma k^2}{\beta^2 - \gamma^2 k^2 }.
\eea
One-dimensional shell spectrum, $ E_u(k) $, is a sum of the modes in a shell of radius $k$.  Therefore, $ E_u(k) $ is proportional to $ k^2 $ for small and moderate $ k $'s, but it gets an upward bend for large $k$'s~\cite{Lee:QAM1952,Kraichnan:JFM1973,Alexakis:PR2018}.  The above forms of energy and kinetic helicity arise due to their conservation.

\begin{figure}[hbtp]
	\centering
	\includegraphics[scale=0.4]{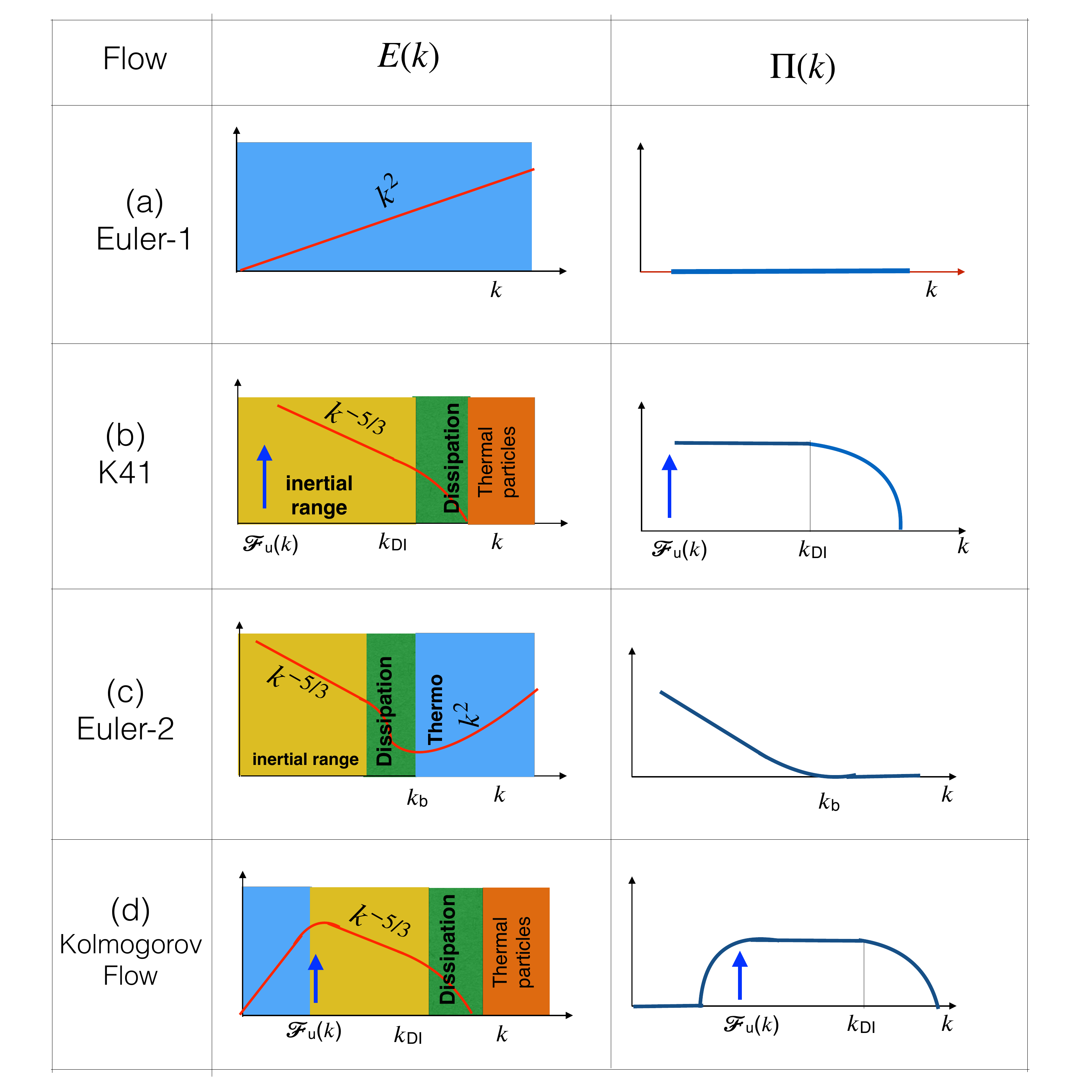}
	\caption{Schematic energy spectra and fluxes for various  flows: (a) Euler-1: Unforced Euler turbulence exhibiting $ k^2 $ for white noise as an initial condition. (b) K41: Turbulent flow of Kolmogorov model. Such flows with small viscosity are forced at large scales. (c) Euler-2: Unforced Euler turbulence with large-scale Taylor–Green vortices as an initial condition. (d) Kolmogorov-flow: A flow that is forced at an intermediate scale. The yellow, green, and blue colours represent inertial, dissipative, and equilibrium ranges respectively. The orange colour represents the thermal or microscopic scales where particles move randomly.  }
	\label{fig:Euler:EkPik}
\end{figure}

Detailed balance is an important property of systems under equilibrium.  Consequently, under equilibrium, we expect no energy exchange among the Fourier modes and, hence, the energy flux $ \Pi_u(k) = 0$ (see Fig.~\ref{fig:Euler:EkPik}(a)).  This observation is consistent with the fact that the phases of  the Fourier modes of a turbulent flow are random~\cite{Verma:PTRSA2020}\footnote{The mode-to-mode energy transfer from mode $ \textbf{u(p)} $ to $ \textbf{u(k)} $ with the mediation of $ \textbf{u(q)} $ in Craya-Herring basis is (see Sec.~\ref{subsec:HK})
\bea
\la S^{u_1 u_1}({\bf k|p|q}) \ra &= & k \sin \bar{\beta}  \cos\bar{\gamma}\la \Im[ u_1(\textbf{p}) u_1(\textbf{q}) u^*_1(\textbf{k})]\ra  \nonumber \\
&=& k\sin \bar{\beta}  \cos\bar{\gamma} \la |  u_1(\textbf{p}) u_1(\textbf{q}) u^*_1(\textbf{k}) |  \sin(\phi_{1p} + \phi_{1q} - \phi_{1k})\ra= 0, \nonumber \\
\la S^{u_2 u_2}({\bf k|p|q}) \ra  &= & -k \sin \bar{\beta}   \la \Im[ u_1(\textbf{p}) u_2(\textbf{q}) u^*_2(\textbf{k})]\ra  \nonumber \\
&=& -k\sin \bar{\beta}  \la |  u_1(\textbf{p}) u_2(\textbf{q}) u^*_2(\textbf{k}) |  \sin(\phi_{2p} + \phi_{1q} - \phi_{2k})\ra= 0, \nonumber 
\eea
where $ \phi_{1k} $ and $ \phi_{2k} $ are the phases of modes $ u_1(\textbf{k}) $ and $ u_2(\textbf{k}) $ respectively.  The energy transfers are zero due to random nature of the phases.}.   Contrast the above property with constant energy flux for Kolmogorov's theory of turbulence where the phases of the Fourier modes are correlated (see Fig.~\ref{fig:Euler:EkPik}(b)).  Hence, the phase space for the Kolmogorov's scenario is not expected to be ergodic.

A number of numerical simulations have been performed to verify the above spectrum. In one such work, Cichowlas et al.~\cite{Cichowlas:PRL2005} simulated Euler turbulence with large-scale Taylor-Green vortices as  initial condition and observed a combination of Kolmogorov’s $k^{-5/3} $ spectrum in the intermediate range and $ k^2 $ spectrum in the dissipative range, as shown in Fig.~\ref{fig:Euler:EkPik}(c).   The energy flux is nonzero for the $ k^{-5/3} $ regime due to the energy cascade from the  large-scale Taylor-Green vortices to the intermediate scales. Here, the $ k^{-5/3} $ regime is out of equilibrium, while the $ k^2 $ regime is equilibrium.  Over time, the $ k^{-5/3} $ regime shrinks at the expense of $ k^2 $ regime.   This is the process of \textit{thermalization}~\cite{Cichowlas:PRL2005,Krstulovic:PRL2011,DAlessio:AdvP2016}.  Note that the predictions of  Kraichnan~\cite{Kraichnan:JFM1973} and Lee~\cite{Lee:QAM1952} are applicable only to the equilibrium regime. 

 Interestingly, random uncorrelated signal, as in white noise, yields $ k^2 $ energy spectrum for the whole range of wavenumbers.  Motivated by this observation, Verma et al.~\cite{Verma:PTRSA2020, Verma:arxiv2020_equilibrium} simulated Euler turbulence using white noise as an initial condition  and observed an equilibrium configuration with  zero kinetic energy flux over the whole range of wavenumbers (also see \cite{Orszag:CP1973,Davis:PRL2001}).  The numerical results are in agreement with the predictions of Kraichnan~\cite{Kraichnan:JFM1973} and Lee~\cite{Lee:QAM1952}; in particular, nonhelical flows yield $k^{2}$ kinetic energy spectrum (see Fig.~\ref{fig:Euler:EkPik}(a)).    It appears that the random initial condition, which  has zero mode-to-mode energy transfer and zero kinetic energy flux, manoeuvres the system towards equilibrium.   Contrast this result with those of   Cichowlas et al.~\cite{Cichowlas:PRL2005} who obtained a combination of $ k^{-5/3} $ and $ k^2 $ spectra for Taylor-Green vortex as an initial condition.   Thus, an initial condition plays a key role in the evolution  of Euler turbulence; some initial condition takes the system to equilibrium state, while some other to a nonequilibrium state. 
  We also remark that the \textit{spectral entropy}, $ -\sum_{\textbf{k}} p(\textbf{k}) \log(p(\textbf{k}))$~\cite{Livi:PRA1985} where $ p(\textbf{k}) = E(\textbf{k})/E $, could be employed to study how the entropy of the system evolves during thermalization. 

Kolmogorov flow, which is forced at intermediate length scales ($k \approx k_f$), also exhibits equilibrium behaviour for $ k \le k_f $.  For such flows,     Prasath et al.~\cite{Prasath:EPL2014},  Dallas et al.~\cite{Dallas:PRL2015}, and  Alexakis and Biferale~\cite{Alexakis:PR2018} showed that $E_u(k) \sim k^2$ and $\Pi_u(k) \approx 0$ for $k < k_f$, and $E_u(k) \sim k^{-5/3}$ and $\Pi_u(k) \approx \mathrm{const} > 0$ for $k > k_f$.  These authors argue that the modes in $k<k_f$ regime are in absolute equilibrium.  See  Fig.~\ref{fig:Euler:EkPik}(d) for an illustration.  Note that in Kolmogorov flow, the large scales, rather than small scales, are in equilibrium.  This is in contrast to flows corresponding to K41 and Euler-2 of Fig.~\ref{fig:Euler:EkPik} where the modes at small scales are in equilibrium.  

Thus, Euler  turbulence provides valuable insights into the thermalization process, which is an important area of research in quantum and classical physics.  In Euler-2 and K41 flows, the energy at large scales cascades to intermediate and dissipation ranges.   The energy flux of the dissipative scale is transferred to the thermal energy of the particles, who move randomly.   In Euler-2 flow,  thermal energy of the particles appear as $ k^2 $ spectrum, but thermal energy is not represented in K41 flow.  We present these regimes in Fig.~\ref{fig:Euler:EkPik} using different colors.  The yellow, green, and blue colours represent the inertial, dissipative, and equilibrium regimes respectively, while the orange colour represents the thermodynamic regime  or  microscopic scales, where the particles move randomly.   Interestingly, $ E_u(k) $ of Cichowlas et al. \cite{Cichowlas:PRL2005} exhibits a small exponential transition regime (similar to that of hydrodynamic turbulence) between the $ k^{-5/3} $ and $ k^2 $ regimes.

Based on these examples, it has been argued that in 3D hydrodynamics, thermalization occurs via a multiscale energy transfer from large and intermediate scales (noequilibrium) to small scales (equilibrium).   Here, the coherent fluid energy of inertial-dissipative regime is converted to the thermal energy of  constituents molecules.     This mechanism  provides a scenario for  the emergence of friction or dissipation in generic systems, including Hamiltonian and quantum systems ~\cite{Verma:EPJB2019}.     The merging of rivers into the ocean, as shown in Fig.~\ref{fig:bottleneck}(a), appears to provide an interesting analogy to the aforementioned thermalization process.  The river flow  represents an out-of-equilibrium system, while ocean represents a system in equilibrium.

The above  formalism for Euler turbulence has been extended to secondary fields advected by the velocity field.  For diffusionless passive scalar turbulence, conservation of $ \sum_\textbf{k} E_\zeta(\textbf{k}) $ leads to the following distribution for $ E_\zeta(\textbf{k}) $:
\be
P(E_{\zeta}(\textbf{k})) =   \beta_\zeta   \exp[-\beta_\zeta   E_{\zeta}(\textbf{k}) ]
\ee
that leads to $ \la E_{\zeta}(\textbf{k}) \ra = 1/\beta_\zeta$~\cite{Lesieur:book:Turbulence,Frisch:book}.  The formulas for MHD turbulence is more complex due the conservation of total energy, cross helicity, and magnetic helicity. The reader is referred to the original papers by Frisch et al.~\cite{Frisch:JFM1975} and Stribling et al.~\cite{Stribling:PP1991}.  For these systems too, the equilibrium spectrum and fluxes are easier to obtain using white noise as an initial condition.  Burgers turbulence too exhibits similar properties  during thermalization~\cite{Ray:PRE2011}.

The energy flux vanishes in the equilibrium regime of hydrodynamic  turbulence.   It is important to contrast this effect from the suppression of the net energy flux   by two opposite fluxes, as observed in quasi-2D stably stratified turbulence~\cite{Boffetta:EPL2011} and in compressible turbulence~\cite{Falkovich:PRF2017}.  In the latter systems,  the detailed balance of energy transfers is broken by  opposing energy fluxes, called  \textit{flux loops}.

In addition to Euler turbulence and other dissipation-less turbulent systems,  many energy-conserving nonequilibrium systems exhibit strong fluctuations or turbulence, which will be described  below.

\subsection{Thermalization in miscellaneous dissipation-less systems}

In this section, we describe some  generic properties of thermalization in Hamiltonian systems and  dissipation-less partial differential equations.  But, first we discuss the dynamics of dissipation-less Burgers equation. 

For a wave  (e.g., $ \sin(x) $)  as an  initial condition,  one-dimensional dissipative Burgers equation  yields shocks~\cite{Burgers:AAM1948}.  Such flows exhibit $ k^{-2} $ spectrum followed by a dissipative spectrum~\cite{Verma:PA2000}.  In contrast,  Frisch et al.~\cite{Frisch:PRL2008} and Ray et al.~\cite{Ray:PRE2011} studied thermalization of Burgers equation by employing a minimal dissipation with small viscosity or hyperviscosity.   In these works, they report a mixed spectrum: $ k^{-2} $ for intermediate wavenumbers (nonequilibrium regime) and $ k^0 $ for large wavenumbers (equilibrium regime),   similar to the findings of Cichowlas  et al.~\cite{Cichowlas:PRL2005} for Euler turbulence.  Note, however, that a random initial condition yields an equilibrium state with  $ E(k) \sim k^0 $ and zero energy flux throughout the wavenumber range.  

For more than half a century, researchers have been studying thermalization in  Fermi-Pasta-Ulam-Tsingou model~\cite{FermiPasta:1955}.  In this model, the energy cascades to higher modes when a large-scale excitation is chosen as an initial condition.  The system eventually reaches a state where the energy is equipartitioned among all the Fourier modes.  However, after some time, the system returns to the initial configuration, consistent with Poincar\'{e} recurrence theorem~\cite {Arnold:book:MathCelestial,Christodoulidi:MiE2019, Benettin:JSP2008}.   On the contrary, for random initial condition, the system reaches an equilibrium state with energy equipartitioned among all Fourier modes~\cite {Christodoulidi:MiE2019, Benettin:JSP2008}.  This observation again demonstrates how random initial condition aids in taking a dynamical system towards thermalization.   Similar behaviour has been observed for KdV equation, which  is related to the Fermi-Pasta-Ulam-Tsingou model~\cite {Benettin:JSP2008}.

Thermalization processes in quantum systems, such as Bose-Einstein condensate and superfluids, are more complex~\cite{Davis:PRL2001,DAlessio:AdvP2016}.  Still, these systems have certain similarities with  Euler turbulence and other related systems.  Thus, Euler turbulence provides valuable insights into the thermalization process.   More work is being carried to the address the following questions among  other:  Do Euler turbulence and related turbulent systems become ergodic under thermalization? What are the consequences of   Poincar\'{e} recurrence theorem on thermalization?  In Euler turbulence with Taylor-Green vortex as an initial condition~\cite{Cichowlas:PRL2005}, would the system return to the initial configuration after thermalization.  Variable energy flux formalism may be a useful tool for answering some of these interesting questions.

\section{Variable energy fluxes in quantum turbulence  and binary-mixtures turbulence}
\label{sec:QT}

Bose-Einstein condensates and  superfluids, which are quantum systems, exhibit turbulent behaviour for some set of parameters~\cite{Madeira:ARCMP2020,Barenghi:arxiv2016,Barenghi:PNAS2014,Sasa:PRB2011}.  Similarly, turbulence is observed in binary-mixtures.  As we show below, variable energy flux formulation  sheds important light into these complex systems.  

\subsection{Quantum turbulence}

To describe quantum systems, we often employ macroscopic wavefunction $ \zeta(x,t) = \sqrt{n} \exp(iS)$, where $ n $ is the  number density of atoms  and $ S $ is the phase.  The time evolution of the wavefunction $ \zeta(x,t)$ is described using Gross-Pitaevskii (GP) equation ~\cite{Tsatsos:PR2016,Barenghi:arxiv2016,Madeira:ARCMP2020,Sasa:PRB2011}: 
\be
i \hbar \frac{\partial \zeta}{\partial t} = \left[ -\frac{\hbar^2}{2 m} \nabla^2 + V(\textbf{r}) + g |\zeta|^2 \right] \zeta,
\label{eq:GP}
\ee
where $ m $ is the mass of the quantum particle, $ V(\textbf{r}) $ is the external potential, and $ g $ is the proportionality constant for the interaction term.  The GP equation exhibits many interesting features, including  entangled quantum vortices, Kelvin waves, etc.  However, in this subsection we focus on the energy fluxes of quantum turbulence.

The velocity of a superfluid flow is given by the gradient of its wavefunction, that is, $ \textbf{u}_s =  (\hbar/m) \nabla S$.  In terms of $ n $ and $ \textbf{u}_s $, the real and imaginary parts of the GP equation are~\cite{Tsatsos:PR2016}
\bea
\frac{\partial n}{\partial t} + \nabla (n \textbf{u}_s) = 0,\\
 \frac{\partial {\bf u}_s}{\partial t} + {\bf (u_s \cdot \nabla) u_s}   = -\frac{1}{mn} \nabla p - \frac{1}{m} \nabla \left(  \frac{\hbar^2}{2 m \sqrt{n}} \nabla^2 \sqrt{n} \right) -\frac{1}{m} \nabla V.
\eea
Note that $ m n=\rho_s $ is the density of the superfluid.  In Helium-4, the superfluid component coexists with the normal fluid, whose velocity field is denoted by $ \textbf{u}_n $.  The equations for  the superfluid and normal-fluid components are~\cite{Lvov:JLTP2006,Roche:EPL2009} 
\bea
   \frac{\partial {\bf u}_s}{\partial t} + ({\bf u}_s \cdot \nabla) {\bf u}_s    &= & - \frac{1}{\rho_s} \nabla p_s -  \frac{\rho_s}{\rho}  {\bf F}_{ns}  , \label{eq:misc:super}  \\
  \frac{\partial {\bf u}_n}{\partial t} + ({\bf u}_n \cdot \nabla) {\bf u}_n  &= &
-\frac{1}{\rho_n}  \nabla p_n + \frac{\rho_n}{\rho}  {\bf F}_{ns}   + \textbf{F}_\mathrm{LS} +  \nu \nabla^2 {\bf u}_n, \label{eq:misc:normal} 
\eea
 where $ \rho_n$ is  the density of the normal fluid, $ \rho = \rho_s+\rho_n$ is the total  density of the fluid, $\textbf{F}_\mathrm{LS}  $ is the large-scale force applied to the normal fluid, and $ {\bf F}_{ns} = (B/2) |\bm{\omega}| ({\bf u}_s - {\bf u}_n) $ is the mutual friction.  Here $\bm{\omega}  $ is the superfluid vorticity, and $ B $ is a constant.  Note that $ p_n = (\rho_n/\rho) p + \rho_s S T $ and $ p_s = (\rho_s/\rho) p - \rho_s S T $ are partial pressures with $ S, T, p $ as the specific entropy, temperature, and pressure respectively.  Also, the relative density $ \rho_n/\rho_s $ increases with temperature, and  Helium-4 becomes a normal fluid at the critical temperature of 2.17 K.   In the following discussion, for simplification,  the fluid densities are assumed to be constant.
 
 Researchers have studied superfluid turbulence using experiments and numerical simulations~(\cite{Tsatsos:PR2016,Barenghi:arxiv2016,Madeira:ARCMP2020,Sasa:PRB2011,Barenghi:PNAS2014} and references therein).  These works show that both superfluid and normal fluid exhibit Kolmogorov-like $ k^{-5/3} $ spectra and nearly constant energy fluxes~\cite{Madeira:ARCMP2020,Barenghi:arxiv2016,Barenghi:PNAS2014,Sasa:PRB2011}.  At very low temperatures, $ \rho_n \approx 0 $, which is observed in Helium-3; for such flows, vortex reconnections  and phonon coupling at small scales provide the necessary dissipation to sustain the $ k^{-5/3} $ spectrum.  Also refer to \cite{Arnol:EPL2020} for a recent work on energy flux in trapped Bose-Einstein condensate.   
 
In this article, we focus on  variable energy flux in quantum turbulence.  In  the turbulent regime, the nonlinear terms of the above equations, $ ({\bf u}_s \cdot \nabla) {\bf u}_s  $ and $ ({\bf u}_n \cdot \nabla) {\bf u}_n  $, induce the respective energy cascades $\Pi_{u,s}(k)$ and $\Pi_{u,n}(k)$ for the two components.  These energy fluxes are affected by the  mutual friction.  The energy injection rates by $ {\bf F}_{ns}$ to the normal and superfluid components are (see Sec.~\ref{sec:VF})
\bea
\mathcal{F}_{u,s}({\bf k})  = - \frac{\rho_s}{\rho} \Re [  {\bf F}_{ns}({\bf k})  \cdot  {\bf u}^*_s({\bf k}) ];~~~
\mathcal{F}_{u,n}({\bf k})  =  \frac{\rho_n}{\rho}  \Re [ {\bf F}_{ns}({\bf k})  \cdot  {\bf u}^*_n({\bf k}) ] .
\eea
Therefore,  the energy fluxes for the two fluids vary with $ k $ as
\bea
\frac{d}{dk} \Pi_{u,s}(k) = \mathcal{F}_{u,s}(k);~~~\frac{d}{dk} \Pi_{u,n}(k)  = \mathcal{F}_{u,n}(k) - 2 \nu k^2 E_n(k),
\eea  
where $ 2 \nu k^2 E_n(k)$ is the dissipation rate for the normal fluid.   Roche et al.~\cite{Roche:EPL2009} and Wacks and Barenghi \cite{Wacks:PRB2011} analyzed the above energy transfers using numerical simulations and observed that $ \mathcal{F}_{u,s} <0 $ and $ \mathcal{F}_{u,n} >0 $.  Note however that these quantities depend on the temperature or $\rho_s/\rho_n  $.  Based on these observations, we expected that $  \Pi_{u,n} $  and $ \Pi_{u,s} $ vary with $k$.  These variations can induce additional $k$ dependence in the energy spectra over  Kolmogorov's   $ k^{-5/3} $ spectrum.  These predictions need to be verified using experiments and numerical simulations.

 Quantum systems are energy conserving.  However, sustenance  of $ k^{-5/3} $ spectrum requires dissipation at small scales. It has been argued that the compressible waves produced during vortex reconnections may provide the required dissipation~\cite{Madeira:ARCMP2020,Barenghi:PNAS2014,Sasa:PRB2011,Barenghi:arxiv2016,Verma:EPJB2019}.  These issues, as well as the energy spectrum and fluxes of GP equations, have been studied using numerical simulations and experiments~\cite{Davis:PRL2001,Krstulovic:PRE2011_GP,Madeira:ARCMP2020,Barenghi:PNAS2014,Sasa:PRB2011,Barenghi:arxiv2016,Shukla:PRE2019}.

The above  discussion illustrates the usefulness of variable energy flux in  quantum turbulence.

\subsection{Variable energy fluxes in binary-mixture turbulence }
\label{subsec:binary_fluids}

In this section, we briefly describe the  energy fluxes of binary fluid mixtures~ \cite{Hohenberg:RMP1977, NovickCohen:PD1984,Ruiz:PRA1981binary, Perlekar:JFM2019}.  We consider a binary mixture with two components whose  relative densities are $\zeta({\bf r})$ and $1-\zeta({\bf r})$.    Researchers describe the dynamics of these fields using \textit{time-dependent Ginzburg-Landau equation} and\textit{ Cahn-Hilliard  equation}~\cite{Cahn:JCP2004,Chaikin:book,Puri:book_edited}.  Here we illustrate the idea of variable energy flux for  the above equations.

The time-dependent Ginzburg-Landau (TDGL) equation, also called \textit{model A}, is~\cite{Puri:book_edited}:
\be
\frac{\partial \zeta}{\partial t} = \zeta - \zeta^3 + \nabla^2 \zeta.
\ee
The corresponding equation for the spectral energy $ E_\zeta(\textbf{k}) = \frac{1}{2} |\zeta(\textbf{k})|^2 $ is
\be
\frac{d}{dt} E_\zeta({\bf k}) =  -\sum_{{\bf k}_1, {\bf k}_2} \Re[\zeta({\bf k}_1) \zeta({\bf k}_2) \zeta({\bf k}_3) \zeta^*({\bf k})]+E_\zeta({\bf k}) -k^2 E_\zeta({\bf k}),
\label{eq:misc:TDGL_E}
\ee
where $ {\bf k}_3 = {\bf k} - {\bf k}_1 -{\bf k}_2 $. In Eq.~(\ref{eq:misc:TDGL_E}), the second term in RHS enhances  $ E_\zeta({\bf k}) $, while the last term dissipates $ E_\zeta({\bf k}) $.  The nonlinear term, the first term in the   RHS of above equation, induces the following energy flux  for a wavenumber sphere of radius $ k_0 $:
\be
\Pi_\zeta(k_0) = - \sum_{k \le k_0} \sum_{{\bf k}_1, {\bf k}_2} \Re[\zeta({\bf k}_1) \zeta({\bf k}_2) \zeta({\bf k}_3) \zeta^*({\bf k})].
\ee  

Numerical simulation and analytical studies of TDGL equation reveal that asymptotically ($ t \rightarrow \infty $), the system exhibits a domain with either $ \zeta = 1 $ or $ -1 $~(see \cite{Puri:book_edited} and references therein). The energy flux provides interesting inputs for understanding the above result.  The arguments for 1D TDGL are as follows. The scalar energy $ E_\zeta $ is dissipated strongly at small scales due to the $ k^2 $ factor in the dissipation rate.  Consequently,  a forward  cascade of $ E_\zeta $ is set up that transfers the energy of large and intermediate scales to small scales, where it is dissipated.    Note however that the mean energy, $ E_\zeta({\bf k}=0) $,  is not dissipated (due to the structure of the dissipation term).  Consequently, only $ k=0 $ mode survives, while the rest of the modes vanish due to the forward cascade and dissipation.  Therefore, the final state is  either $ \zeta = 1 $ or $ -1 $ with zero energy flux.  The dynamics for 2D and 3D is expected to be similar.  Thus,  the energy flux provides useful insights into the dynamics of TDGL equation.

 Similar analysis is applicable to  \textit{model B} or Cahn-Hilliard (CH) equation~\cite{Puri:book_edited}, which is
\be
\frac{\partial \zeta}{\partial t} = -\nabla^2 (\zeta - \zeta^3 + \nabla^2 \zeta).
\ee
The evolution equation for the modal energy of CH equation is
\be
\frac{d}{dt} E_\zeta({\bf k}) = -k^2 \sum_{{\bf k}_1, {\bf k}_2} \Re[\zeta({\bf k}_1) \zeta({\bf k}_2) \zeta({\bf k}_3) \zeta^*({\bf k})]+k^2 E_\zeta({\bf k})  - k^4 E_\zeta({\bf k}),
\label{eq:misc:CH_E}
\ee
while the corresponding scalar energy flux is
\be
\Pi_\zeta(k_0) = - k^2 \sum_{k<k_0} \sum_{{\bf k}_1, {\bf k}_2} \Re[\zeta({\bf k}_1) \zeta({\bf k}_2) \zeta({\bf k}_3) \zeta^*({\bf k})].
\ee 
The energetics of CH equation is very similar to that of TDGL: $ k^2 E_\zeta({\bf k}) $ term feeds energy into the system, while $k^4 E_\zeta({\bf k})$ term
 dissipates the energy.  The first term in the RHS of Eq.~(\ref{eq:misc:CH_E}) creates forward energy cascade.  The cascaded energy gets dissipated at small scales.  Asymptotically, only $ k = 1$ mode survives because the steady state solution has zero energy flux (or zero nonlinearity) and $ k^2 - k^4 = 0 $.

Inclusion of hydrodynamic effects into  Cahn-Hilliard equation yields the following equations for the velocity field and $\zeta$~\cite{Cahn:JCP2004}:
\bea
\frac{\partial {\bf u}}{\partial t} + ({\bf u} \cdot \nabla) {\bf u}  =
- \frac{1}{\rho} \nabla p +  a \zeta \nabla \nabla^2 \zeta + {\bf F}_\mathrm{LS}+  \nu \nabla^2 {\bf u}, \label{eq:misc:bin_u} \\
\frac{\partial \zeta}{\partial t} + ({\bf u} \cdot \nabla) \zeta  = -\nabla^2 (\zeta - \zeta^3 + \nabla^2 \zeta), \label{eq:misc:bin_zeta} \\
 \nabla \cdot \mathbf{u} = 0, \label{eq:misc:bin_div_u_eq0}
\eea
with corresponding forces as $ {\bf F}_u  =  a \zeta \nabla \nabla^2 \zeta $  and $ F_\zeta  =  \nabla^2 \zeta^3 $.   The energy spectra and fluxes of the above system have been studied in detail by many researchers  (see  \cite{Ruiz:PRA1981binary, Perlekar:JFM2019} and references therein).   

The nonlinear terms $({\bf u} \cdot \nabla) {\bf u} $ and $({\bf u} \cdot \nabla) \zeta$ generate the respective fluxes $\Pi_u(k)$ and $\Pi_\zeta(k)$ for the  kinetic  and scalar energies.  The force ${\bf F}_u $  induces variations in $ \Pi_u(k) $ due to the energy injection rate:
\bea
\mathcal{F}_u({\bf k}) & = & \Re[{\bf F}_u({\bf k}) \cdot {\bf u^*(k)}] 
= \sum_{\bf p} \Im [  p^2  \zeta({\bf k-p})  \zeta({\bf p})  \{ {\bf p \cdot  u^*(k)} \} ].
\eea 
Refer to Eq.~(\ref{eq:misc:CH_E}) for the expression of $ \mathcal{F}_\zeta({\bf k})  $.  Interestingly, hydrodynamic version of CH equation too exhibits coarsening~\cite{Ruiz:PRA1981binary, Perlekar:JFM2019}.  Note that the scalar field appears to exhibit forward cascade.  Thus, the coarsening process in binary fluid is not due to any inverse cascade (as in 2D turbulence), but it is due to the energy injection term of the equation.

In summary, the scalar energy  flux provides valuable inputs to the dynamics of  coarsening.    However,  more work is required for definitive conclusions.


\section{Third-order structure function and energy flux}
\label{sec:struct_fn}
So far, we have discussed energy transfers and flux in Fourier space. Note, however, that there are many interesting connections between energy flux and field correlations in real space.  In one of the path-breaking works, Kolmogorov~\cite{Kolmogorov:DANS1941Structure,Kolmogorov:DANS1941Dissipation} showed a relationship between the energy flux and the third-order structure functions.   Later, Kolmogorov's theory was generalized to 2D hydrodynamic turbulence,   scalar turbulence,  MHD turbulence, anisotropic turbulence, and other related systems.  We will briefly discuss these results in this section.
%

Kolmogorov considered statistically homogeneous, isotropic, and steady 3D turbulence, with forcing employed at  large  scales.  In such a flow, he considered  two real-space points $ \textbf{r} $ and $ \textbf{r+l} $ where the velocities are $ \textbf{u} $ and $ \textbf{u}' $ respectively.   Under the limit of infinite Reynolds number, for the inertial range, Kolmogorov~\cite{Kolmogorov:DANS1941Structure,Kolmogorov:DANS1941Dissipation} showed that  the third-order longitudinal structure function  $ S_3(l) $ is  
\bea
 \la [{\bf (u'-u)} \cdot \hat{\textbf{l}}]^3 \ra =  -\frac{4}{5} \epsilon_u l,
 \label{eq:K41:S3_bar_4by3_rule}
 \eea
where $\hat{\textbf{l}}$ is the unit vector along  vector ${\bf l}$, and $\epsilon_u$ is the viscous dissipation rate that equals the kinetic energy flux in the inertial range.  A useful variant of the above relation is
\bea
 \la |  {\bf u'-u }|^2 [{\bf (u'-u)} \cdot \hat{\textbf{l}}] \ra  = -\frac{4}{3} \epsilon_u l.
\label{eq:K41:S3_l}
\eea
The above relations are borne out in many experiments and numerical simulations, which are too numerous to be listed here.  We refer the reader to Frisch~\cite{Frisch:book}, Lesieur~\cite{Lesieur:book:Turbulence}, and references therein for details.  

For a passive secondary scalar field $\zeta$ advected in turbulent flow and forced at large scales, in the inertial range~\cite{Yaglom:DANS1949,Monin:book:v1, Monin:book:v2},
\bea
 \la   { (\zeta'-\zeta) }^2 [{\bf (u'-u)} \cdot \hat{\textbf{l}}] \ra  = -\frac{4}{3} \epsilon_\zeta l,
 \label{eq:struct_fn:secondary}
\eea
where $ \epsilon_\zeta  $ is the dissipation rate or the energy flux of the scalar field.  The same relation holds for passive vector and tensorial fields that are forced at large scales.  However, the relationship of Eq.~(\ref{eq:struct_fn:secondary}) does not hold for the velocity and magnetic fields of MHD turbulence because  MHD equations have additional terms, $ ({\bf u \cdot \nabla} )\bm{\zeta} $ and $ (\bm{\zeta}  \cdot \nabla ) {\bf u}$, compared to passive vector (see Eqs.~(\ref{eq:mhd:MHDu_fluct}, \ref{eq:mhd:MHDb_fluct})).   However, the derivation of passive scalar can be generalized to Els\"{a}sser variables $ {\bf z}^{\pm} $  due to the similarities in the forms of the equations (see Eq.~(\ref{eq:MHD_formalism:MHDzpm})).  Following this approach, Politano and Pouquet~\cite{Politano:PRE1998}   derived that
\bea
S^{z^+}_3(l) & = &  \la |{\bf z^{+'} -z^+}|^2  [({\bf z^{-'} -z^-})  \cdot \hat{ \bf{l}}]  \ra = -  \frac{4}{3} \epsilon_{z^+} l, 
\label{eq:struct_Sz+} \\
S^{z^-}_3(l) &= &  \la |{\bf z^{-'} -z^-}|^2  [ ({\bf z^{+'} -z^+})  \cdot \hat{ \bf{l}} ] \ra = - \frac{4}{3} \epsilon_{z^-} l,  
\label{eq:struct_Sz-}
\eea
where $\epsilon_{z^\pm}$ are the dissipation rates or the energy fluxes of ${\bf z^\pm}$ fields.    Politano and Pouquet~\cite{Politano:PRE1998}  also derived relations for $ {\bf u} $ and $ \bm{\zeta} $,  but these relations  are more complex than Eqs.~({\ref{eq:struct_Sz+}, \ref{eq:struct_Sz-}}).

The structure functions of 2D turbulence are more complex than above functions due to dual scaling: the kinetic energy cascades backward in the wavenumber band  $ k<k_f $, while the enstrophy cascades forward in the band  $ k>k_f $. Here, $ k_f $ is the forcing wavenumber band (see Sec.~\ref{sec:2D}).  For homogeneous and isotropic 2D turbulence, in the inertial range of $k<k_f$ regime~\cite{Lindborg:JFM1999, Bernard:PRE1999, Lindborg:JFM1999},
\bea
\la |  {\bf u'-u }|^2 [{\bf (u'-u)} \cdot \hat{\textbf{l}}] \ra  = 2 \epsilon_u l;~~~ \la [{\bf (u'-u)} \cdot \hat{\textbf{l}}]^3 \ra =  \frac{3}{2} \epsilon_u l.
\eea
However, in the inertial range of  $ k>k_f $ regime~\cite{Lindborg:JFM1999},
\bea
\la  |\omega-\omega'|^2 [({\bf u'-u})\cdot \hat{ l}] \ra = -2  \epsilon_\omega l;~\la  |{\bf u'-u}|^2 [({\bf u'-u})\cdot \hat{ l}] \ra = -\frac{1}{8} \epsilon_\omega l,
\eea
where $ \epsilon_\omega $ is the enstrophy dissipation rate that equals the enstrophy flux. These relations have been verified in several numerical simulations (see \cite{Boffetta:PRE2000,Boffetta:ARFM2012,Tabeling:PR2002} and references therein).

Isotropy is a major assumption  in the  derivation of Eq.~(\ref{eq:K41:S3_bar_4by3_rule}).   Researchers have attempted to generalise Kolmogorov's relation  to anisotropic turbulence.  For example,  Biferale and Procaccia~\cite{Biferale:PR2005} employed symmetry groups to decipher anisotropic structure functions.  Arad et al.~\cite{Arad:PRL1998} computed the anistropic structure functions in atmospheric surface layer.     Danaila et al.~\cite{Danaila:PD2012b} constructed anisotropic structure functions for axisymmetric anisotropic turbulence by invoking scale-by-scale energy budget.   Ching~\cite{Ching:book} and  Bhattacharya et al.~\cite{Bhattacharya:PF2019:SF}  generalized Eq.~(\ref{eq:K41:S3_bar_4by3_rule}) for turbulent thermal convection.  As described in Sec.~\ref{subsec:anisotropic_turb}, pressure plays an important role in anisotropic  spectral energy transfer.  It will be interesting to relate the Fourier-space anisotropic energy-transfer formulas to the anisotropic structure functions.  We also remark that  the triple product in the left-hand-side of Eq.~(\ref{eq:K41:S3_l})  is related to the  mode-to-mode energy transfer formula, $ S(\textbf{k}| \textbf{p}| \textbf{q}) $, of Eq.~(\ref{eq:ET:Suu_kpq}).   Refer to Verma~\cite{Verma:book:ET} for details.

The relations discussed so far in this section are related to the third-order structure function.  There have been valiant efforts  to derive analytical relations for higher-order structure functions, but these efforts have not yielded  the final results.  These works are related to the \textit{intermittency effects} in turbulence, a topic which is beyond the scope of this review.   We refer the reader to  Frisch~\cite{Frisch:book},  Dubrulle~\cite{Dubrulle:JFM2019},   Stolovitzky and Sreenivasan~\cite{Stolovitzky:RMP1994}, and references therein for details.  Some of the notable field-theoretic works regarding intermittency are  \cite{Belinicher:JSP1998,Lvov:PRE1995I,Das:EPL1994} and references therein.  It has been argued that the higher-order structure functions are related to the fluctuations in the energy flux.

\section{ Summary and conclusions}
\label{sec:conclusions}

The energy flux is an important quantity in turbulence.  In hydrodynamic turbulence with large-scale forcing, the inertial-range energy flux is constant.   However, in the presence of inertial-range energy injection ($ \mathcal{F}_u(k)$) and dissipation ($D_u(k)$), the kinetic energy flux becomes scale-dependent and  is described by $d\Pi_u(k)/dk = \mathcal{F}_u(k) -D_u(k) $.   In this review, we show how the variable energy flux formalism provides valuable insights into the dynamics of  many turbulent systems, especially in determining their energy spectra and fluxes.     A summary of the results presented in the review is as follows.
\begin{itemize}
\item {\bf  Buoyancy driven turbulence}:   In stably stratified turbulence with moderate stratification, the kinetic energy  is transferred to the potential energy, hence $ \mathcal{F}_u(k) < 0$. Therefore,  $\Pi_u(k)$ decreases with $k$ in the inertial range itself; in particular,   $\Pi_u(k) \sim k^{-4/5}$ and  $E_u(k) \sim k^{-11/5}$, which is steeper than $k^{-5/3}$ spectrum~ \cite{Bolgiano:JGR1959,Obukhov:DANS1959}.  However, in  turbulent thermal convection, $ \mathcal{F}_u(k) > 0$  for small and moderate thermal Prandtl numbers (Pr).  Hence, $E_u(k)$ for thermal convection is expected to be shallower than $k^{-5/3}$.  Yet, turbulent thermal convection with $ \mathrm{Pr} \lessapprox 1 $ has behaviour similar to hydrodynamic turbulence (nearly constant energy flux and $k^{-5/3}$ kinetic energy spectrum  in the inertial range).  This is due to the fact that for $ \mathrm{Pr} \lessapprox 1 $, $ \mathcal{F}_u(k)$ decreases sharply with $k$ and  is quite weak  in the inertial range, similar to that in Kolmogorov's model for hydrodynamic turbulence.

\item {\bf MHD and Polymeric turbulence}:  In MHD turbulence, the nonlinear interactions between the velocity and magnetic fields cause energy transfers from the velocity field to the magnetic field.  This conversion mechanism is responsible for the growth of  the magnetic field in astrophysical objects, and for making  $\Pi_u(k) $ a decreasing function of $k$.  Similar energy transfers are observed in turbulent flows with polymers.  The  suppression of $\Pi_u(k)$ or the nonlinear term ${\bf (u \cdot \nabla) u}$ is one of the primary causes of  drag reduction in such flows.  Using variable energy flux formalism we  also derive identities relating various fluxes.

\item {\bf Dissipation}:  In turbulence,  viscous dissipation suppresses the energy flux in the dissipation range.  Using several assumptions, Pao~\cite{Pao:PF1965, Pao:PF1968} showed that  in the inertial-dissipation range of hydrodynamic turbulence, the energy flux and normalized energy spectrum vary as $ \exp(-(k/k_d)^{4/3}) $, where $ k_d $ is Kolmogorov wavenumber.   Ekman friction that acts at all scales steepens the inertial-range energy spectrum further than the $ k^{-5/3} $ power-law.  In quasi-static MHD turbulence, the Joule dissipation provides similar steepening of the kinetic energy  spectrum.  Pao's model has been generalized to 2D hydrodynamic turbulence with Ekman friction.

\item {\bf Energy flux of a secondary flow $\zeta$}: The advection term  of a secondary field $ \zeta $, ${\bf (u \cdot \nabla)} \zeta$, has an associated secondary energy flux $\Pi_\zeta$.  This flux too exhibits variability: $d\Pi_\zeta(k)/dk = \mathcal{F}_\zeta(k) -D_\zeta(k) $, where $ \mathcal{F}_\zeta(k)$ is the secondary energy injection rate, and $D_\zeta(k) $ is the diffusion rate of the secondary field.   Such flux variations are present in MHD turbulence, stably stratified turbulence, binary-mixture turbulence, and related complex flows.
\end{itemize}

 \begin{table}
  \begin{center}
    \caption{Table illustrating 3D  turbulent systems along with their forces  (${\bf F}_u({\bf k})$), the kinetic energy  injection rates by the forces ($\mathcal{F}_u({\bf k})$), and the nature of kinetic energy fluxes in the inertial range.  For the meaning of the symbols, refer to the discussion in the paper.  IC stands for initial condition.}
    \label{tab:conclusions:table}
    \begin{tabular}{lcccc} 
    \hline\noalign{\smallskip}
    System & ${\bf F}_u({\bf k})$ &  $\mathcal{F}_u({\bf k})$ & Nature of  $\Pi_u(k)$ \\  
     \noalign{\smallskip}\hline\noalign{\smallskip}
     Kolmogorov's K41 law &  0  &  0 & Const \\ 
     QS MHD turbulence & $- B_0^2 \cos^2\theta {\bf u(k)}$ & $- 2 B_0^2 \cos^2\theta E_u(k)<0$ 
     & Decreases \\
          Ekman friction (3D) & $-\alpha {\bf u}({\bf k})$ & $- 2 \alpha E_u({\bf k})< 0$ & Decreases \\
      Stably stratified turbulence & $  - N \zeta({\bf k})$ & $- N \Re[\zeta({\bf k}) u_z^*({\bf k})]<0$  
     & Decreases \\
     Thermal convection& $  \alpha g  \zeta({\bf k})$ & $ \alpha g  \Re[\zeta({\bf k}) u_z^*({\bf k})] > 0$  
     & Marginally increases \\
     Unstably  stratified turbulence &    $g \zeta({\bf k})$ & $g  \Re[\zeta({\bf k}) u_z^*({\bf k})] >0$  
     & Increases \\
    MHD (Dynamo) &  ${\bf [J \times B}]({\bf k})$  &  $\Re\{ [{\bf J \times B}]({\bf k}) \cdot {\bf u^*(k)}\} < 0$
    & Decreases  \\
     Dilute polymer &  $\frac{\mu}{\tau_p} \partial_j (f   \zeta_{ij})$  &   Complex convolution $<0 $
    & Decreases  \\
    Shear turbulence & Shear force & Positive  & Increases \\
        Euler turbulence &  0 &  0 & zero or mixed  \\
        &&& depending on IC \\
       \noalign{\smallskip}\hline
    \end{tabular}
  \end{center}
\end{table}

Variable energy flux is also useful for modelling shear turbulence, stably and unstably stratified turbulence, Euler turbulence, quantum turbulence, etc.    Variable energy flux formalism provides valuable insights for understanding quantum turbulence and the coarsening processes in  time-dependent Ginzburg-Landau and Cahn-Hilliard equations.  Interestingly, the money supply in a free market economy too exhibits a cascade  across various income groups; this cascade  has similarities with the energy flux in turbulence~\cite{Verma:book_chapter:2019}. It is also important to note that the nature of variable energy flux depends on the space dimensionality. In Table~\ref{tab:conclusions:table}, we summarise the  variable  energy fluxes discussed in this review. 

For random (white noise) initial condition, truncated Euler turbulence exhibits equilibrium behaviour with vanishing kinetic energy flux and nearly $ k^2 $ energy spectrum.  However, for orderly initial condition, such as Taylor-Green vortex, Euler turbulence yields a mix of nonequilibrium and equilibrium behaviour with a combination of $ k^{-5/3} $ and $ k^2 $ spectra. The $ k^{-5/3} $ regime shrinks at the expense of $ k^2 $ regime.   Dissipation-less MHD turbulence and passive-scalar turbulence also exhibit similar behaviour. These features of dissipation-less turbulence provide very  valuable insights for modelling  thermalization of complex systems. 

Thus, variable energy flux provides a unifying platform for modelling  many turbulent systems.    Note, however, that there are many unresolved issues in this framework, e.g., variable energy fluxes in anisotropic turbulence, quantum turbulence, binary-mixture turbulence, etc.  The present review does not cover many important turbulent flows, such as  electron magnetohydrodynamics~ \cite{Biskamp:PRL1996}, shell model of turbulence~ \cite{Plunian:PR2012}, compressible turbulence~ \cite{Banerjee:PRE2017}, weak turbulence~ \cite{Nazarenko:book:WT}, and plasma turbulence~ \cite{Teaca:NJP2017}, where variable energy fluxes are present.   We have also omitted discussions on the field-theoretic treatment of energy flux in turbulence.  In summary, variable energy flux is a very useful tool for understanding many turbulent systems.

\section*{Acknowledgments}

For writing this review I drew heavily from the discussions and idea exchanges I had with my collaborators, namely, Franck Plunian, Rodion Stepanov, Ravi Samtaney,  Daniele Carati, Stephan Fauve, Jai Sukhatme, Sanjay Puri, K. R. Sreenivasan,  Alexandros Alexakis,  Gaurav Dar,  Vinayak Eswaran, and Bernard Knaepen.   I am very grateful to them for the same.  In addition, I received critical inputs and ideas from many  past and present doctoral and master students---Abhishek Kumar, Shashwat Bhattacharya, Roshan Samuel,  Mohammad Anas, Shadab Alam, Soumyadeep Chatterjee, Pankaj Mishra, Satyajit Barman, Manohar Sharma,  Shubhadeep Sadhukhan,  Olivier Debliquy, Bogdan Teaca,   Thomas Lessinness, Valerii Titov, Ambrish Pandey, Sandeep Reddy, Anando Chatterjee, Arvind Ayyer, and V. Avinash.  I also thank J. K. Bhattacharjee, Peter Frick,  Annick Pouquet, Arnab Rai Choudhuri, Avinash Khare, P. K. Yeung, Diego Donzis, Xavier Albets, Itamar Procaccia, Maurice Rossi, Andrei Teimurazov, Andrei Sukhanovskii, Marc Brachet, Gregory Eyink, Luca Moriconi, Sagar Chakraborty, Supratik Banerjee, Sonakshi Sachdev, Akanksha Gupta, and Luca Biferale for useful discussions. I gratefully acknowledge the support of   Indo-French projects  4904-1 and 6104-1 from CEFIPRA, IFCAM project MA/IFCAM/19/90, Indo-Russian project  INT/RUS/ RSF/P-03 from Department of Science and Technology India, and SERB project SERB/F/3279/2013-14  that made the collaborative work and idea exchanges possible.  Some of the results presented in the review have been generated using SHAHEEN II of KAUST (project K1052) and  HPC2013 of IIT Kanpur.

\section*{References}

 \bibliographystyle{iopart-num}
 \bibliography{/Users/mkv/Dropbox/docs-pub/bib/journal,/Users/mkv/Dropbox/docs-pub/bib/book,/Users/mkv/Dropbox/docs-pub/bib/book_edited,/Users/mkv/Dropbox/docs-pub/bib/book_chapter,/Users/mkv/Dropbox/docs-pub/bib/conf,/Users/mkv/Dropbox/docs-pub/bib/preprint,/Users/mkv/Dropbox/docs-pub/bib/thesis,/Users/mkv/Dropbox/docs-pub/bib/report,/Users/mkv/Dropbox/docs-pub/bib/web} 
\end{document}